\documentclass[12pt]{article}
\usepackage{amsmath, amsfonts, bbm}
\usepackage{graphicx}
\usepackage{enumerate}
\usepackage{multicol}
\usepackage{multirow}
\usepackage[font=small,labelfont=bf]{caption}
\usepackage{subcaption}
\usepackage{color, bm}
\usepackage[linesnumbered,ruled,vlined]{algorithm2e}
\usepackage{array}
\usepackage{bbold}
\usepackage{ulem}
\usepackage{hyperref}
\usepackage{booktabs}
\usepackage{enumerate}
\usepackage{longtable}
\usepackage{setspace}
\usepackage{natbib}
\usepackage{geometry}
\setcitestyle{authoryear, open={(},close={)}}

\title{
Inferring a population composition from survey data with nonignorable nonresponse: Borrowing information from external sources
}
\author{\large Veronica Ballerini$^1$, Brunero Liseo$^2$}
\date{\normalsize $^1$ University of Florence - veronica.ballerini@unifi.it \\
$^2$ Sapienza University of Rome - brunero.liseo@unifi.it}

\begin{document}
\maketitle

\doublespacing

\begin{abstract}
We introduce a method to make inference on the composition of a heterogeneous population using survey data, accounting for the possibility that capture heterogeneity is related to key survey variables. 
To deal with nonignorable nonresponse, we combine different data sources and propose the use of Fisher's noncentral hypergeometric model in a Bayesian framework. 
To illustrate the potentialities of our methodology, we focus on a case study aimed at estimating the composition of the population of Italian graduates by their occupational status one year after graduating, stratifying by gender and degree program. 
We account for the possibility that surveys inquiring about the occupational status of new graduates may have response rates that depend on individuals' employment status, implying the nonignorability of the nonresponse.
Our findings show that employed people are generally more inclined to answer the questionnaire. 
Neglecting the nonresponse bias in such contexts might lead to overestimating the employment rate. 
\end{abstract}


\section{Introduction}
\label{sec:intro}
Many social or socioeconomic phenomena, such as voting intentions and opinions in general, can only be detected via surveys. 
However, different biases, such as social desirability and nonresponse biases, can likely affect survey data.
The \textit{social desirability bias} is caused by an individual's decision not to disclose sensitive or socially undesirable information and manifests itself in an incomplete or biased response; that is, the variables collected may be missing or affected by errors. 
For example, consider income surveys; individuals whose incomes belong to the extremes of the distribution, particularly the right tail, are more likely to lie, under or over-reporting their income, or fail to answer questions about it \citep{tourangeau2007sensitive, neri2023total}. 
Furthermore, consider questionnaires that investigate opinions on sensitive issues, both personal, such as sexuality, harassment, abortion \citep{peytchev2010measurement}, and social, such as intention to vote, immigration, and integration; in these cases, shame can cause individuals to provide answers that are considered socially more acceptable, but that differ from their genuine opinions.

Often correlated to measurement error due to desirability bias, there is a nonresponse issue \citep{neri2023total, tourangeau2010}.
\textit{Nonresponse} may or may not be related to the key variables of interest investigated by the survey; in the former case, nonresponse is said to be nonignorable, and the data are said to be missing ``not at random'' \citep[MNAR,][]{rubin1976, little2002statistical}. 
For instance, consider a survey designed to evaluate the effects of a public policy; it is genuine to expect a higher response rate among those who have benefited from that policy. 
Likewise, in a survey aimed at estimating how many young people find an occupation immediately after graduation, one may speculate that those who are promptly employed will be more likely to respond.
The latter example is the specific motivation for this work.

Nonignorable nonresponse in survey data is generally dealt with multiple imputation \citep{rubin1986multiple, glynn1993multiple, gelman1998not} and sometimes requires ad-hoc statistical solutions \citep[e.g.,][]{phipps2012analyzing,horton2014adjusting}.
Existing methodologies are mainly tailored to handling individual-level information \citep[see, e.g.,][]{ibrahim1996parameter,little2002statistical}, usually estimating an individual propensity to respond. 
In surveys, auxiliary data sources related to some marginal distribution can be used to address unit and/or item nonresponse or, more generally, to guide multiple imputation techniques.
Recent examples of this approach can be found in \cite{akande2021leveraging} and \cite{tang2024using}.

The problem of nonignorable nonresponse in the presence of aggregated data has been less debated.
This paper aims to provide a method to make inference on the composition of a population using aggregate survey data in the presence of nonignorable nonresponse.
Our motivation comes from the need to correct nonresponse bias in Almalaurea surveys\footnote{\url{https://www.almalaurea.it/en/our-data/almalaurea-surveys/graduates-employment-status}}, which are surveys inquiring about the occupational status of people who have recently graduated. 
Our aim is to estimate the size of the Italian employed and unemployed graduates by gender and degree program. 
We exploit genuine extra-experimental information from administrative data and provide estimates for different cohorts of graduates, starting with people who achieved their degrees in 2011.
We assume that the decision not to disclose their occupation status leads individuals to not respond to the questionnaire rather than to lie; namely, individuals' responses are not affected by social desirability bias.

To achieve our goal, we exploit the underused Fisher's noncentral hypergeometric (FNCH) distribution from a Bayesian perspective, which allows us to easily combine information from different sources.
FNCH describes a biased urn problem: some colored balls are independently drawn from an urn, and the probability of extracting a specific ball depends not only on the total number of balls of each color but also on the relative odds, or weights, of the colors.
Assume that a sample survey partially enumerates a heterogeneous population. 
The coverage probabilities may vary among the sub-groups, which is equivalent to observing different colored balls drawn according to their weights.
Despite its strong adaptability, such distribution is not popular in survey statistics.
While it is relatively easy to generate samples from a FNCH distribution, the main reason behind its poor spread is probably the computational burden given by its probability mass function \citep{fog2008, liao2001fast}; as a consequence, it has been mainly used as a tool for implementing simulation-based methods, like for example permutation tests \citep{epstein2012}. 
See also \cite{fisher, agresti1992survey} for the analysis of 2 by 2 contingency tables.
Yet, we are not aware of likelihood-based or Bayesian approaches to inference about its parameters. 

This article rediscovers FNCH distribution, making it a suitable model to estimate the population composition leveraging biased samples.
Here, we adopt a Bayesian perspective and, exploiting Markov chain Monte Carlo (MCMC) methods, we can overcome the computational issues and make the methodology easily accessible to the final user.
Furthermore, we believe that the Bayesian approach is probably the most natural one to deal with data integration \citep[among others, see][]{wisniowski2020integrating, sakshaug2019supplementing,schifeling2019data}.

In the next section, we present the case study that motivates our work and describe the data set. 
Section \ref{sec:problemform} details the methodology by introducing the FNCH statistical model and proposing a Bayesian approach to inference.
Section \ref{sec:results} illustrates the results; the discussion follows.

\section{The case study: Italian graduates} \label{sec:casestudy}
\subsection{Key question and main data sources} \label{datasources1}
People enrolled in the Italian university system are tracked in the Students' National Register\footnote{Anagrafe Nazionale Studenti, URL: \url{http://ustat.miur.it/}} (SNR) of the Ministry of University and Research. 
Every year, the SNR provides data on the number of graduates.
We aim to estimate the composition of the population of newly graduated in terms of their occupational status.

To this aim, we investigate the possibility of relying on survey data, which collects information on individuals who recently graduated.
In particular, we look at the data collected by
the Italian Inter-university Consortium ``Almalaurea'', which yearly conducts the Graduates' Employment Status Survey and sends a questionnaire to all individuals who graduated from Italian Universities in the previous solar year.
The Almalaurea survey collects information on the employment condition of the interviewed via CAWI (Computer-Assisted Web Interview) and CATI (Computer-Assisted Telephone Interview) methodologies.
The data are integrated with the universities' administrative archives involved in the investigation, providing additional information such as gender, date of birth, and degree program.

The Almalaurea survey response rate is never $100\%$; we speculate that the propensity to participate in the survey for purely statistical purposes might differ between those employed and those who have not found a job yet.
For instance, an unemployed person may be less likely to fill out a questionnaire about their employment condition; in such a case, nonresponse would be nonignorable. 
We define ``nonrespondents'' as those who do not return the Almalaurea survey. 
The share of those who respond but for whom the value of occupational status is missing can be considered negligible.

\begin{figure}
    \centering
    \includegraphics[width=.8\linewidth]{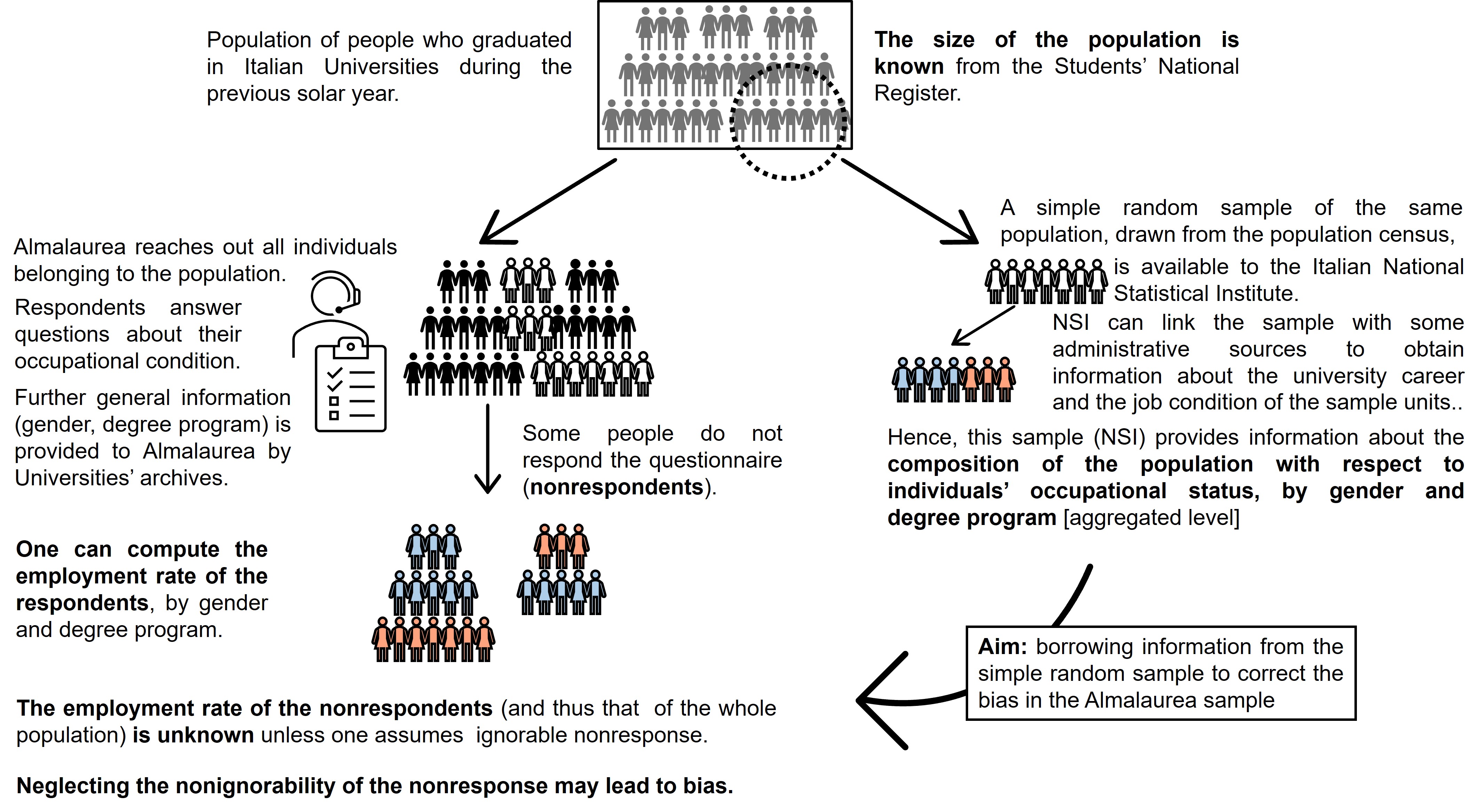}
    \caption{Key question at a glance.}
    \label{fig:sketch1}
\end{figure}

We aim to test the equality between the response rates of employed and unemployed individuals. 
Without borrowing additional information, inference would not be possible in this case.
Indeed, although we observe the number of employed and unemployed respondents, we cannot compare their response rates without information on the corresponding population sizes.
Hence, we need to exploit external information. 
Such a proposal is coherent with the current approach of National Statistical Institutes, which are moving towards a statistics production based on integrated sources.
Figure \ref{fig:sketch1} sketches the problem and anticipates the strategy described in the next paragraphs.

\subsection{Borrowing information from external data sources} \label{datasources2}
The last decennial census of the Italian National Statistical Institute (NSI, henceforth) dates back to 2011. 
It collected a broad set of information about individuals residing in Italy.
The NSI can link the census data with data coming from several administrative sources: beyond the Students' National Register mentioned above, further information comes from administrative lists of the Ministry of Economics and Finance, the National Institute for Insurance against Accidents at Work, and the National Social Security Institute.
Linking data with the SNR enriches individual information by including variables about the academic degree.
Furthermore, integrating data from other administrative sources makes individual employment information available to the NSI unless the individual emigrates or does not have a regular contract.
Table \ref{tablaureati} lists the variables available to the NSI.
\begin{table}[t]
\caption{Classification variables for the NSI sample and their original source among the administrative registers of the other Italian institutions.}
\begin{center}
\footnotesize
\begin{tabular}{ll}
\toprule
Variables  &  Original source \\ 
\midrule
$\cdot$ Anagraphic data & $\cdot$ 2011 census by NSI\\
$\cdot$ Degree's classification & $\cdot$ SNR\\ 
$\cdot$ Degree's class & $\cdot$ SNR \\
$\cdot$ Degree's achievement date  & $\cdot$ SNR  \\ 
\begin{tabular}[c]{@{}l@{}}
     $\cdot$ Starting date of a job contract or \\
     opening date of a VAT number \\ \textcolor{white}{white row} \end{tabular}  &  \begin{tabular}[c]{@{}l@{}} $\cdot$ Ministry of Economics and Finance and/or \\ $\color{white}\cdot$ National Institute for Insurance against Accidents at Work and/or \\ $\color{white}\cdot$ National Social Security Institute \end{tabular}\\      \bottomrule
\end{tabular}
\end{center}
\label{tablaureati}
\end{table}

NSI does not disclose individual data. 
According to its policy, a small simple random sample (about 3.5\%) of such census data can be publicly available at an aggregated level. 
The integration of the census with the administrative sources allows the data to be aggregated by gender, degree program, and occupational status. 
This way, we can match the information contained in the Almalaurea sample.

We exploit aggregated NSI information (hereafter referred to as the ``NSI sample'') to estimate the total number of 2011 graduates employed one year after graduation. 
If such an NSI sample were available yearly, one would obtain unbiased estimates of the employment rates. 
Yet the sample is available for 2011 only.
Hence, we exploit the 2011 NSI data to estimate the bias, namely the difference in the response rates, in the 2012 Almalaurea survey \citep[the 2012 survey refers to the 2011 cohort of graduates;][]{almalaurea2012}.

We make the following adjustments and assumptions to make the populations targeted by the two sources consistent.
First, we consider the Almalaurea interviewed who declared to work without a regular contract as unemployed. 
Then, since the survey did not provide any information about the geographical area where the respondents worked until the 2015 wave, we do not have the chance to detect the number of emigrated workers among the 2011 cohort.
Hence, we assume the percentage of new graduates who decide to work abroad within the first year after graduation to be negligible.
Finally, we address the issue of possible time lags. 
Indeed, the survey procedure envisaged three reminders; 
then, those who did not respond to the online questionnaire were contacted by telephone. 
To avoid a bias due to the possible time lag, we adopt a conservative approach and consider all those in the Istat sample who started a job within the first 18 months after graduation as employed within one year.

\subsection{Target population} \label{targetpop}
We include in the analysis only those individuals who achieved a degree of the so-called \textit{Nuovo Ordinamento}, i.e., programs in effect after the Bologna Process in 1999. 
The Bologna process has been a reform process aimed at unifying the European higher education systems. 
One of the significant reformations for the Italian programs consisted of the adoption of the ``3+2'' system:
four to six years programs split into 3-year and 2-year single programs, i.e., \textit{corso di Laurea Triennale} and \textit{corso di Laurea Specialistica/Magistrale}, the equivalent of a bachelor's and a master's degree, respectively.
There are some exceptions; the so-called \textit{Laurea Specialistica/Magistrale a ciclo unico} (literally ``single-cycle Master's degree'') preserved their duration (e.g., Medicine, Law).

Despite the formal adjustment, in Italy, many occupations are rarely held by those with a bachelor's degree.
Since we are interested in the employment level of those who have concluded their education path and are ready for the labor market, we exclude from the sample those who achieved a bachelor's degree in 2011.

Excluding bachelor's programs, the initial number of degree classes, i.e., 309, reduces to 213.
We organize the classes into programs, mainly according to a classification made by the Ministry of University and Research in 2020, disaggregating the groups when too heterogeneous; e.g., we split ``Political Science, Sociology, and Communication'' into ``Political Science,'' ``Sociology, and Anthropology,'' and ``Communication and Publishing.''

\subsection{Final dataset}
The NSI sample consists of $n^{\text{NSI}} = 2717$ individuals, while the Almalaurea survey records $n = 53015$ interviews for the same cohort of graduates. 
The total number of Master's students who graduated in 2011 is $N = 87649$.

\begin{table}
\footnotesize
\centering
\caption{Percentages of people who obtained \textit{Laurea Specialistica/Magistrale (a ciclo unico)}, namely a Master's degree, in Italian universities in 2011 and got employed within a year after their graduation, by gender and degree program, according to the NSI and Almalaurea samples.
Columns 2-3 and 7-8 report the percentages in the two samples for males and females, respectively. 
The larger the difference between the NSI and the Almalaurea percentages, the larger the expected bias.
Columns 4-6 report the sizes of the NSI and Almalaurea samples and the Students' National Register (SNR) for males; columns 9-11 are the respective for females.}
\label{tab:dati}
\begin{tabular}{@{}lccccc|ccccc@{}}
\toprule
\multirow{3}{*}{}                              & \multicolumn{5}{c|}{M}        & \multicolumn{5}{c}{F}        \\ \cmidrule(l){2-11} 
 &
  \multicolumn{2}{c}{Employed (\%)} &
  \multicolumn{3}{c|}{Total} &
  \multicolumn{2}{c}{Employed (\%)} &
  \multicolumn{3}{c}{Total} \\ \cmidrule(l){2-11} 
 &
  \multicolumn{1}{l}{NSI} &
  \multicolumn{1}{l}{Almal.} &
  \multicolumn{1}{l}{NSI} &
  \multicolumn{1}{l}{Almal.} & 
  \multicolumn{1}{l|}{SNR} &
  \multicolumn{1}{l}{NSI} &
  \multicolumn{1}{l}{Almal.} &
  \multicolumn{1}{l}{NSI} &
  \multicolumn{1}{l}{Almal.}&
  \multicolumn{1}{l}{SNR} \\ \midrule
Agricultural and Forestry sciences & \textit{48.6} & \textit{51.7} & 37  & 810 & 1189 & \textit{42.1} & \textit{42.6} & 38  & 841 & 1215\\
Architecture and Eng.                   & \textit{35.6}& \textit{51.0}  & 87  & 949 & 2528 & \textit{27.5} & \textit{42.8} & 91  & 1224 & 2956\\
B\&A, Economics, Finance                      & \textit{36.7} & \textit{55.34} & 237 & 4051 & 7384 & \textit{33.1} & \textit{52.5} & 266 & 4958 & 8232\\
Communication and Publishing                   & \textit{45.5} & \textit{58.8}  & 33  & 799 & 1220 & \textit{28.0}& \textit{54.3} & 75  & 1246 & 2412\\
Industrial and Information Eng.         & \textit{47.8} & \textit{72.4}  & 251 & 4794& 7417& \textit{44.2} & \textit{62.7} & 52  & 1020 & 1629\\
Law and Legal Sciences                         & \textit{33.3} & \textit{20.9}  & 165 & 3007& 6734 & \textit{18.5}& \textit{15.3} & 270 & 5352 & 9120\\
Literature and Humanities                      & \textit{39.3} & \textit{42.5}  & 56  & 1099& 1809& \textit{52.5} & \textit{41.6} & 120 & 2462 & 3720\\
Medicine, Dentistry, Pharmacy                  & \textit{47.2} & \textit{50.3}  & 144 & 3372& 5108 & \textit{44.9} & \textit{50.7} & 321 & 6582 & 9438\\
Political Science                              & \textit{56.0}  & \textit{56.5}  & 75  & 1379 & 2364 & \textit{30.4} & \textit{45.9} & 79  & 1755 & 2669\\
Science and IT                                 & \textit{33.1} & \textit{47.0} & 127 & 2982& 4487 & \textit{34.7} & \textit{36.6} & 193 & 4333 & 6018\\ \midrule 
Total & & & 1212 & 23242 & 40240 & & &1505 & 29773 & 47409
\\ \bottomrule
\end{tabular}
\end{table}

Table \ref{tab:dati} shows the percentages of units recorded as employed one year after graduation by the two sources, classified by gender and macro classification of the degree programs. 
The Table also reports the total sample sizes and the total number of graduates recorded by the Students' National Register (SNR).

Let the number of employed graduates recorded by the NSI and Almalaurea samples be realizations $\{x_{hie}\},\{y_{hie}\}$ of the random variables $\{X_{hie}\},\{Y_{hie}\}$, respectively, with $h = \text{M, F}$ and $i$ denoting the degree's program. 
Similarly, we denote with $\{x_{hiu}\},\{y_{hiu}\}$ the number of not yet employed graduates in the two samples.
To lighten the notation, we will discard the discipline's and gender's subscripts hereafter.
We assume
\begin{equation} \label{firsteq}
\begin{split}
    X_{j} & \sim \text{Binom}(M_{j}, \zeta^{\text{NSI}}_{j}) \\
    Y_{j} & \sim \text{Binom}(M_{j}, \zeta_{j})\; , \quad \quad j = e, u
\end{split}
\end{equation}
where $M_{j}$ is the total number of 2011 graduates who are in the $j^{th}$ employment condition one year after, $j = e, u$; $\zeta_{j}^{\text{NSI}}$ and $\zeta_{j}$ are the capture probabilities in the NSI sample and the Almalaurea survey addressing the 2011 cohort of graduates, respectively.
An implicit assumption in the Binomial specification is that units belonging to the same group share the same probability of being listed in a specific source. 

In the next section, we will argue that $Y_{e} \mid Y_{e} + Y_{u} = n$ (and also $X_{e} \mid X_{e} + X_{u} = n^{\text{NSI}}$) follows a known distribution, namely Fisher's noncentral hypergeometric distribution (FNCH).
Such a distribution depends on a \textit{weight} parameter, which can be expressed in terms of the odds ratio
 
\begin{equation}\label{weight_def}
   w = \dfrac{{\zeta_{e}}/{(1-\zeta_{e})}}{{\zeta_{u}}/{(1-\zeta_{u})}} \;.
\end{equation}
Our aim is to test the hypothesis of ignorable nonresponse in the Almalaurea survey, i.e., $H_0: w = 1$. However, this is not possible by leveraging the Almalaurea dataset alone. 
In the next section, we show how to make inference on the FNCH parameters by exploiting extra experimental information from the NSI sample.
\section{Model setting}\label{sec:problemform}
\subsection{Fisher's noncentral hypergeometric distribution to infer a population composition} \label{sec:fisher}
In 2008, Agner Fog clarified the distinction between two distributions, both known in the literature as ``the'' noncentral hypergeometric distribution \citep{fog2008appendix, fog2008}. 
He solved the nomenclature issue, naming them \textit{Wallenius'} and \textit{Fisher's}, after the persons who first proposed them \citep{fisher,wallenius}.
The main difference between the two distributions resides in the dependence structure of the draws.
Assume an urn of size $N$ contains $M_c$ balls of color $c, c = 1, \dots, C$, with $N = \sum\limits_{c}{M_c}$.
Wallenius' noncentral hypergeometric distribution describes a situation in which the balls are drawn without replacement until a prespecified number $n$ of balls are sampled, and the probability of sampling $Y_c$ balls of color $c$ depends on the colors' relative weights. 
It is said to describe a biased urn experiment since the weight associated with each color can be seen as the probability of retaining a ball of that color when drawn \citep[as suggested by][]{chesson}. 

Instead, Fisher's noncentral hypergeometric distribution describes an urn experiment where the balls are drawn independently, without replacement, and without fixing $n$ in advance.
It can be seen as the conditional distribution of independent Binomial distributions given their sum \citep{mccullagh1989generalized, harkness}.
For each group $c, \; c = 1, ..., C$, assume
\begin{equation}
    Y_c \sim \text{Binom}(M_c, \zeta_c) \;
\end{equation}
and denote with $w_c$ the odds ${\zeta_c}/{(1-\zeta_c)}, \, \forall c$.
Hence, conditional on the elements' sum, the vector $\bm{Y} = (Y_1, \dots, Y_C)$ is distributed as a multivariate FNCH with parameters $\bm{M} = (M_1, ..., M_C)$, $n$ and $\bm{w} = (w_1, ..., w_C)$ and the probability mass function is
\begin{equation} \label{multifisher}
    P\left(\bm{Y} = \bm{y}|\sum\limits_{c=1}^{C}{Y_c}=n\right) = \dfrac{\prod\limits_{c=1}^{C}{\binom{M_c}{y_c}{w_c}^{y_c}}}{\sum\limits_{\bm{z}\in \mathcal{Z}}{\prod\limits_{c=1}^{C}{\binom{M_c}{z_c}{w_c}^{z_c}}}}
\end{equation}
where 
\begin{equation} \label{2pmfmvF}
    \mathcal{Z} = \left\{\bm{y} \in {\mathbb{N}_0}^C : \bigg[\sum\limits_{c=1}^{C}{y_c} = n \bigg] ~ \cap ~ \bigg[0 \leq y_c \leq M_c\bigg], \forall c \right\}
\end{equation}
\citep{fog2008}. 
The weights $w_c$ are defined up to a positive constant $k$; then, FNCH distribution is identified by the odds ratio $w_{cc'} = {w_{c}}/{w_{c'}}$.
Note that the sum at the denominator of \eqref{multifisher} makes evaluating the likelihood challenging, especially as $N$ and the number of different categories in the population increase.

To our knowledge, FNCH has not been used in survey statistics to handle nonresponse or quantify uncaptured population units. 
However, it would be natural to think $\bm{Y}$ as the vector of numbers of units belonging to $C$ different groups or cells captured in a list of size $n$. 
Hence, $\bm{M}$ would be the vector of groups' total sizes, or total sampled cases in the $C$ cells, and $w_c$ would inform about the \textit{exposure} of the group $c$ in the sampling with respect to a reference category $c'$.

In the following subsection, we introduce a Bayesian approach to inference for FNCH. 
For the sake of simplicity, we describe only the univariate model, which is suitable for our case study. 
In the supplementary material, we provide details on the multivariate FNCH, which is useful for multicategorical and/or compositional data.

\subsection{Bayesian inference for the univariate FNCH}\label{sec:bayesinf}
Following equation \eqref{multifisher}, when $C = 2$ the probability mass functions simplifies into
\begin{equation}
    P\left(Y_1=y_1|Y_1+Y_2=n\right) = \dfrac{\binom{M_1}{y_1}\binom{M_2}{n-y_1}{w}^{y_1}}{\sum\limits_{z\in \mathcal{Z}}{\binom{M_1}{z}\binom{M_2}{n-z}{w}^{z}}}
\end{equation}
with $\mathcal{Z}$ given by \eqref{2pmfmvF} when $C=2$.
Then,
\begin{equation}\label{formula8}
    Y_{1}|Y_{1}+Y_{2}=n \sim \text{FNCH}(M_{1},M_{2},n,w)\; ;
\end{equation}
note that this is exactly the situation described at the end of Section \ref{sec:casestudy}.
Since $M_2 = N - M_1$ and $y_2 = n - y_1$, the formulation above is equivalent to:
\begin{equation}
    Y_1 | n \sim \text{FNCH}(M_1, N, n, w) \; .
\end{equation}
We will interchangeably use the two parameterizations throughout this work.
All parameters $M_1$, $N$, and $w$ may be unknown quantities; under a Bayesian approach, we need to elicit a prior distribution $\pi(M_1,N,w) = \pi(M_1\mid N,w)\pi(N\mid w)\pi(w)$. 
In the survey statistics framework, it is sensible to assume that the relative exposure in the sampling is independent of the groups' sizes, i.e., $w \perp \!\!\! \perp M_1,M_2$.
Hence, $\pi(M_1,N,w) = \pi(M_1\mid N)\pi(N)\pi(w)$. 
We generally write
\begin{equation}
    M_{1}|N \sim \pi(\cdot; \bm{\theta}^{M_{1}}, N)\;,
\quad    N \sim \pi(\cdot; \bm{\theta}^{N})\;,
\quad    w \sim \pi(\cdot; \bm{\theta}^{w})\;,
\end{equation}
where $\pi(\cdot)$ denotes a generic distribution depending on some parameters $\bm{\theta}^{*}$. 

The model is not identifiable unless we include genuine prior information. 
For instance, one may have some prior information on one of the groups; such a situation is common when dealing with administrative data. 
Indeed, consider a sample of resident (group 1) and non-resident (group 2) persons living in a city; given the reliable information contained in the municipal registries, including genuine prior information on $M_1$ would be legitimate.
Alternatively, consider employed (group 1) and yet not employed (group 2) young graduates whose respective sizes are unknown, as in the case study object of this work. 
National registers generally provide the annual total number of graduates, $N$; thus, the associated error can be assumed negligible. 
We would subjectively elicit a concentrated prior distribution for $N$ in such a case.
Subjective elicitation is a debated issue since the attribute ``subjective'' is often perceived as including personal beliefs in a negative sense. 
Instead, we consider the elicitation process a rational way to incorporate experts' knowledge and take advantage of their experience; for a deep and detailed discussion about the probabilities' elicitation process, see \citeauthor{berger} (\citeyear{berger}, Ch. 3) and \cite{ohagan}.

Let us denote the likelihood function with $L(y;M_{1},N,n,w)$.
Hence, the joint posterior distribution is
\begin{equation} \label{posterior}
    \pi(M_{1},N,w|y_{1},n) \propto L(y;M_{1},N,n,w)\pi(M_{1}|N)\pi(N)\pi(w) \; ;
\end{equation}
it can be easily computed via MCMC methods, e.g., using a Metropolis-within-Gibbs algorithm.
\section{Analysis and results}\label{sec:results}
Our estimation procedure is divided into three steps: 
(i) the estimation of the number of graduates who were employed one year after their graduation among the 2011 cohort, exploiting the NSI sample and the National Students' Register values; 
(ii) the estimation of the propensity to participate in the 2012 Almalaurea survey exploiting the results at step (i); 
(iii) the correction of the employment rates of the new graduates from the 2012 to 2020 cohorts (according to the available data), assuming the response rate remains constant over the years.

Note that the steps described in this Section are performed stratifying by gender and degree program; for the sake of brevity, here we show results for some of the categories.
The interested reader can find results for all degree programs in the supplementary material.

\subsection{Modeling details}
Using Equation \eqref{formula8}, we say that the number of employed individuals captured by a list, conditionally on the total number of individuals captured by that list, is Fisher's noncentral hypergeometrically distributed:
\begin{equation}\label{refmod_NSI}
        X_{e} \mid X_{e} + X_{u} = n^{\text{NSI}} \sim \text{FNCH}(M_{e},M_{u},n^{\text{NSI}},w^{\text{NSI}}) \;,
\end{equation}
\begin{equation}\label{refmod}
        Y_{e} \mid Y_{e} + Y_{u} = n  \sim \text{FNCH}(M_{e},M_{u},n,w) \; .
\end{equation}
As introduced in Section \ref{sec:casestudy}, $X_e$ ($X_u$) is the number of employed (unemployed) people among the $n^{\text{NSI}}$ graduates captured by the NSI sample. 
Similarly, $Y_e$ ($Y_u$) is the number of employed (unemployed) among the $n$ captured by the Almalaurea survey addressing the same cohort.
Then, $M_{e}$ ($M_{u}$) is the total number of employed (unemployed) graduates. 
Finally, $w^\text{NSI}, w$ may be interpreted as the \textit{bias} in the NSI sample and the Almalaurea survey, as defined in Section \ref{sec:fisher}.

\paragraph{Step (i)}
Here, we focus on the 2011 cohort of graduates. 
We can leverage the simple random sample drawn from the census; thanks to the auxiliary information integrated from the administrative registers, we know the proportion of employed people in that sample. 
We also have prior information about the total number of graduates in Italy that year from the National Students' Register. 
Hence, it can be dealt with using a hypergeometric model. In this model, we have strong prior information on the size of the urn and want to estimate its composition; we adopt a Bayesian to account for residual uncertainty.

In step (i), the model setting is that in Equation \eqref{refmod_NSI}; we refer to the NSI sample.
We need to elicit the joint prior $\pi(M_{e},N,w^{\text{NSI}}) = \pi(M_{e}\mid N)\pi(N)\pi(w^{\text{NSI}})$.

We use a discrete uniform distribution for $M_{e}\mid N$:
\begin{equation}
    M_{e} \mid N \sim \text{Unif}(a_{M_{e}} = x_{e} + 1, b_{M_{e}} = N - x_{u} - 1) \;.
\end{equation}

Given the accuracy of the Students' National Register (SNR) data, we assume that $N$ follows the following left-truncated Poisson distribution:
\begin{equation}
    N \sim \text{ltruncPois}(N^{\text{SNR}}, a_{N})
\end{equation}
where the mean parameter $N^{\text{SNR}}$ is the SNR value, and the lower bound $a_N = (x_{e} + 1) + (x_{u} + 1)$.

The NSI sample is a simple random sample; 
it amounts to assuming that the inclusion probability in the NSI sample is independent of the occupational status; thus, $\zeta^{\text{NSI}}_{j} = \zeta^{\text{NSI}}$.
Consequently, $w^{\text{NSI}} = 1$, i.e., we assume a degenerate prior for $w^{\text{NSI}}$. 
This way, FNCH turns out to be a hypergeometric distribution.

The final goal of this first step is to estimate the posterior $\pi(M_{e}, N \mid x_{e}, x_{u})$.

\begin{figure}[!t]
    \centering
\subfloat[][Agriculture and Forestry, Veterinary]{
    \includegraphics[width=.5\linewidth]{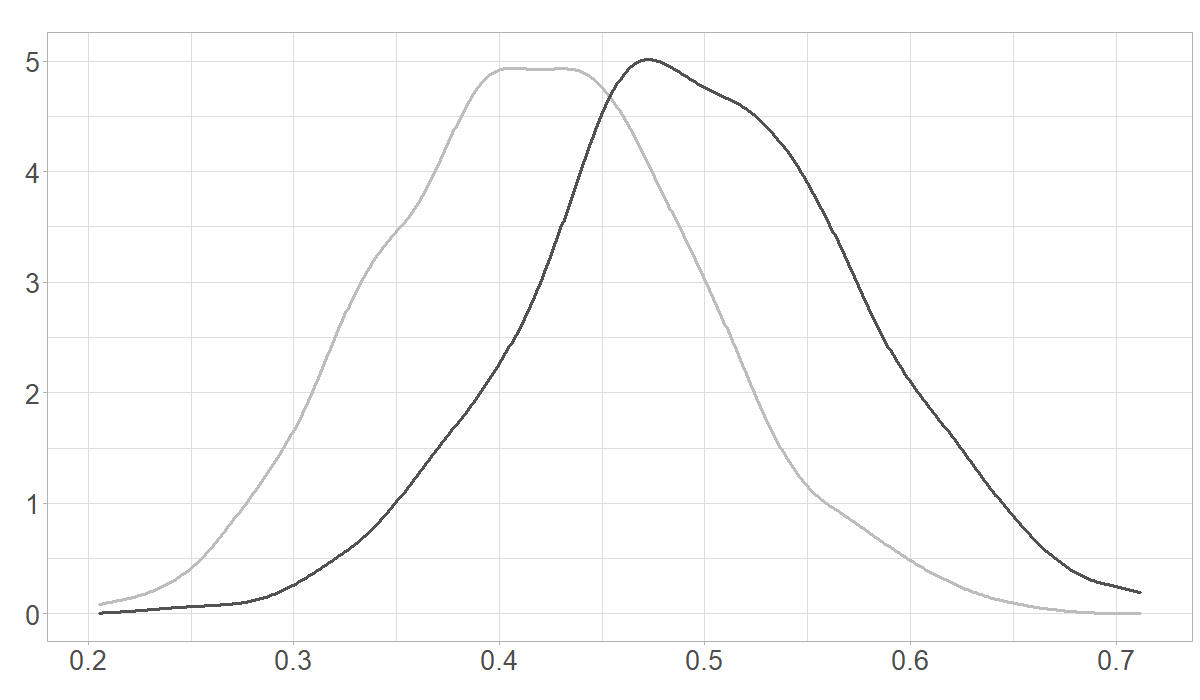}}
\subfloat[][Architecture and Engineering]{
    \includegraphics[width=.5\linewidth]{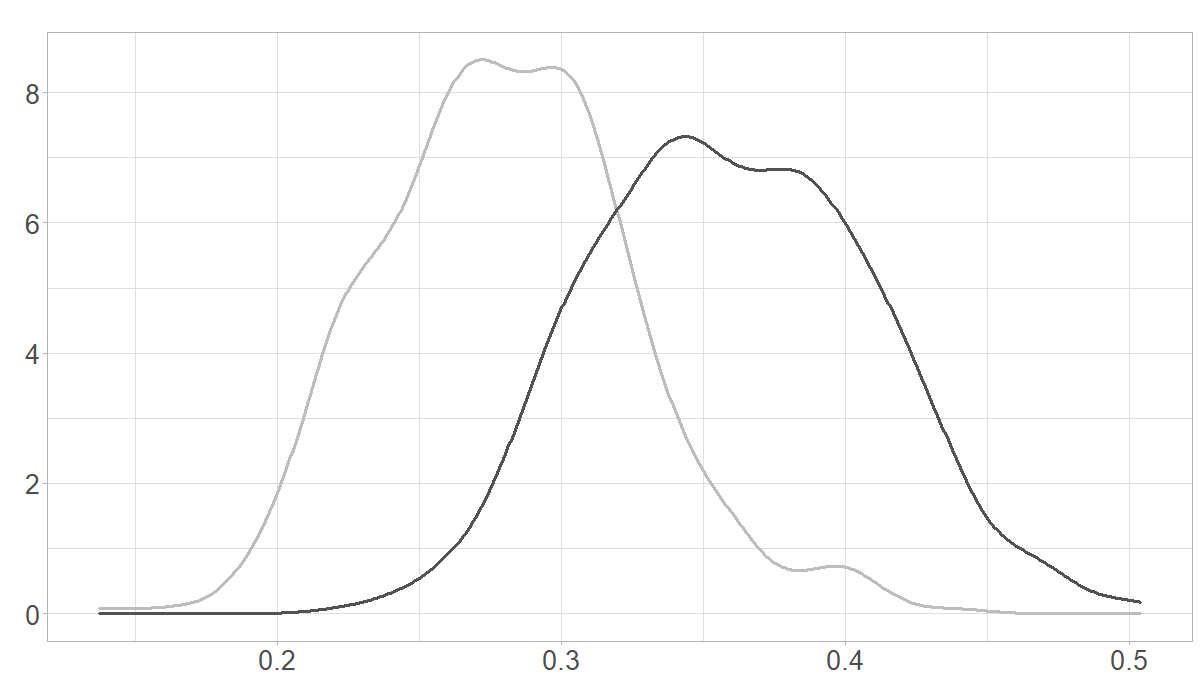}}
\qquad
\subfloat[][Law and Legal sciences]{
    \includegraphics[width=.5\linewidth]{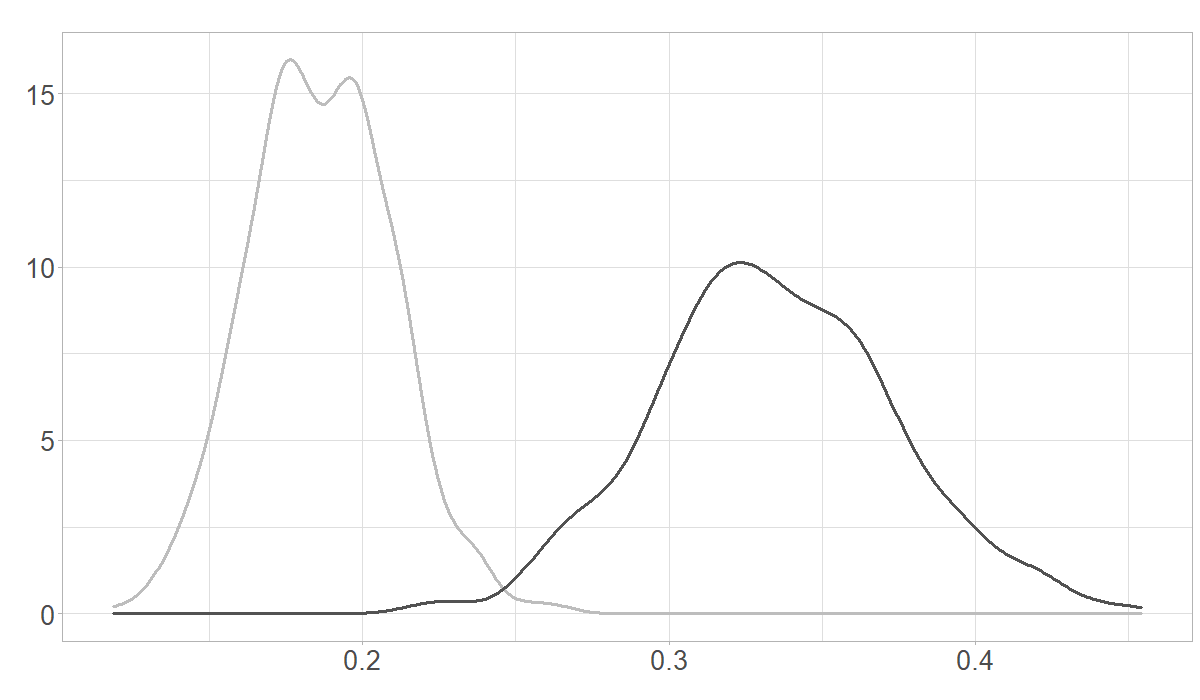}}
\subfloat[][Medicine, Dentistry, Pharmacy]{
    \includegraphics[width=.5\linewidth]{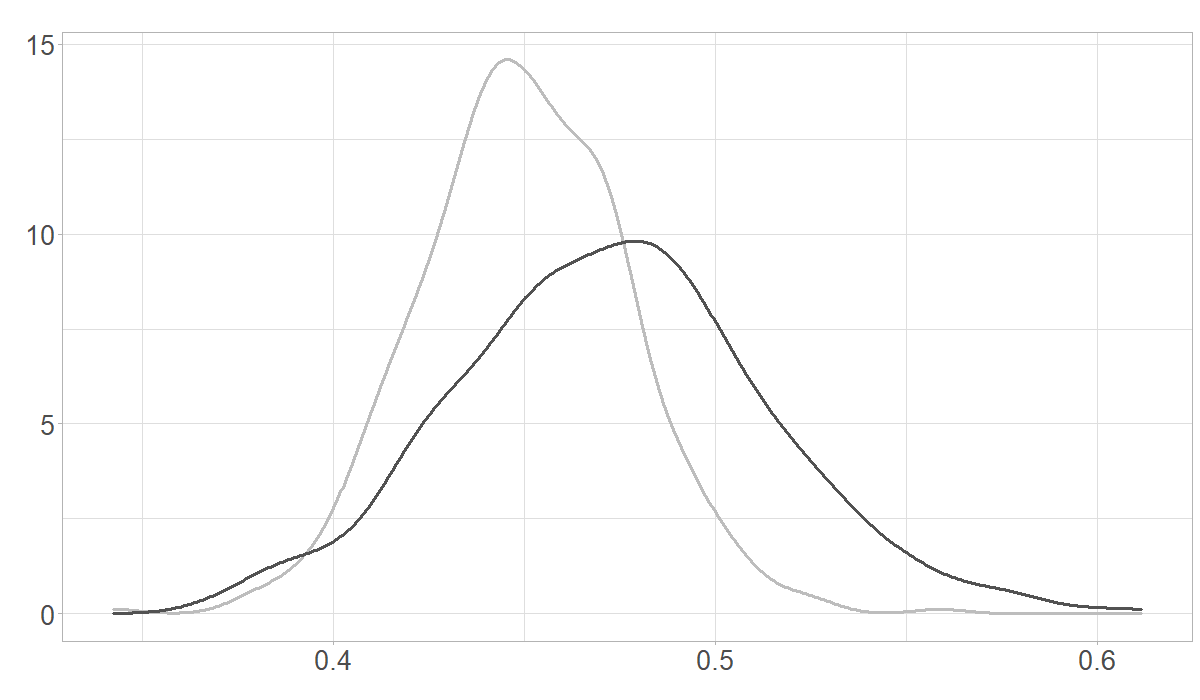}}
\caption{Posterior distributions of $\{M^{11}\}$, i.e., the sizes of 2011 graduates who were employed one year after their graduation, divided by $N^{11}$ posterior mean estimated at step (i), by gender (grey: females, dark grey: males) and degree programs, obtained using the 2011 NSI sample.}
\label{fig:step1}
\end{figure}

\paragraph{Step (ii)}
In the previous step, we derived the joint posterior distribution of the composition of the 2011 graduates population. 
Now, we can estimate the response bias of the Almalaurea survey.

Our model setting is that in Equation \eqref{refmod}; now, we refer to the Almalaurea survey data for the 2011 cohort.
Similarly to the previous step, we must elicit $\pi(M_{e},N,w) = \pi(M_e, N)\pi(w)$.

For $(M_e, N)$, we use a bivariate Normal distribution whose hyperparameters are set equal to the posterior values derived in the previous step.
Concerning the ``exposure'' of the employed in the Almalaurea survey, 
we use a weakly informative prior for the log odds ratios $\log(w) = \log({w_{e}}/{w_{u}})$, i.e., $\log(w) \sim \text{N}(\mu,\tau)$. 
Using a symmetric prior (on the log scale) seems to be a sensible noninformative choice, independent of our speculations about the expected sign of the bias.
In our implementation, we fix $\mu = 0$, which reasonably implies setting the a priori odds median equal to $1$, and we test the sensitivity of the results to different values of $\tau$.

The final goal of step (ii) is to estimate the marginal posterior $\pi(w\mid y_e, y_u)$.

\paragraph{Step (iii)}
In the previous step, we estimated the response bias in the 2012 Almalaurea survey, which addressed the 2011 cohort one year after they graduated. 
It is now possible to adjust the employment rate.
However, we want to move a step forward: assuming that the response bias is constant over time (but different among genders and degree programs), we adjust the employment rate series until 2020.

Our model setting is still that in Equation \eqref{refmod}.
We leverage $\pi(w\mid y_e, y_u)$ drawn in the previous step to elicit a prior distribution for $w^{t}, t = 2012, \dots, 2020$. 
We opt for a Normal approximation of the posterior sample drawn in step (ii).

Finally, we use a discrete uniform distribution for $M^{t}_{e}, t = 2012, \dots, 2020$: 
\begin{equation}
    M^{t}_{e} \mid N^{t} \sim \text{Unif}(y_{e}^{t} + 1, N^{\text{SNR},t} - y_{u}^{t} - 1) \;.
\end{equation}

At this step, we consider $N^{t}$ as known and equal to the SNR record for that year.

\subsection{Results for the response bias in the 2011 cohort}
Once the sizes of employed and unemployed graduates by degree programs and gender are estimated (see Figure \ref{fig:step1}) as described in the previous Section, step (i), we can estimate the relative exposure of employed people in the Almalaurea survey for the 2011 cohort, i.e., $w$.
This quantity is informative on whether the employed people are more likely to respond to the questionnaire than the unemployed.
Figure \ref{fig:step2} shows the posterior distribution of $w$ by gender for some degree programs; computations were made setting $\tau=1$. 
Results for the other disciplines are available in the supplementary material; see Section \ref{sec:sens} for further discussion on sensitivity to prior assumptions.

With a few exceptions (see, e.g., ``Agricultural and Forestry sciences, Veterinary'' in Figure \ref{fig:step2}), employed and unemployed people are generally far from being equally exposed.
As suspected, employed people almost always have a higher propensity to answer the questionnaire, namely, $w>1$.

However, some degree programs show the opposite behavior. 
For instance, we estimate a higher exposure of the \textit{unemployed} for 
who graduated in ``Law and Legal Sciences.''
An intuitive explanation for the latter degree program could be related to the ``practicum,'' a practice the would-be attorneys must go through.
Although the practicum is often unpaid, and the trainees do not fall into the employed category, it is so standard that new graduates may not fear declaring themselves practitioners.

\begin{figure}[!b]
    \centering
\subfloat[][Agriculture and Forestry, Veterinary]{
    \includegraphics[width=.5\linewidth]{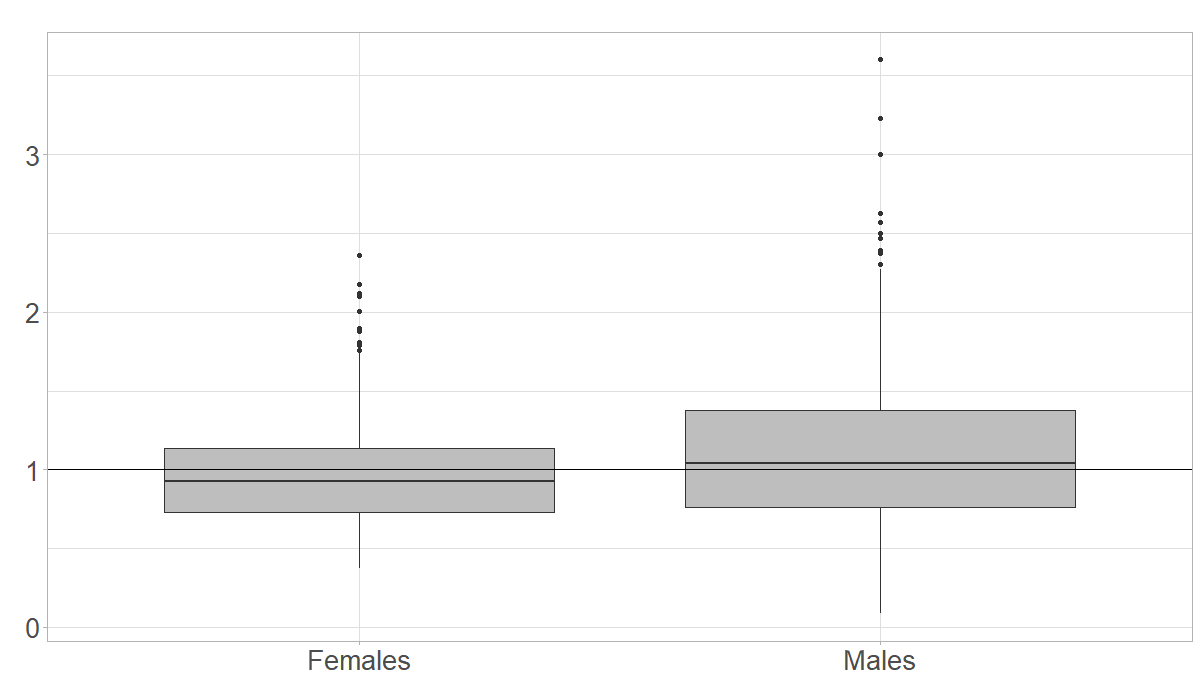}}
\subfloat[][Architecture and Engineering]{
    \includegraphics[width=.5\linewidth]{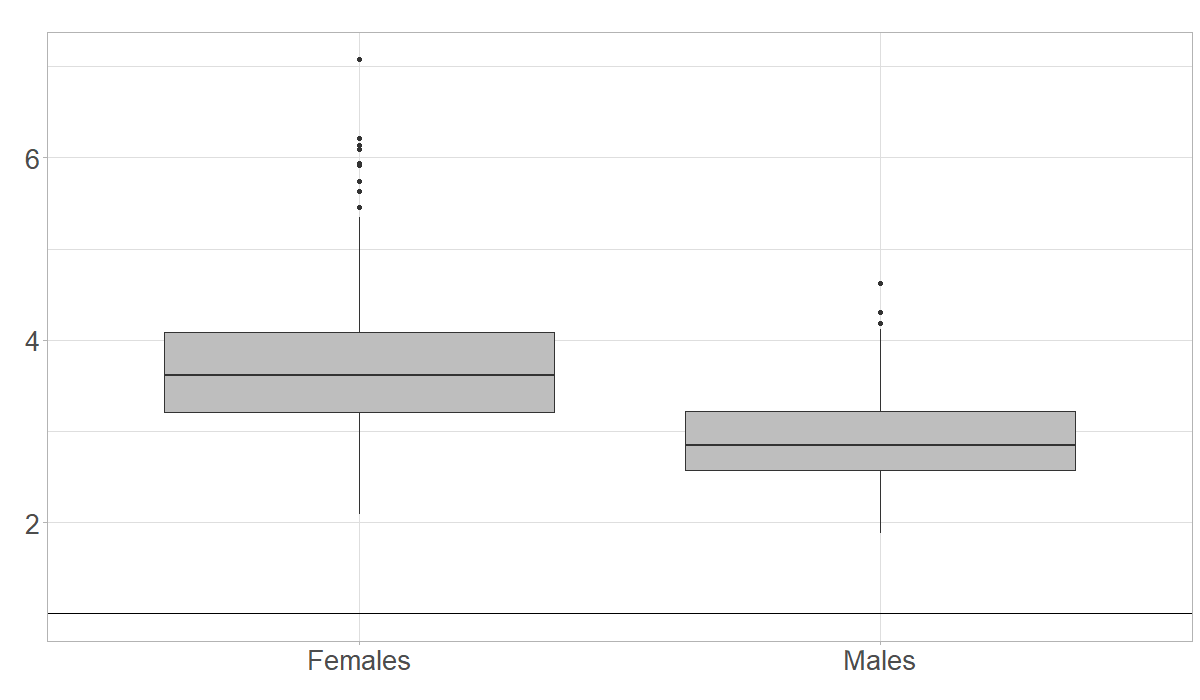}}
\qquad
\subfloat[][Law and Legal sciences]{
    \includegraphics[width=.5\linewidth]{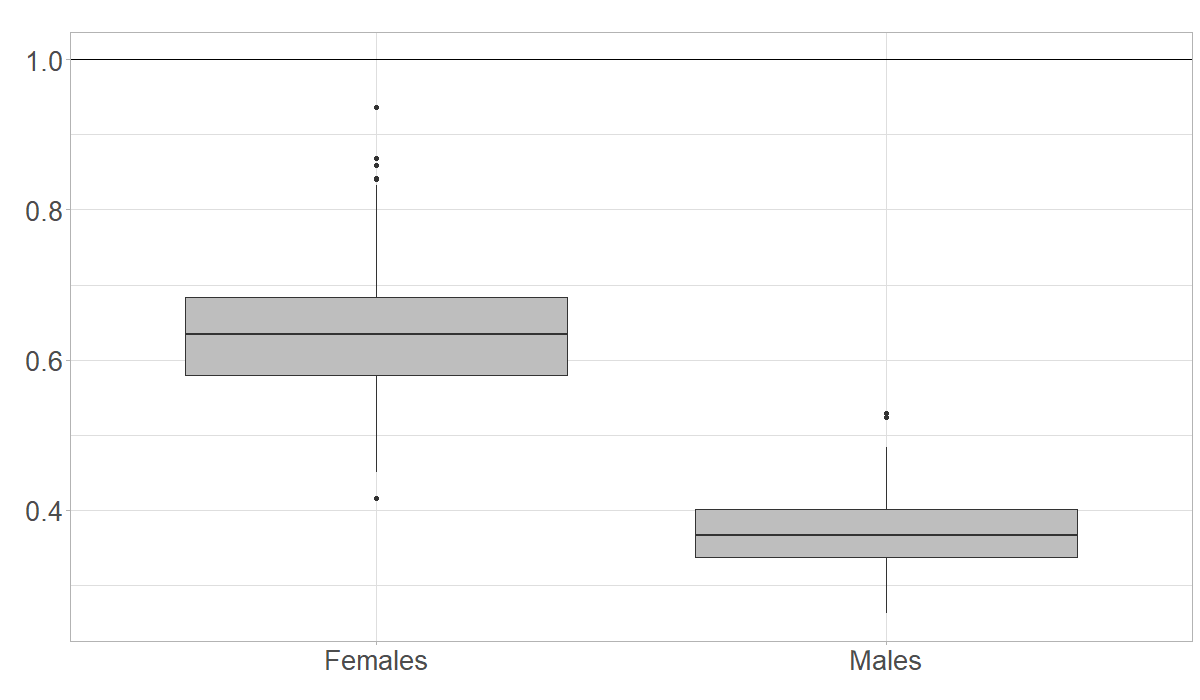}}
\subfloat[][Medicine, Dentistry, Pharmacy]{
    \includegraphics[width=.5\linewidth]{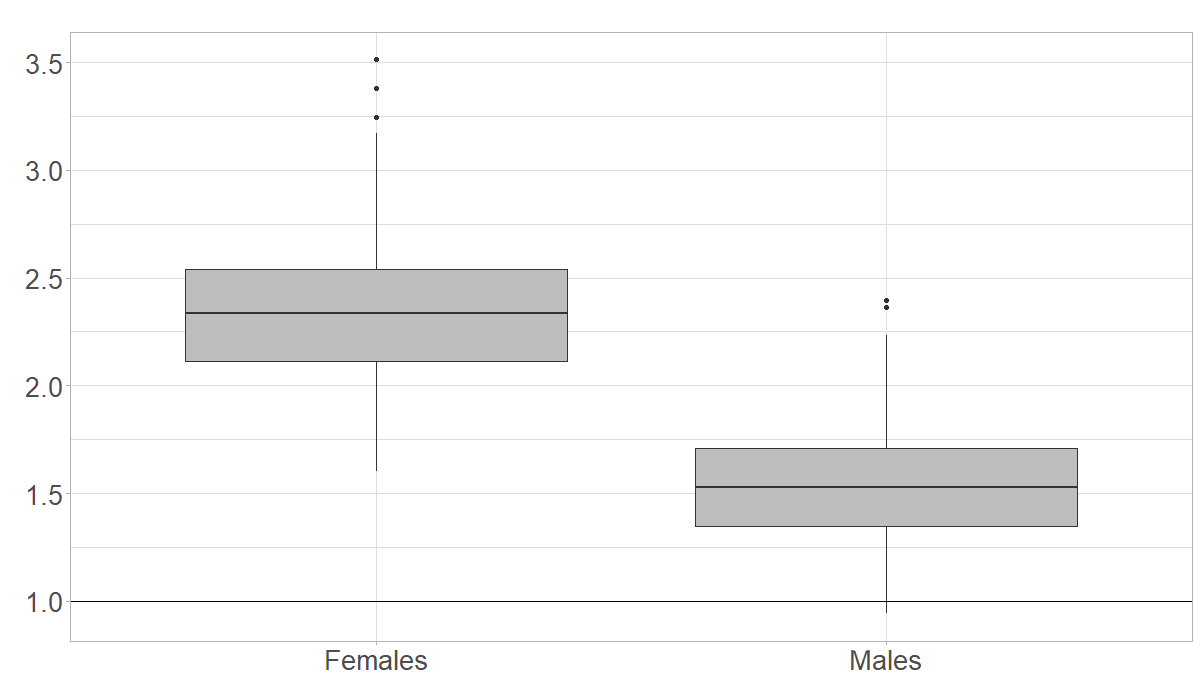}}
\caption{P}
\label{fig:step2}
\end{figure}

Table \ref{tab:step2} shows the results in terms of posterior summaries. 
It emerges that the odds ratios are quite homogeneous within groups.

\begin{table}[!t]
\centering
\small
\caption{Posterior summaries of the odds ratios $w$, for females ($w_F$) and males ($w_M$) and by degree program, obtained using Almalaurea survey data.}
\label{tab:step2}
\begin{tabular}{@{}lcccccc@{}}
\toprule
                                                                                                & \multicolumn{3}{c}{$w_F^{11}$}                & \multicolumn{3}{c}{$w_M^{11}$}                \\ \midrule
                                                                                                & \begin{tabular}[c]{@{}c@{}}Posterior \\ mean\end{tabular} & \begin{tabular}[c]{@{}c@{}}Posterior \\ median\end{tabular} & \begin{tabular}[c]{@{}c@{}}Posterior \\ sd\end{tabular} & \begin{tabular}[c]{@{}c@{}}Posterior \\ mean\end{tabular}  & \begin{tabular}[c]{@{}c@{}}Posterior \\ median\end{tabular}  & \begin{tabular}[c]{@{}c@{}}Posterior \\ sd\end{tabular}  \\ \cmidrule(l){2-7} 
\begin{tabular}[c]{@{}l@{}}Agricultural and forestry \\ sciences,  Veterinary\end{tabular}   & 0.97        & 0.92      & 0.34                      & 1.13       & 1.04    & 0.52                     \\ \midrule
\begin{tabular}[c]{@{}l@{}}Architecture and Engineering\end{tabular}                         & 3.72        & 3.62      & 0.72                      & 2.91       & 2.85    & 0.47                      \\ \midrule
\begin{tabular}[c]{@{}l@{}}Law and Legal sciences\end{tabular}                               & 0.64        & 0.63      & 0.08                      & 0.37       & 0.37    & 0.05                   \\ \midrule
\begin{tabular}[c]{@{}l@{}}Medicine, Dentistry, Pharmacy\end{tabular}                        & 2.34        & 2.33      & 0.31                      & 1.54       & 1.53    & 0.27                    \\ \midrule
\end{tabular}
\end{table}

\subsection{Employment rates estimates from 2012 to 2020}\label{sec:timeseries}
Assuming the response rates are constant over the years, we estimated the employment rates one year after the graduation of the 2012-2020 cohorts of new graduates. 
Figure \ref{fig:step3} shows the posterior means and the 95\% highest posterior density intervals of the employment rates. 
It also includes the employment rates computed using the raw Almalaurea data for comparison.
Independently of the estimation method, a positive trend emerges, which is coherent with the Italian history of the last decade.
However, it is clear from Figure \ref{fig:step3} that ignoring the MNAR mechanism leads to a (generally upward) bias in estimating the employment rate.
Among the reported degree programs, the 95\% highest posterior density intervals cover the employment rates computed using raw Almalaurea data at each time only for ``Agriculture and Forestry, Veterinary''.

\begin{figure}[!ht]
    \centering
\subfloat[][Agriculture and Forestry, Veterinary - females]{
    \includegraphics[width=.5\linewidth]{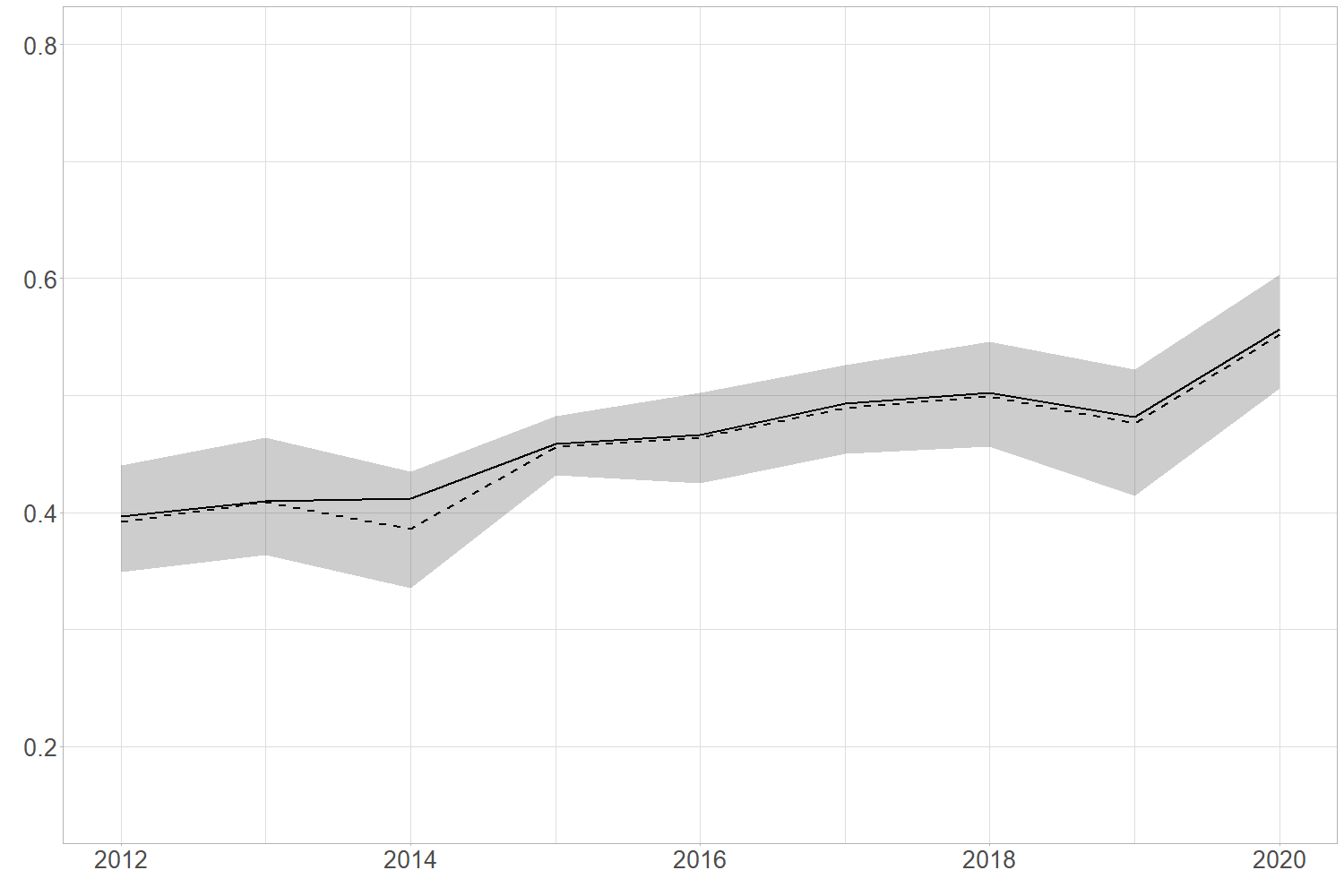}}
\subfloat[][Architecture and Engineering - females]{
    \includegraphics[width=.5\linewidth]{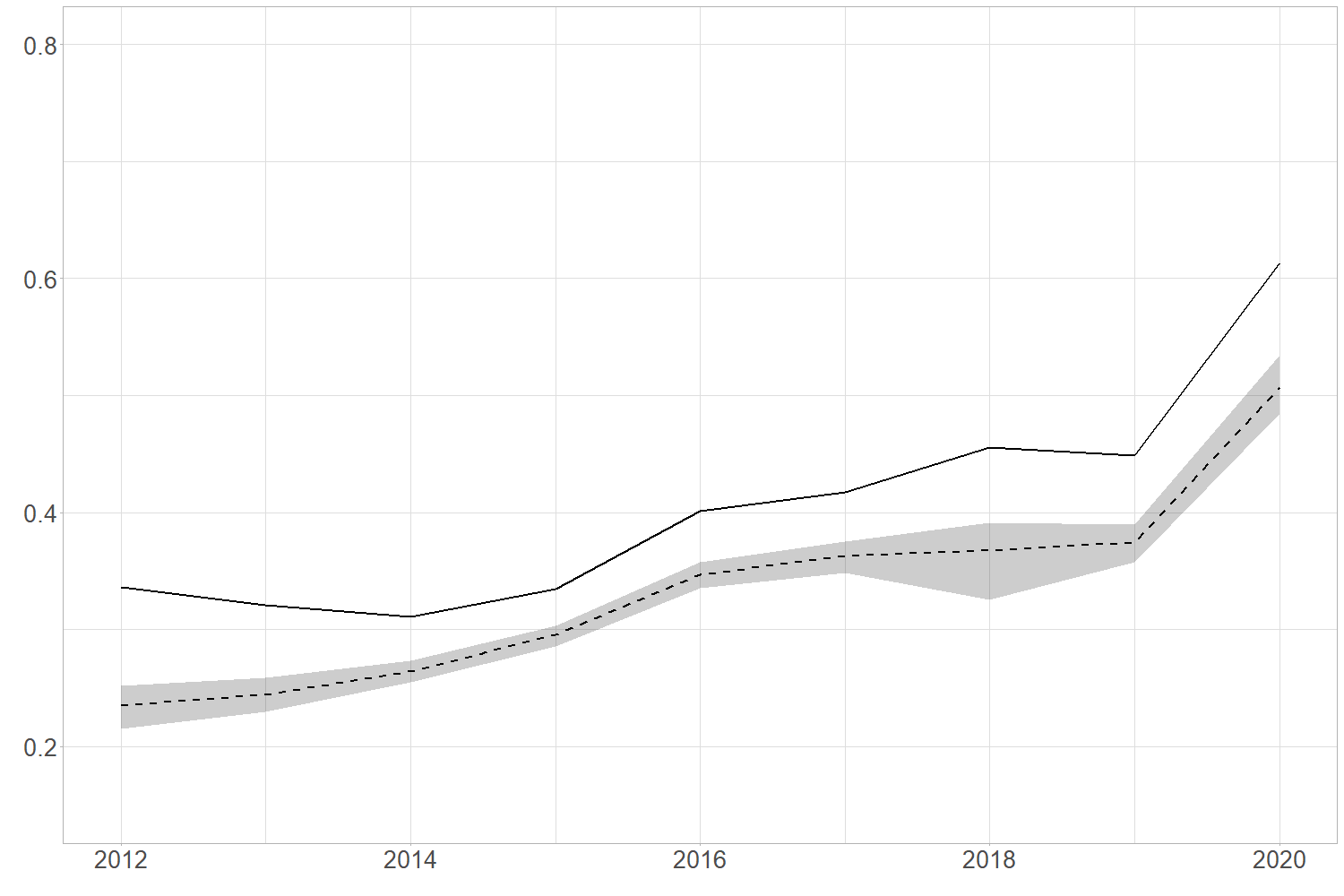}}
\qquad
\subfloat[][Law and Legal sciences - females]{
    \includegraphics[width=.5\linewidth]{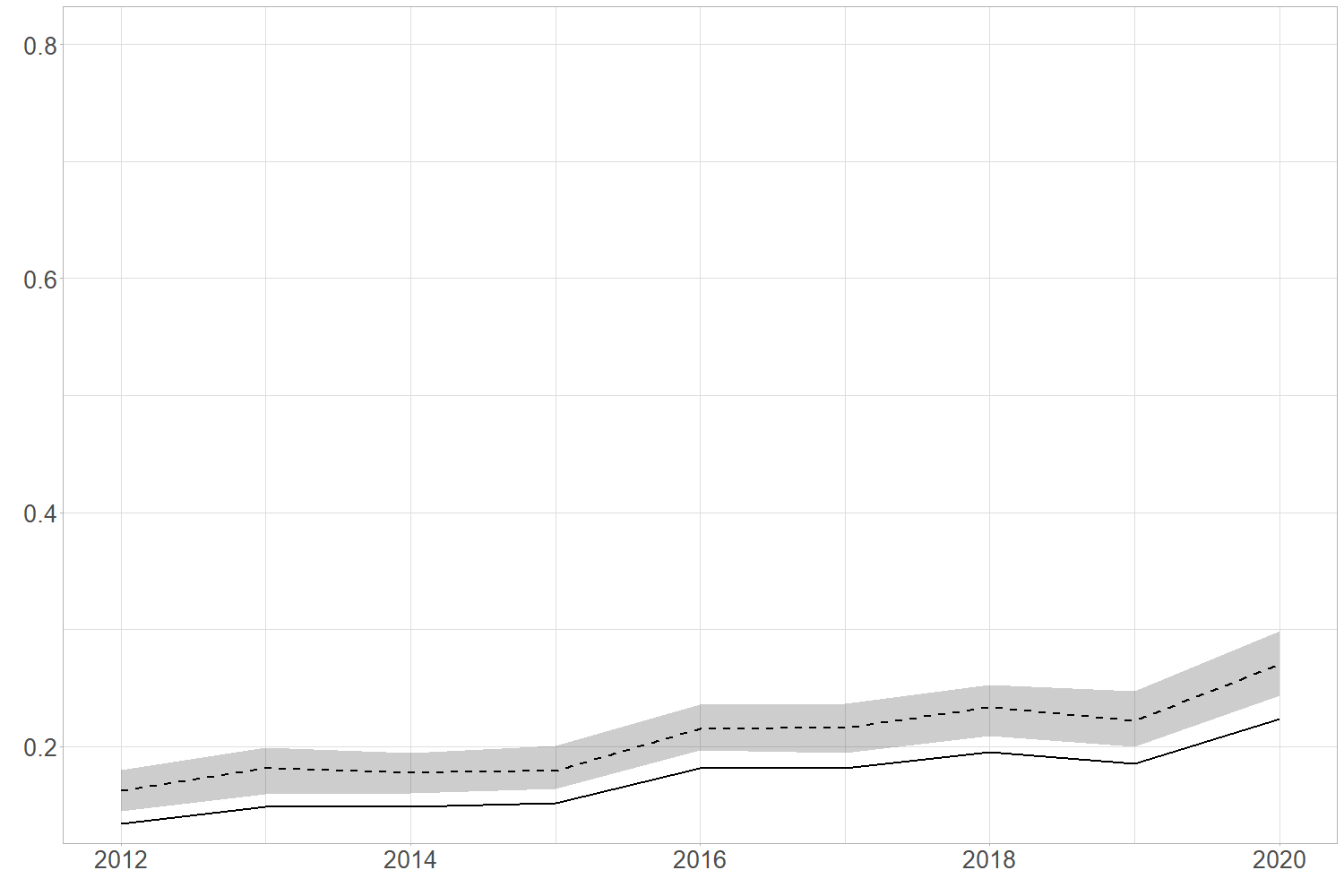}}
\subfloat[][Medicine, Dentistry, Pharmacy - females]{
    \includegraphics[width=.5\linewidth]{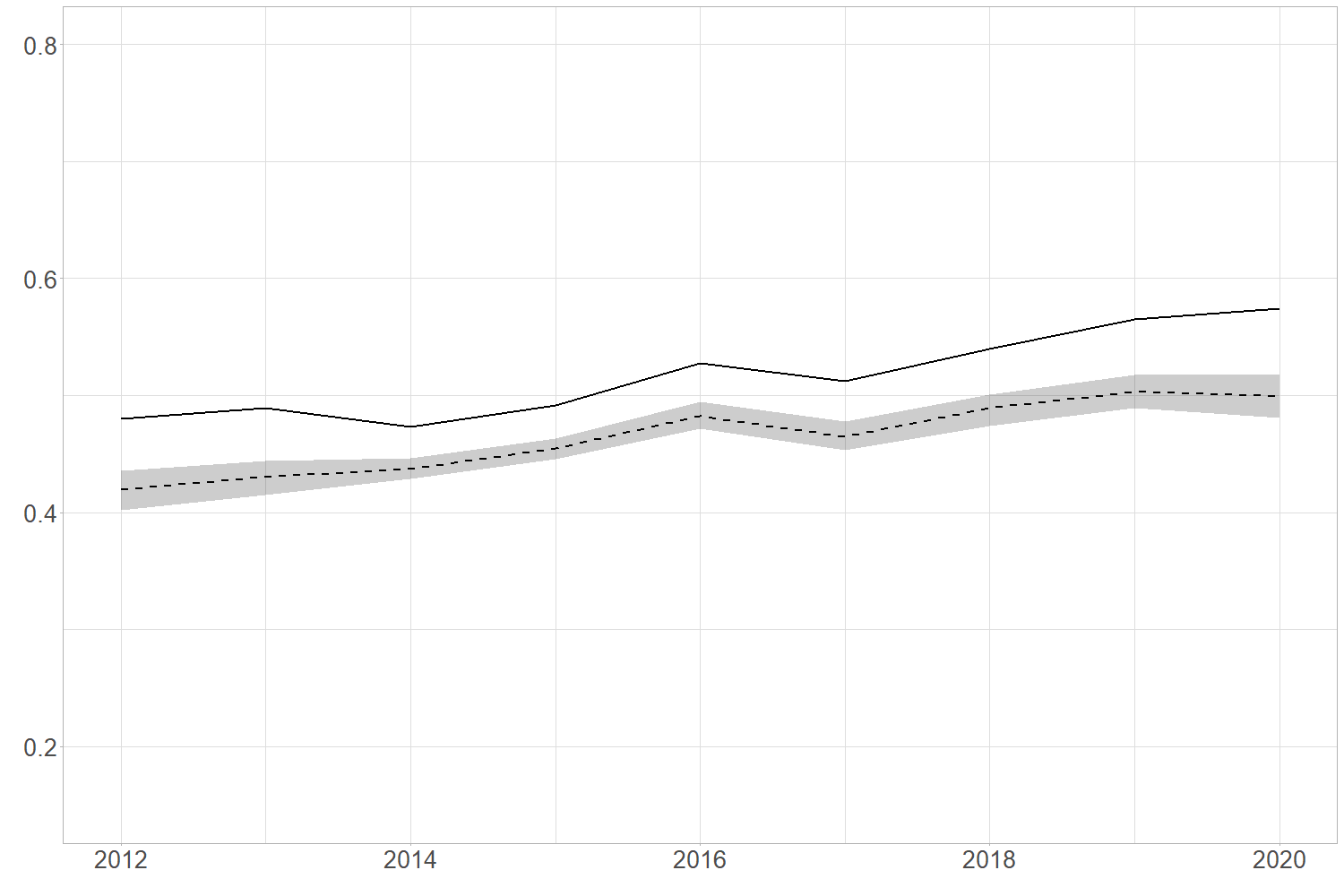}}
\caption{Posterior mean (dashed line) and 95\% highest posterior density interval of the females' employment rates for the years between 2012 and 2020, by degree program, estimated at step (iii), versus the employment rates computed using raw Almalaurea data (solid line).}
\label{fig:step3}
\end{figure}

\subsection{Sensitivity to the assumptions}\label{sec:sens} 
To assess the robustness of our results, we check the sensitivity of the estimates to prior elicitation in two different ways.

First, we test the robustness of the odds ratio estimates at step (ii), namely for the year 2011, to different specifications of its prior standard deviation, using $\tau = 3, 4, 5$.
The results generally align with those in Figure \ref{fig:step2}. 
As expected, when a substantial conflict between different data sources occurs, the impact of the prior is more evident, and the results should be interpreted with caution; these are a few exceptional cases reported in the supplementary material. 

Then, we assess the robustness of the results in time, namely of the results obtained in step (iii); we consider the following prior for $w^{t}, t>2011$:
\begin{equation}
\log(w^{t}) \sim 
\text{N}\left ({q_{\alpha}^{11}}, 
{\tau^{11}}\right),
\end{equation} 
where ${q_{\alpha}^{11}}$ is the $\alpha$-th quantile of the $w^{11}$ marginal posterior, and ${\tau^{11}}$ is its standard deviation.
We estimate the posteriors for $\alpha = 0.25, 0.75$; the results are very close to those shown in Figure \ref{fig:step3}.
More details and results for all disciplines and genders can be found in the supplementary material.

\section{Discussion} \label{sec:conc}
In this study, we have addressed the challenge of estimating the composition of a population when only aggregated survey data are available and there is a nonignorable nonresponse issue. 
Our focus has been on estimating the employment rates of new graduates in Italy using survey data. 
To address the nonidentifiability issue, we have proposed a borrowing information strategy in a Bayesian framework.
In particular, we have used extra experimental information to calibrate a Fisher's noncentral hypergeometric model (FNCH).
FNCH is a kind of biased urn model, particularly suitable for describing situations where the probability of observing a ball of a specific color depends not only on the composition of the urn but also on the relative \textit{exposure} of that color with respect to the others.
To our knowledge, this is the first use of FNCH in survey statistics and in a Bayesian framework.

The versatility of our methodology extends beyond this specific application, as it holds the potential for estimating population composition in various scenarios involving aggregated survey data and nonignorable nonresponse. 

The ``not-at-random missing data mechanism'' we have considered in this paper is indeed commonly observed in many surveys investigating individuals' socioeconomic aspects in modern countries with developed national statistical systems. 
For instance, similar situations arise in electoral surveys, where people's inclination to disclose their political opinions may vary across different political parties. 
The utility of the Bayesian approach would be twofold. 

On the one hand, it would make it natural to leverage historical data or auxiliary information to learn about the nonresponse bias.
Auxiliary information could be incorporated by assuming that the weights are functions of some covariates. 

On the other hand, although the method presented in this work deals with simple random samples, the flexibility of the Bayesian approach would allow one to consider more complex survey designs.
This could be managed in two alternative ways.
First, one could think of the sizes $M_c$'s as coming from a specific sampling design. 
In this case, a Bayesian model should incorporate an additional layer of the hierarchy to account for the uncertainty related to $M_c$'s. 
The estimated odds ratio would incorporate such uncertainty, and one could still interpret it as a simple relative exposure of the $c$-th category with respect to a reference category.
Second, one could incorporate the sampling probabilities in the model \eqref{multifisher} via a known correction factor to the $w_c$'s. 
It would be more convenient to reparametrize the model in terms of the probabilities $\zeta_c$'s to facilitate the interpretation.

Both alternatives to introduce complex designs and further extensions of the methodology would imply an affordable increment of computational complexity.

From a more general perspective of population size estimation, our approach could be interpreted as an example of a capture method with single capture information enriched by some partial prior information on at least one subgroup of the population. 
This interpretation opens the way to several different applications where the multivariate version of the FNCH will be necessary.
The interested reader can find details in the supplementary material.

\section*{Supplementary material}
\begin{enumerate}
    \item Bayesian inference for multivariate FNCH.
    \item Simulation study performed to compare the proposed methods for multivariate inference.
    \item Additional results for all disciplines and genders.
    \item Details of sensitivity analysis mentioned in section \ref{sec:sens}.
\end{enumerate}

\bibliography{arxiv}

\newpage

\appendix
\section*{Supplementary Material}

\section{Bayesian inference for the multivariate FNCH} 
\label{sec:multiF}

When one deals with compositional data, the applications often require a multivariate approach. 
Here, we generalize the method described in the work to the multivariate case, exploiting the conditional structure of FNCH and showing that it is often possible to rely on an MCMC. 
As $n$ and $N$ increase, it may become computationally expensive to evaluate the likelihood function several times at each iteration; thus, we also consider a likelihood-free alternative based on Approximate Bayesian Computation (ABC) methods.

As in the univariate case, we need to introduce some genuine prior information on at least one of the $M_c$'s (or on $N$); for convenience and without loss of generality, we refer to such parameter as $M_1$. 
The hierarchical model in the multivariate case will be:
\begin{equation}
    \bm{Y}|\sum\limits_{c}{Y_c}=n \sim \text{mvFNCH}(\bm{M},n,\bm{w})
\end{equation}
where $M_1, \dots, M_C$ are mutually independent with
\begin{equation}
    M_c \sim \pi(\cdot; \bm{\theta}^M_c)
\end{equation}
The vector $\bm{w}$ can be either known or unknown. 
For brevity, we fix $\bm{w}$ in this section, but the extension to the case of unknown weights is straightforward and similar to the univariate case.

For $n$ and $N$ sufficiently large, any method involving repeated evaluation of the likelihood function becomes computationally expensive.
Below, we propose exploiting the conditional structure of FNCH to draw from the posterior $\pi(\bm{M}\mid \cdot)$ via MCMC and ABC methods.

\subsection{Posterior computation: MCMC method} \label{sec:MCMC}
As underlined by Fog (2008), the conditional distribution of any component $Y_c$ given the remaining ones is univariate FNCH; 
we exploit this fact to obtain the posterior distribution $\pi(\bm{M}|\bm{y})$ via Metropolis-within-Gibbs algorithm.
At each iteration $t$, we first propose $M_1^*$ from $q_t(\cdot|M_1^{t-1})$; the acceptance ratio is
\begin{equation} \label{ratio}
    \min \left(1; \dfrac{\text{mvFNCH}(\bm{y}|M_1^*,M_2^{t-1},...,M_C^{t-1},n)\pi(M_1^*)}{\text{mvFNCH}(\bm{y}|M_1^{t-1},M_2^{t-1},...,M_C^{t-1},n)\pi(M_1^{t-1})} \times \dfrac{q_t(M_1^{t-1}|M_1^*)}{q_t(M_1^*|M_1^{t-1})}\right)\;.
\end{equation}
The probability mass function of $\bm{Y}$ can be written as 
\begin{equation}\label{univref}
\begin{split}
P\bigg(\bm{Y}=\bm{y}|\sum\limits_{c=1}^{C}{Y_c}=n\bigg) & = P\bigg(Y_1=y_1, Y_2=y_2, ..., Y_C=y_c|\sum\limits_{c=1}^{C}{Y_c}=n\bigg) \\
& = P\bigg(Y_1=y_1, Y_{c'}=y_{c'}|\bm{Y}_{-(1,c')}, \sum\limits_{c=1}^{C}{Y_c}=n\bigg) \\ 
& \; \times P\bigg(\bm{Y}_{-(1,c')}=\bm{y}_{-(1,c')} | \sum\limits_{c=1}^{C}{Y_c}=n\bigg) \\
& = P\bigg(Y_1=y_1,Y_{c'}=y_{c'}|Y_1+Y_{c'} =
n-\sum\limits_{c, -(1,c')}{Y_c}\bigg) \\ 
& \; \times P\bigg(\bm{Y}_{-(1,c')}=\bm{y}_{-(1,c')} | \sum\limits_{c=1}^{C}{Y_c}=n\bigg)
\end{split}
\end{equation}
where $c'$ can be any $c \neq 1$. 
The first element of the last expression of \eqref{univref} is the probability mass function of a univariate FNCH. 
The ratio in \eqref{ratio} then simplifies into
\begin{equation}
        \dfrac{\text{FNCH}(y_1,y_{c'}|M_1^*,M_{c'}^{t-1},n_{1c'})\pi(M_1^*)}{\text{FNCH}(y_1,y_{c'}|M_1^{t-1},M_{c'}^{t-1},n_{1c'})\pi(M_1^{t-1})}
        \times \dfrac{q_t(M_1^{t-1}|M_1^*)}{q_t(M_1^*|M_1^{t-1})}
\end{equation}
where $n_{1c'} = y_1 + y_{c'}$. 
We sample the remaining $M_c$, $c \neq 1$ in the same fashion, always setting $M_{c'} = M_1$.

\subsection{Posterior computation: ABC method} \label{sec:ABC}
To avoid a massive evaluation of the likelihood function, we also explore the use of Approximate Bayesian Computation (ABC) methods. 
The first ABC algorithms date back to Tavaré et al. (1997) and Pritchard et al. (1999), and for the last two decades, ABC methods have spread enormously thanks to their flexibility. 
Such methods replace the evaluation of the likelihood with the simulation of a synthetic data set $\bm{x}$ and the computation of a summary statistics $\eta(\bm{x})$; then, $\eta(\bm{x})$ is compared to $\eta(\bm{y})$, namely, the statistics relative to the observed data, based on some distance metric $\rho(\eta(\bm{y}),\eta(\bm{x}))$.
In the most basic version of the ABC algorithm, namely the ``ABC rejection", the synthetic data are simulated from the prior predictive; if the distance between the synthetic and the observed data is smaller than a certain threshold $\varepsilon$, the value of the parameter that generated those data is \textit{accepted}.
For comprehensive reviews of such methods, see Sisson et al. (2018, Ch. 1) and Karabatsos and Leisen (2018). 

ABC methods are particularly suitable for noncentral hypergeometric distributions since evaluating the likelihood is costly but we can easily draw samples from the generating model (Fog, 2008).
Grazian et al. (2019) used an ABC rejection to estimate the weights of a Wallenius noncentral hypergeometric distribution.
In this context, we propose using a more efficient algorithm, i.e., the ABC-Gibbs by Clarté et al. (2021). 
Such a componentwise ABC combines the advantage of avoiding the computation of the likelihood function with the efficiency of the dimensionality reduction brought by the conditional structure of the Gibbs sampler; the synthetic data are simulated from the conditional posterior predictive.
In our case, the ABC-Gibbs requires the introduction of a group-specific summary statistic $\eta_c(\cdot)$ to be compared to a group-specific threshold $\varepsilon_c$.

We may define  
\begin{equation} \label{sumstat}
    \eta_c(\bm{y}) = \frac{y_{c}}{n}; \quad \eta_c(\bm{x}) = \frac{x_{c}}{n} 
\end{equation}
where $x_c$ is a count randomly drawn from a univariate FNCH.
Then, to compare the two statistics, we employ the following metric: 
\begin{equation}\label{metric}
\rho(\eta_c(\bm{y}),\eta_c(\bm{x})) = |\eta_c(\bm{y}) - \eta_c(\bm{x})| = \frac{1}{n}|y_c - x_c| \; ,\end{equation}
that is the absolute difference between the synthetic and the observed proportion of group $c$.
One could also employ the relative differences.

Finally, the thresholds $\varepsilon_c$'s can be chosen to be quantiles of the distances computed between the observed data and a large sample of synthetic data generated from the conditional priors. 
In the simulations in the next section, we will use the $2^{nd}$ percentiles.

Algorithm \ref{GibbsABC} describes the ABC-Gibbs we use to estimate $\bm{M}$. 
According to the results in section \ref{sec:MCMC}, the conditional distribution we use to simulate the synthetic data is still a univariate FNCH.

\begin{algorithm}[t]
\SetAlgoLined
Set $\bm{M}^0 = (M_1^0, ..., M_C^0)$ \;
    \For{$t\leftarrow 1$ \KwTo $T$}{
    \For{$c\leftarrow 1$ \KwTo $C$}{
    \Repeat{$\rho(\eta_c(\bm{y}),\eta_c(\bm{x})) < \varepsilon_c$}{draw $M_c^*$ from its conditional prior distribution $\pi(M)$ \;
    simulate $x_c \sim \text{FNCH}(x_c|M_c^*,M_{-c,1}^{t-1},M_{-c}^{'t})$}
    $M_c^t = M_c^*$}}
\caption{Gibbs-ABC for FNCH}
\label{GibbsABC}
\end{algorithm}

\section{ABC-Gibbs vs. Gibbs: A comparison}
We present the results of a simulation study aiming to estimate the total population size $N$ in the presence of $C = 5$ subgroups. 
We set $N = N^* = 10000$, and generate 100 samples as follows.
We simulate the compositional structure of the population from a Dirichlet$(\bm{\alpha}=\bm{1})$, and the propensity to be captured for each group from a Beta$(a=1,b=1)$.
Then, for each group we simulate 100 counts $y_c$ from a Binomial$(M^*_c, \zeta^*_c)$. 
More formally,

\begin{algorithm}[!htbp]
\SetAlgoLined
Set $N^*$ \;
    draw $\bm{p}^* = \bm{M}^*/N^* \sim$ Dirichlet$(\bm{\alpha}=(1,\dots,1))$ \;
    \For{$c\leftarrow 1$ \KwTo $C$}{
    draw $\zeta^*_c \sim$ Beta$(1,1)$ \;
    draw $y_c \sim$ Binomial$(M^*, \zeta_c^*)$
    }
\caption{Samples simulation}
\label{simstud}
\end{algorithm}

We implement both the methodologies described in Section 2 of the main article.
In particular, we assume
\begin{equation}
    M_1 \sim \text{Pois}(M_1^*)
\end{equation}
\begin{equation}
    M_c \sim \text{Unif}(y_c+1, M_\text{upp}), \quad c = 2, \dots, C,
\end{equation}
where $M_\text{upp} = 2\times10^4$. 
Coherently with the description in the main article, here we assume $\bm{w}$ to be fixed and $w_c = {\zeta^*_c}/{(1-\zeta^*_c)}$ for each $c = 1, \dots, C$.
Concerning the proposal distributions adopted in the Metropolis-within-Gibbs, we use an independent sampler for $M_1$ based on its prior distribution and random walk proposals for the other $M_c$'s based on Normal distributions with standard deviations tuned to reach acceptance rates between $0.25$ and $0.5$.

\begin{figure}[!htbp]
    \centering
    \includegraphics[width=.7\linewidth]{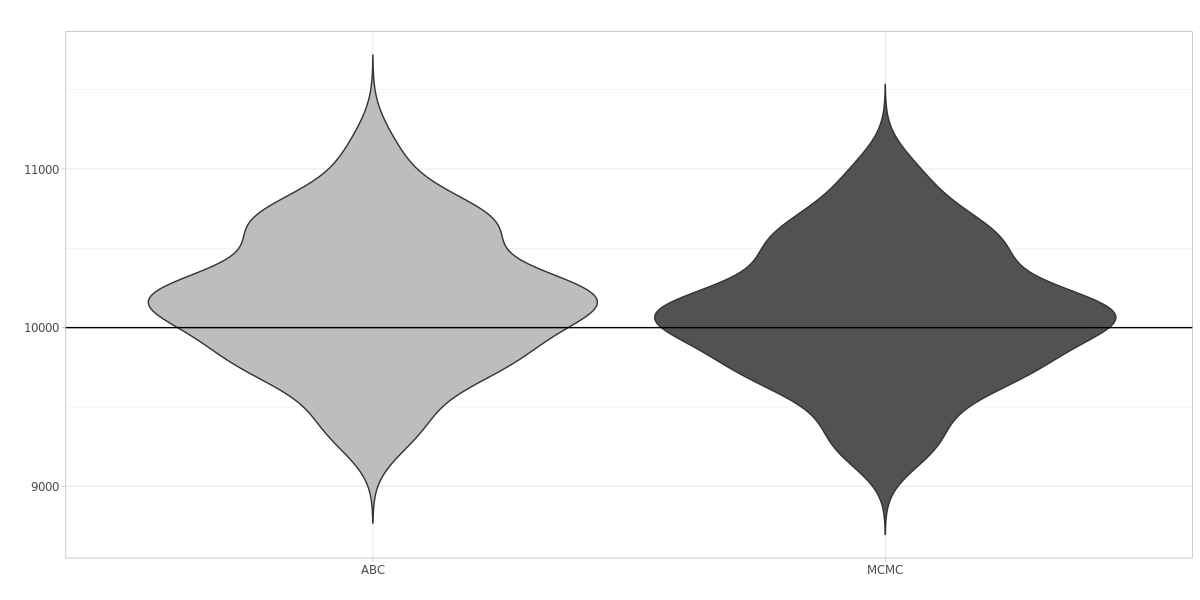}
    \caption{Distribution of the posterior mean of $N$ simulated via ABC (left) and MCMC (right), 100 samples}
    \label{fig:cfr5N}
\end{figure}

Figures \ref{fig:cfr5N} and \ref{fig:cfr5_m} show the distributions of the $N$ and $M_c$'s posterior mean across 100 samples. 
The MCMC approach shows a better ability to estimate the parameters' posterior distribution. 
Tables \ref{tab:princlusion-N5} and \ref{tab:princlusion-m5} report the frequentist coverages of such parameters'; the ABC approach is the worst in approximating the tails of the posterior, while the MCMC better estimates posterior uncertainty.
Therefore, MCMC methods should always be preferable when feasible due to their unmatched ability to estimate the posterior distribution. 
However, when the iterative evaluation of the likelihood function burdens an MCMC algorithm significantly, ABC methods offer a viable alternative.

\begin{table}[!htbp]
\centering
\caption{The 95\% Highest posterior density intervals for $N$ include the true value, frequencies over 100 samples.}
\begin{tabular}{@{}ll@{}}
\toprule
 & $N$ \\ \midrule
MCMC & 0.92 \\
ABC & 0.99 \\ \bottomrule
\end{tabular}
\label{tab:princlusion-N5}
\end{table}

\begin{table}[!htbp]
\centering
\caption{The 95\% Highest posterior density intervals for $\bm{M}$ include the true values, frequencies over 100 samples.}
\begin{tabular}{@{}lllllllllll@{}}
\toprule
 & $M_1$ & $M_2$ & $M_3$ & $M_4$ & $M_5$ \\ \midrule
MCMC & 1.00 &  0.99 &  0.99 &  1.00 &  0.98 \\
ABC & 1.00 &  1.00 &  1.00 &  1.00 &  1.00 \\ \bottomrule
\end{tabular}
\label{tab:princlusion-m5}
\end{table}

\begin{figure}[!htbp]
    \centering
\subfloat[][$M_1$]{
    \includegraphics[width=.7\linewidth]{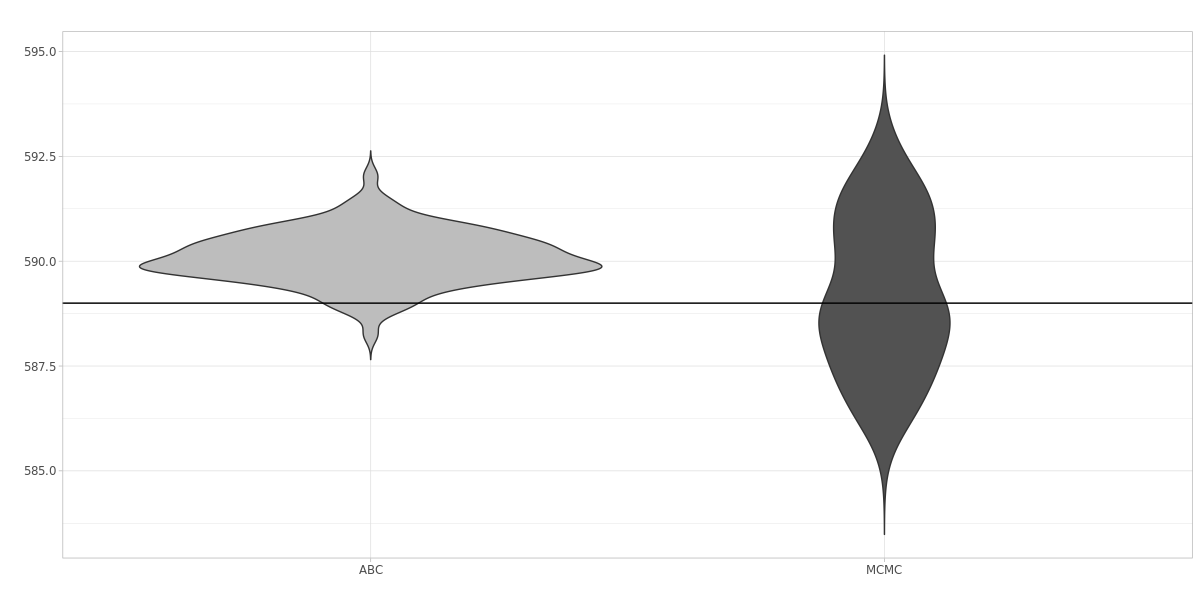}}
\qquad
\subfloat[][$M_2$]{
    \includegraphics[width=.7\linewidth]{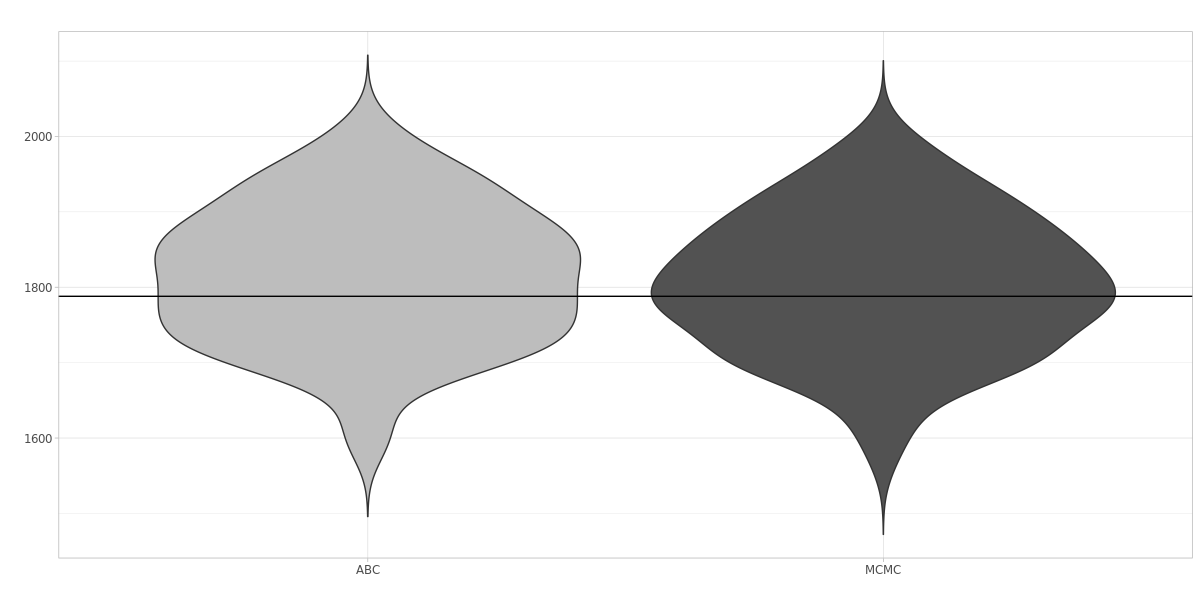}}
\end{figure}
\begin{figure}\ContinuedFloat
    \centering
\subfloat[][$M_3$]{
    \includegraphics[width=.7\linewidth]{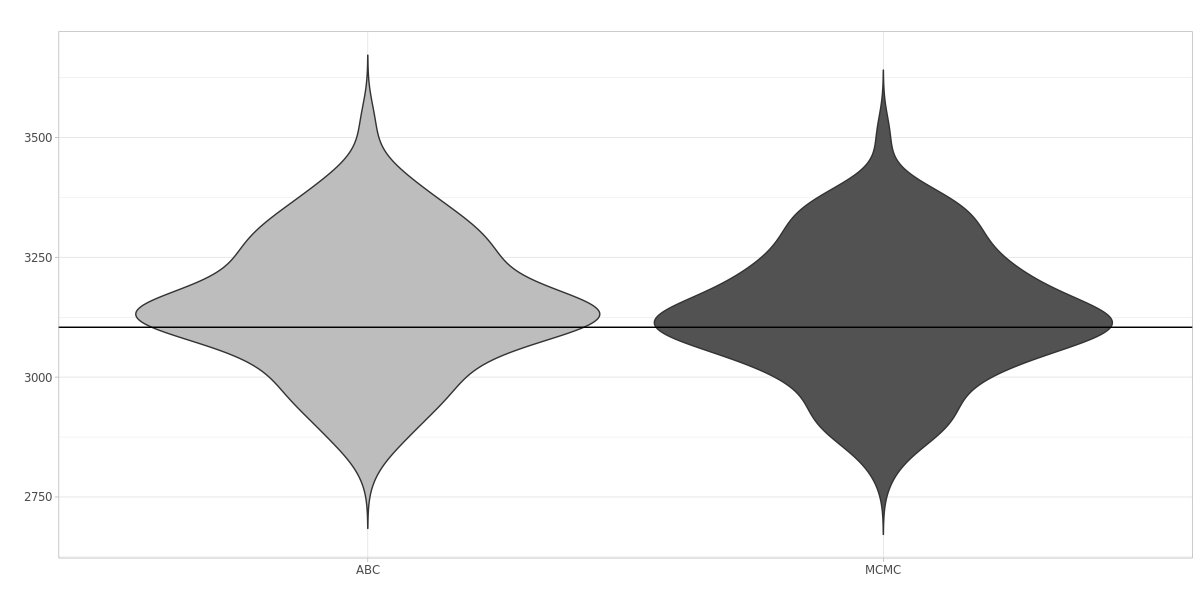}}
\qquad
\subfloat[][$M_4$]{
    \includegraphics[width=.7\linewidth]{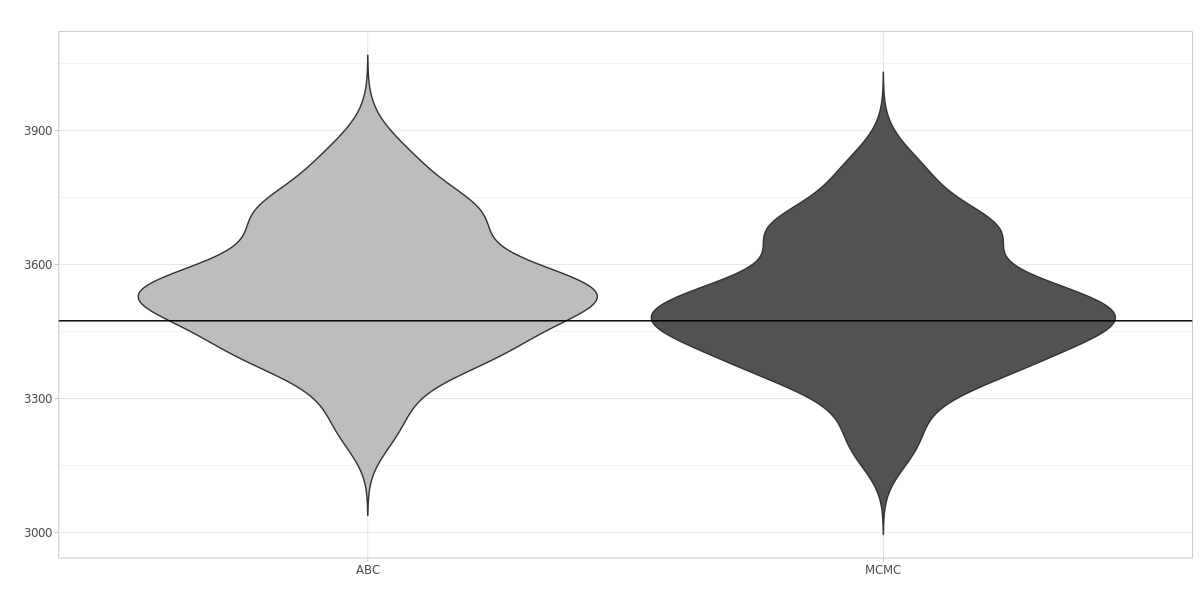}}
\qquad
    \subfloat[][$M_5$]{
    \includegraphics[width=.7\linewidth]{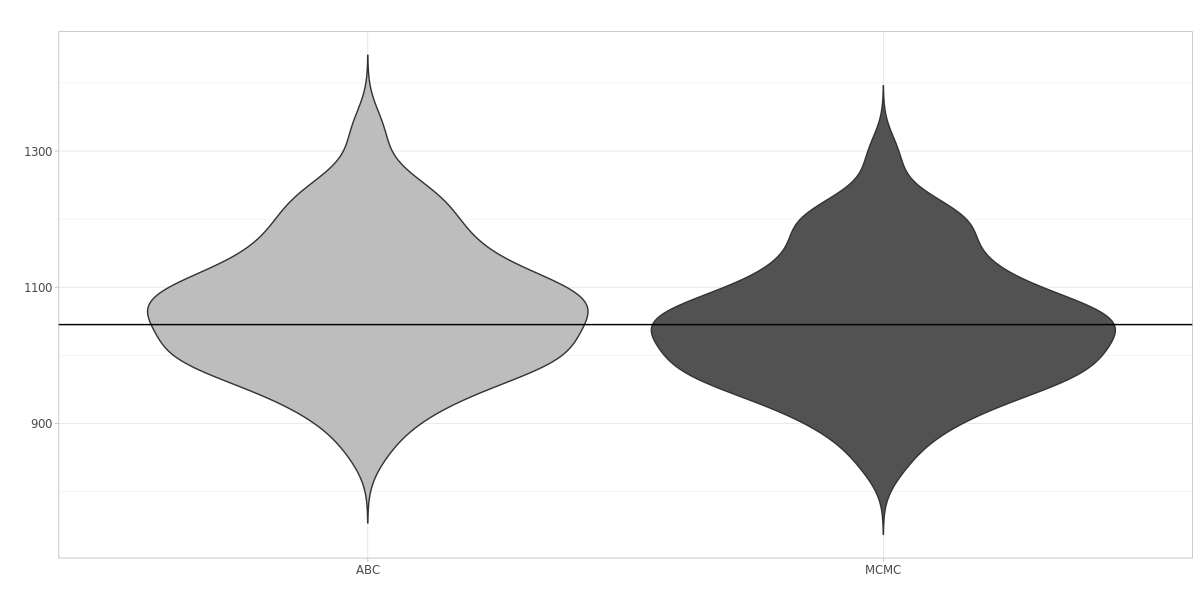}}
\caption{Posterior means of $\bm{M}$ simulated via ABC (left) and MCMC (right), 100 samples.}
\label{fig:cfr5_m}
\end{figure}

\clearpage
\section{Results for all disciplines and genders}
Figures \ref{fig:step1}-\ref{fig:step3} show results of steps (i)-(iii), respectively, for all disciplines and genders. 
We do not report the disciplines and genders already included in the main text.

\begin{figure}[!t]
    \centering
\subfloat[][B\&A, Economics, Finance]{
    \includegraphics[width=.5\linewidth]{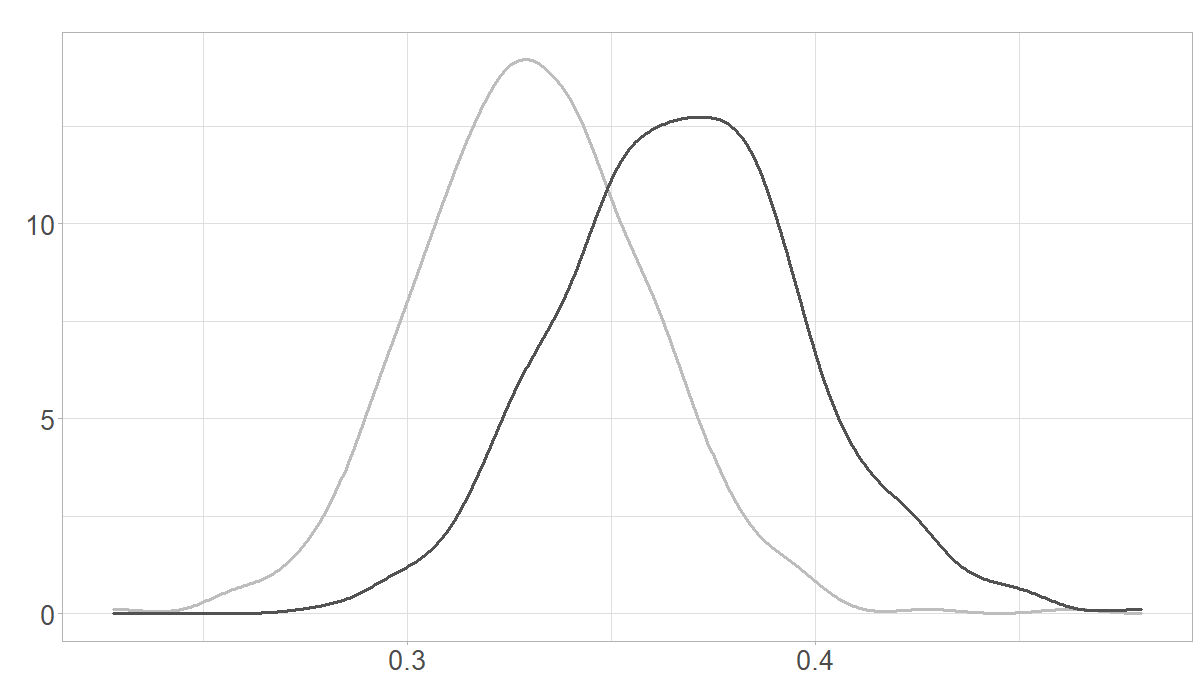}}
\subfloat[][Communication and Publishing]{
    \includegraphics[width=.5\linewidth]{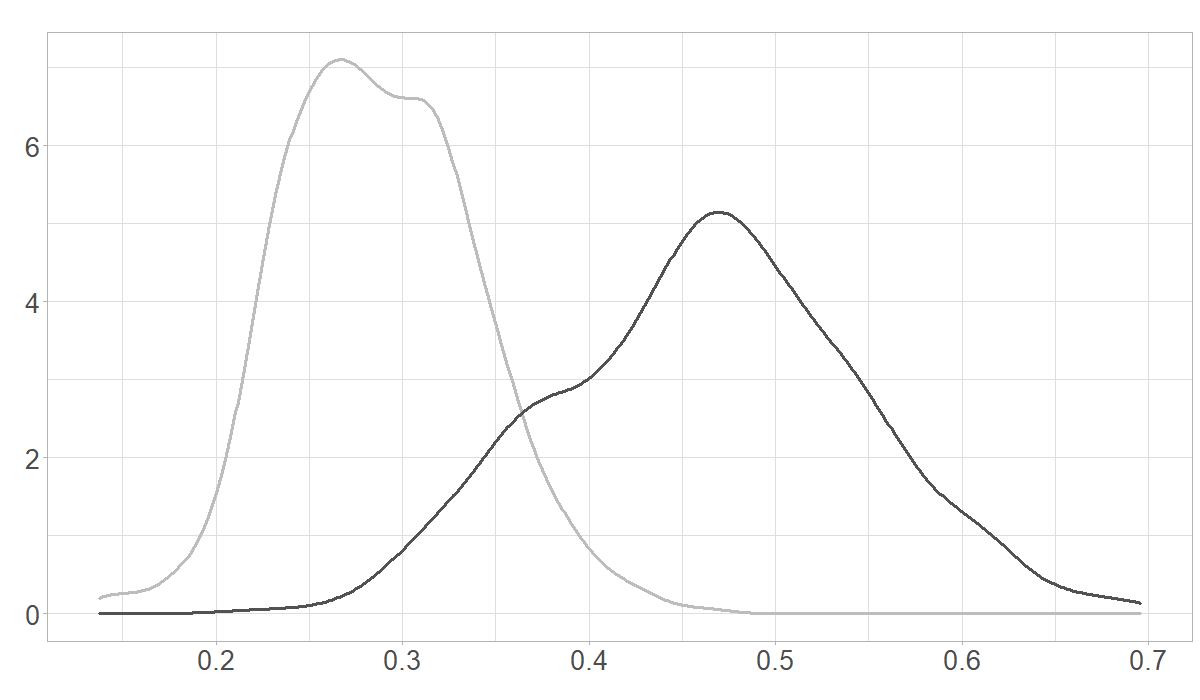}}
\qquad
\subfloat[][Industrial and Information Engineering]{
    \includegraphics[width=.5\linewidth]{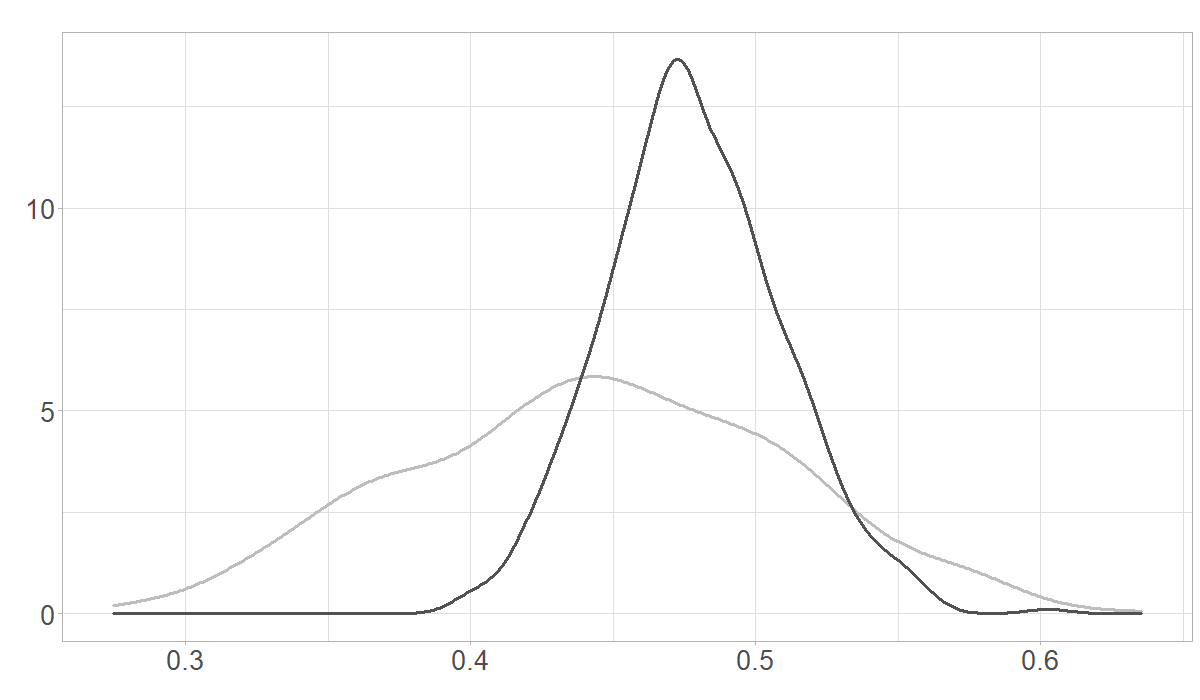}}
\subfloat[][Literature and Humanities]{
    \includegraphics[width=.5\linewidth]{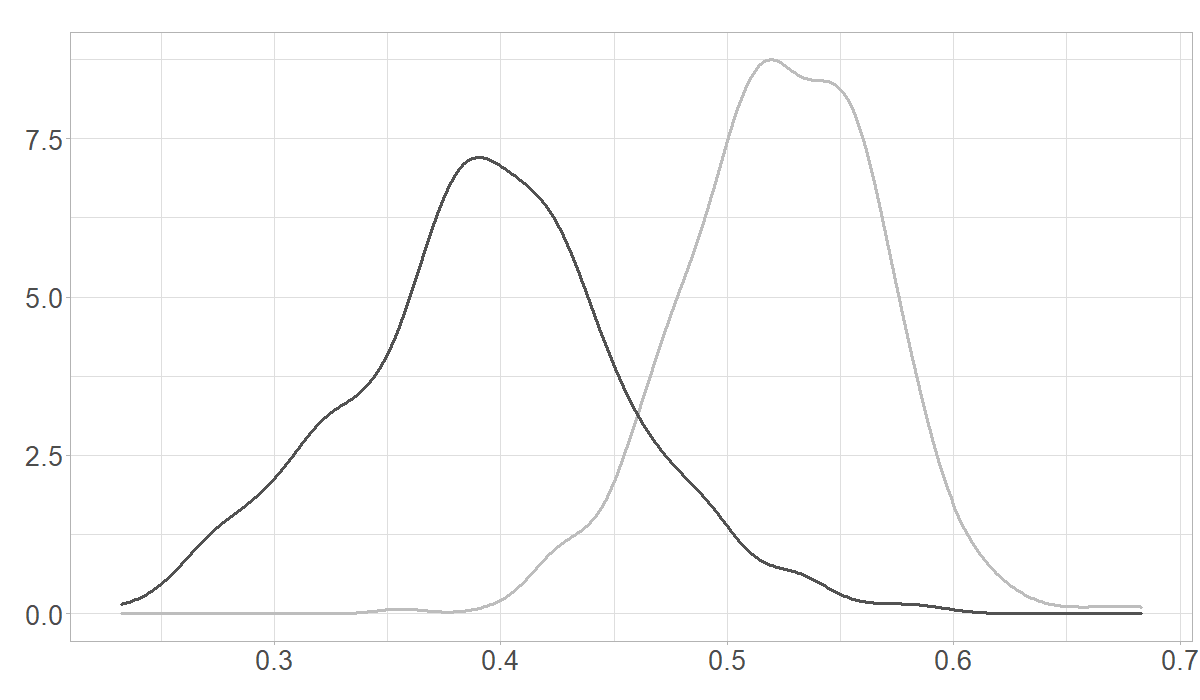}}
\qquad
\subfloat[][Political Science]{
    \includegraphics[width=.5\linewidth]{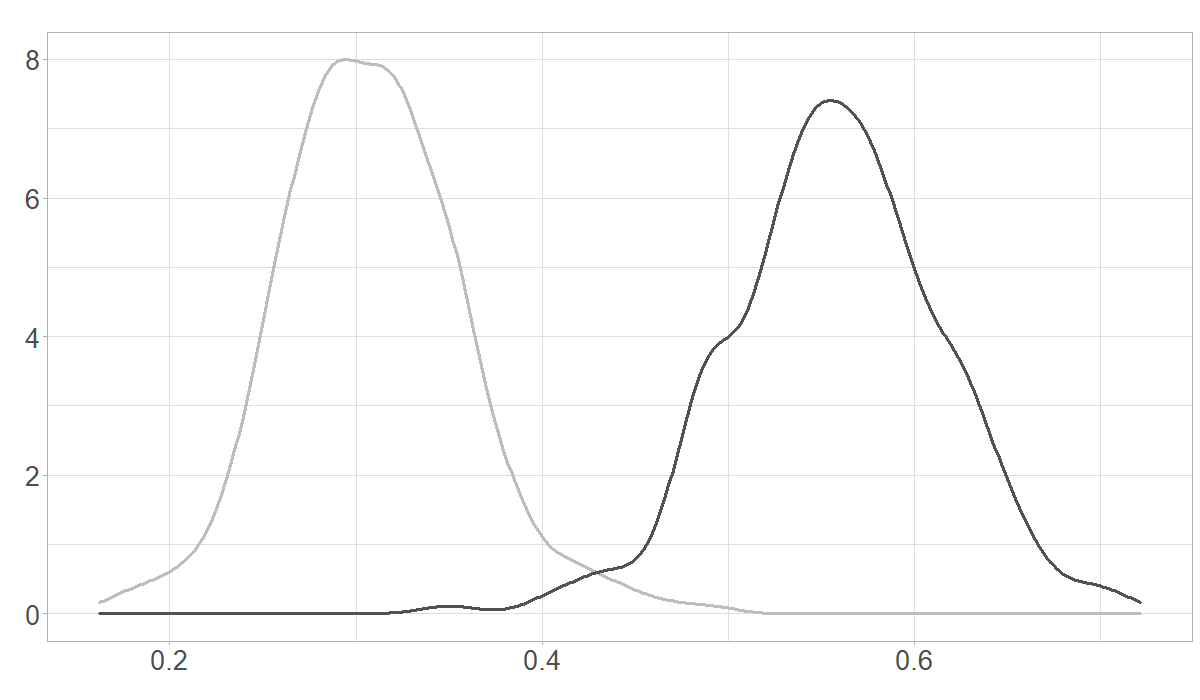}}
\subfloat[][Science and IT]{
    \includegraphics[width=.5\linewidth]{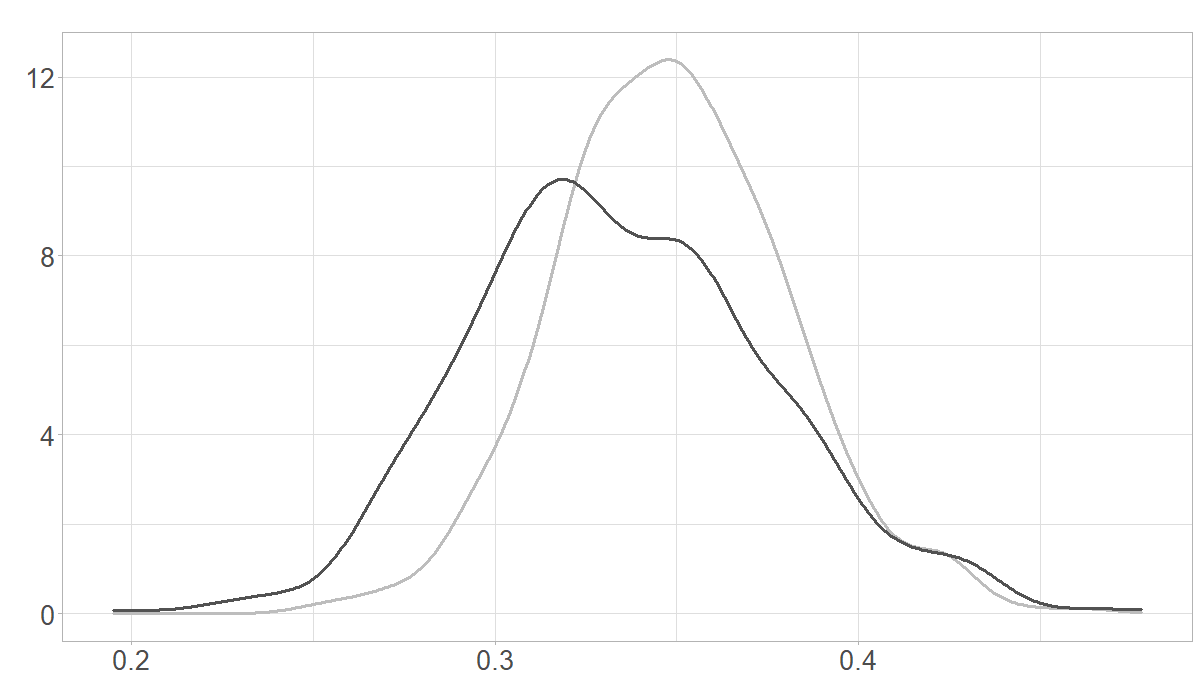}}
\caption{Posterior distributions of $\{M^{11}\}$, i.e., the sizes of 2011 graduates who were employed one year after their graduation, divided by $N^{11}$ posterior mean estimated at step (i), by gender (grey: females, dark grey: males) and degree programs, obtained using the 2011 NSI sample.}
\label{fig:step1}
\end{figure}

\begin{figure}[!b]
    \centering
\subfloat[][B\&A, Economics, Finance]{
    \includegraphics[width=.5\linewidth]{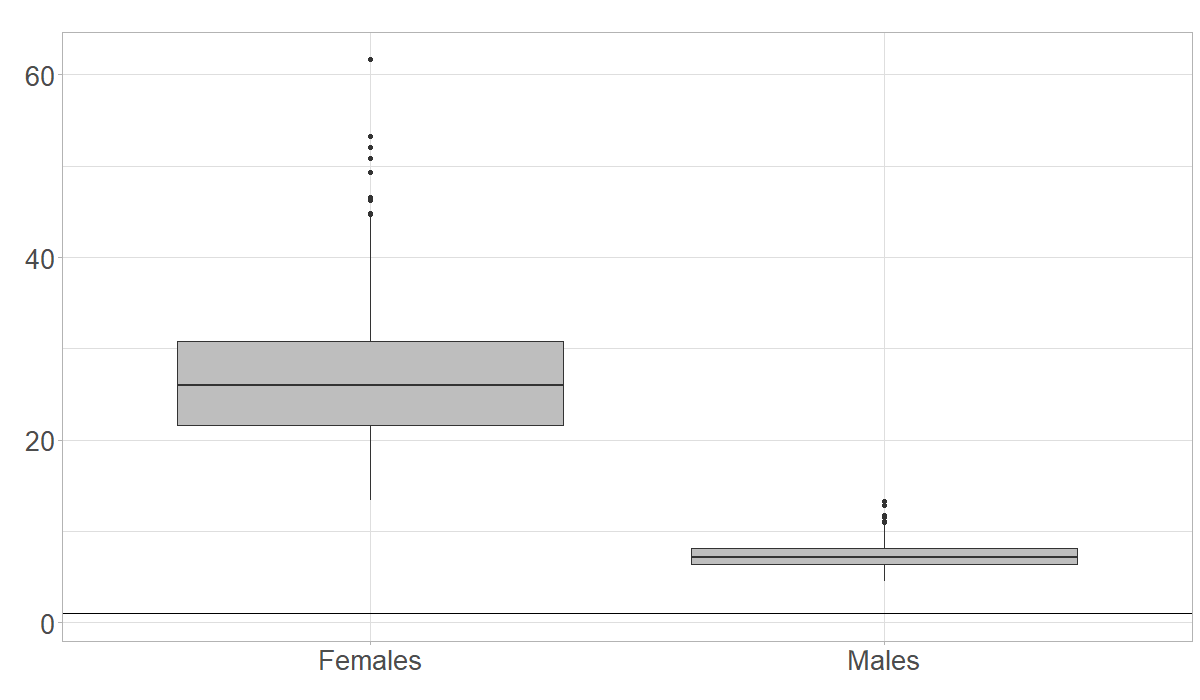}}
\subfloat[][Communication and Publishing]{
    \includegraphics[width=.5\linewidth]{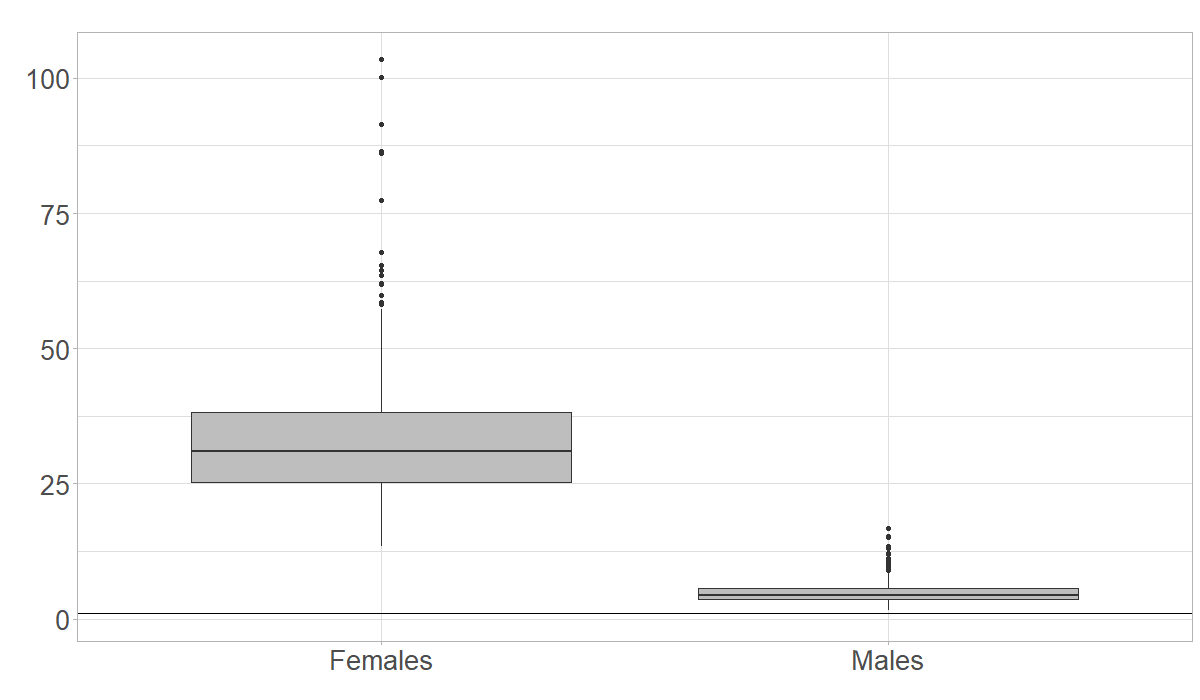}}
\qquad
\subfloat[][Industrial and Information Engineering]{
    \includegraphics[width=.5\linewidth]{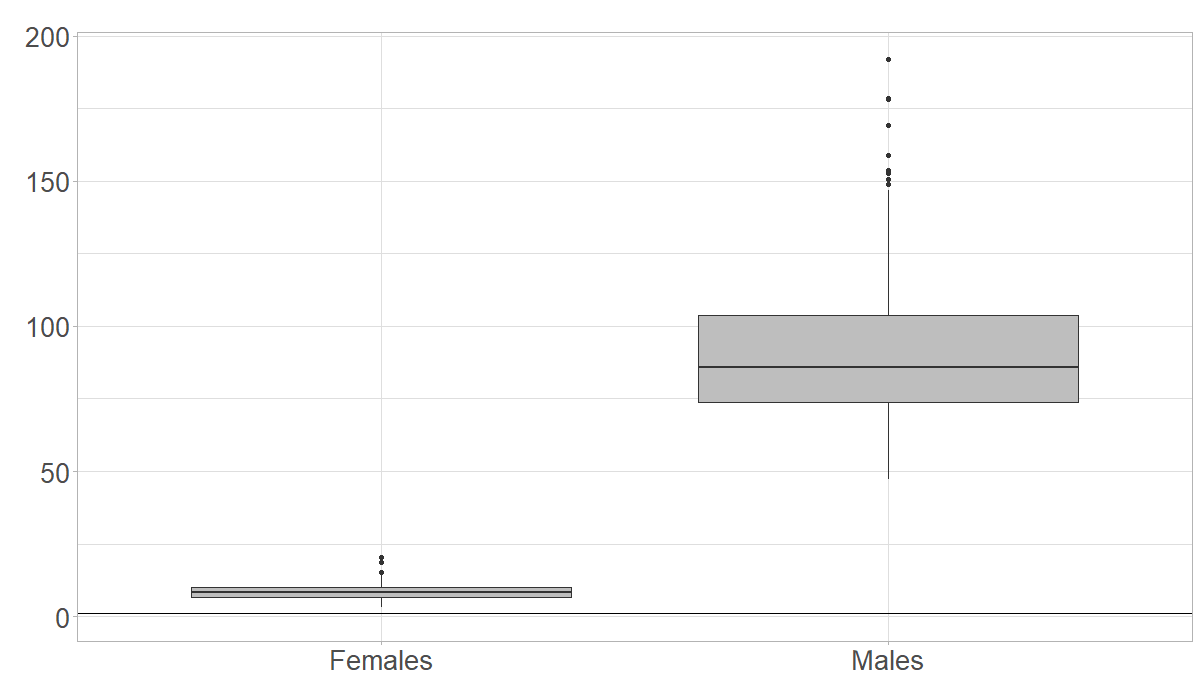}}
\subfloat[][Literature and Humanities]{
    \includegraphics[width=.5\linewidth]{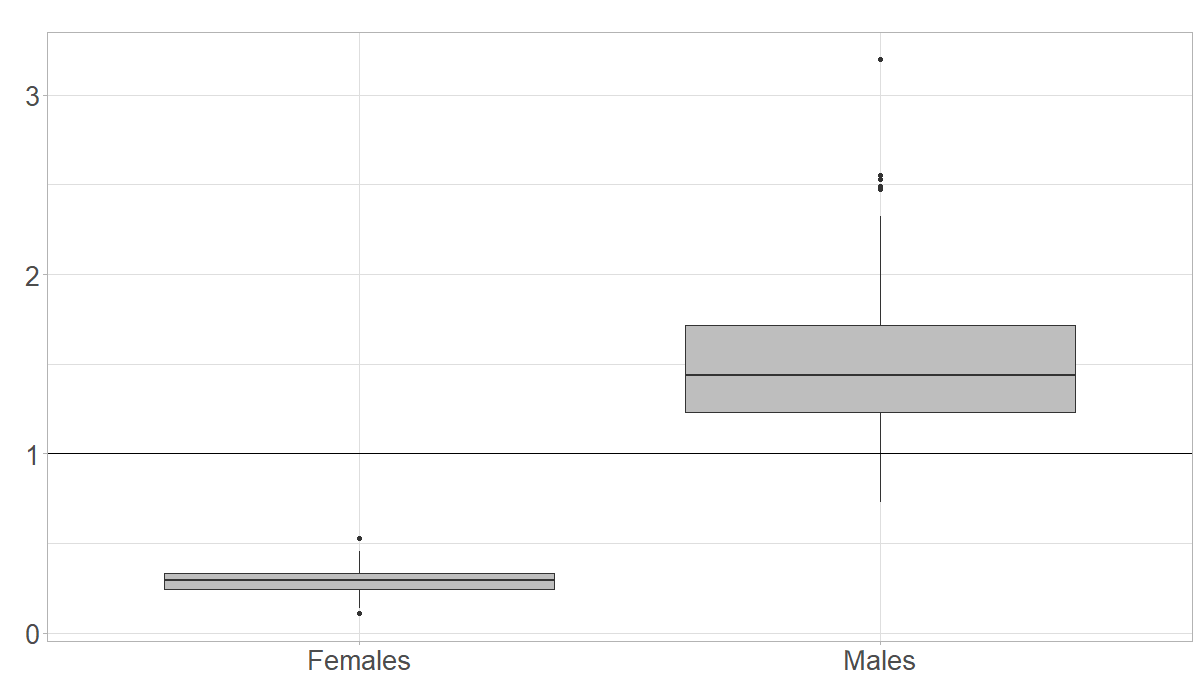}}
\qquad
\subfloat[][Political Science]{
    \includegraphics[width=.5\linewidth]{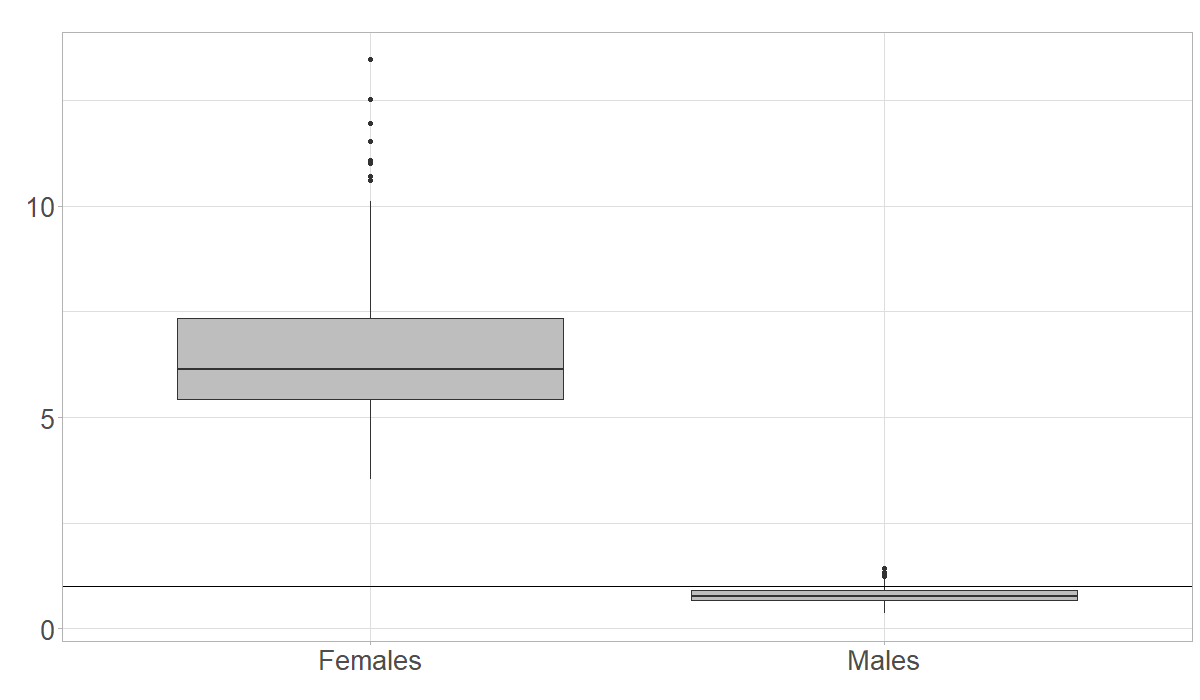}}
\subfloat[][Science and IT]{
    \includegraphics[width=.5\linewidth]{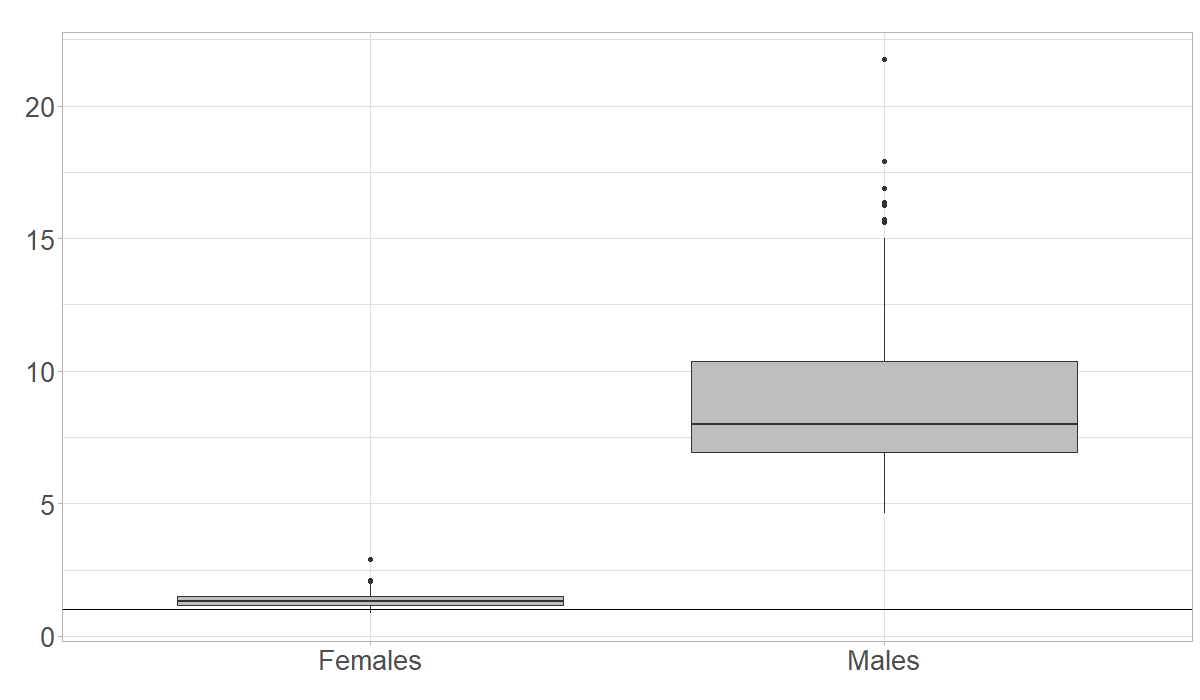}}
\caption{Posterior distributions of $\{w\}$, i.e., the Almalaurea survey's response bias for the 2011 cohort, by degree program and gender (females on the left), estimated at step (ii), obtained using Almalaurea survey data.}
\label{fig:step2}
\end{figure}

\begin{table}[!t]
\centering
\small
\caption{Posterior summaries of the odds ratios $w$, for females ($w_F$) and males ($w_M$) and by degree program, obtained using Almalaurea survey data.}
\label{tab:step2}
\begin{tabular}{@{}lcccccc@{}}
\toprule
                                                                                                & \multicolumn{3}{c}{$w_F^{11}$}                & \multicolumn{3}{c}{$w_M^{11}$}                \\ \midrule
                                                                                                & \begin{tabular}[c]{@{}c@{}}Posterior \\ mean\end{tabular} & \begin{tabular}[c]{@{}c@{}}Posterior \\ median\end{tabular} & \begin{tabular}[c]{@{}c@{}}Posterior \\ sd\end{tabular} & \begin{tabular}[c]{@{}c@{}}Posterior \\ mean\end{tabular}  & \begin{tabular}[c]{@{}c@{}}Posterior \\ median\end{tabular}  & \begin{tabular}[c]{@{}c@{}}Posterior \\ sd\end{tabular}  \\ \cmidrule(l){2-7} 
\begin{tabular}[c]{@{}l@{}}B\&A, Economics, Finance\end{tabular}                             & 27.05       & 25.95     & 7.70                      & 7.37       & 7.13    & 1.52                     \\ \midrule
\begin{tabular}[c]{@{}l@{}}Communication and Publishing\end{tabular}                         & 33.44       & 31.04     & 12.80                     & 5.01       & 4.44    & 2.25                      \\ \midrule
\begin{tabular}[c]{@{}l@{}}Industrial Engineering\end{tabular}                               & 8.47        & 8.17      & 2.51                      & 90.62      & 86.00   & 24.35                    \\ \midrule
\begin{tabular}[c]{@{}l@{}}Literature and Humanities\end{tabular}                            & 0.29        & 0.29      & 0.06                      & 1.48       & 1.44    & 0.36                      \\ \midrule
\begin{tabular}[c]{@{}l@{}}Political Science\end{tabular}                                    & 6.40        & 6.14      & 1.54                      & 0.78       & 0.75    & 0.18                      \\ \midrule
\begin{tabular}[c]{@{}l@{}}Science and IT\end{tabular}                                       & 1.36        & 1.33      & 0.27                      & 8.77       & 8.00    & 2.66                     \\ \bottomrule
\end{tabular}
\end{table} 

\clearpage
\begin{figure}[!h]
    \centering
\subfloat[][Agriculture and Forestry, Veterinary - males]{
    \includegraphics[width=.5\linewidth]{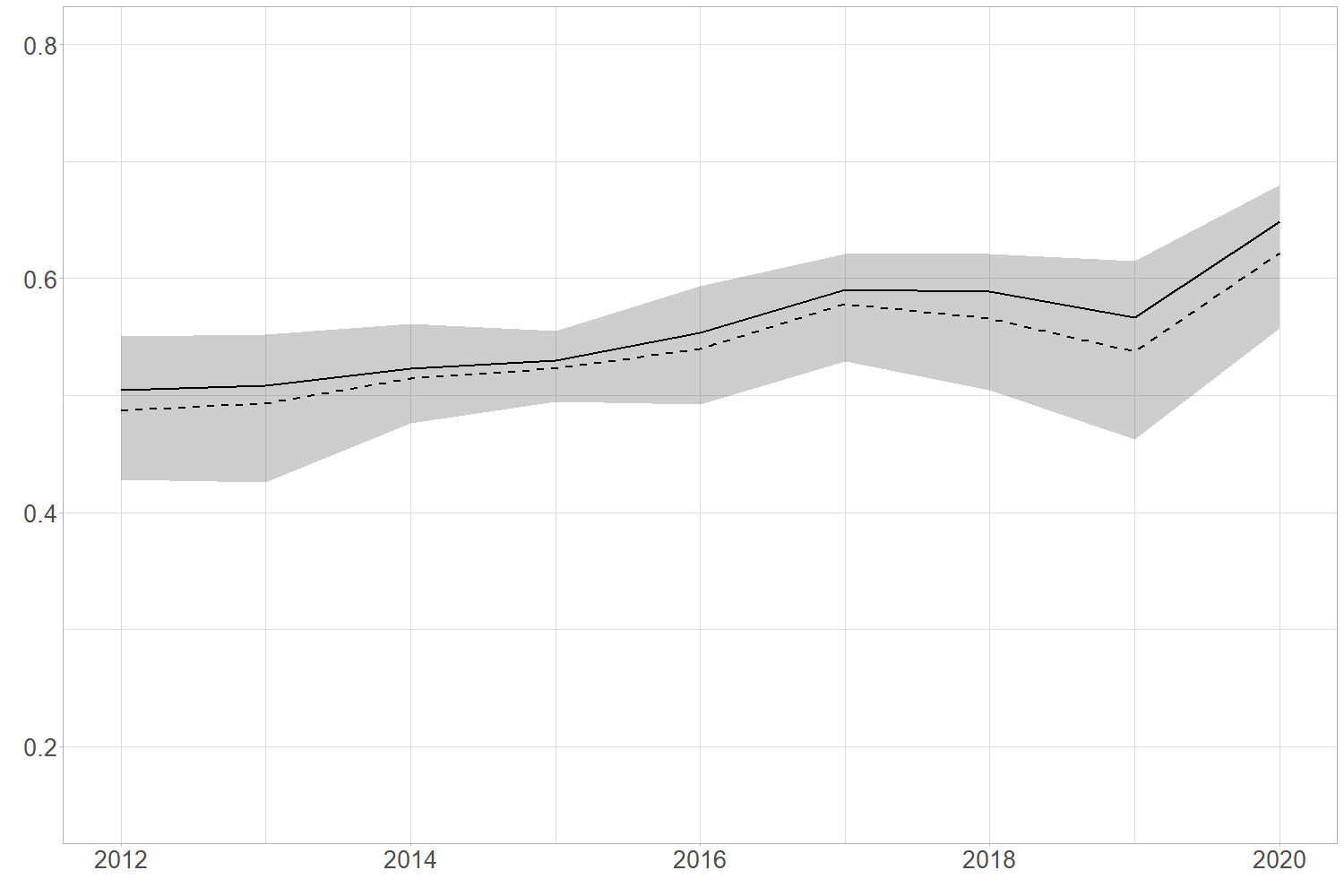}}
\subfloat[][Architecture and Engineering - males]{
    \includegraphics[width=.5\linewidth]{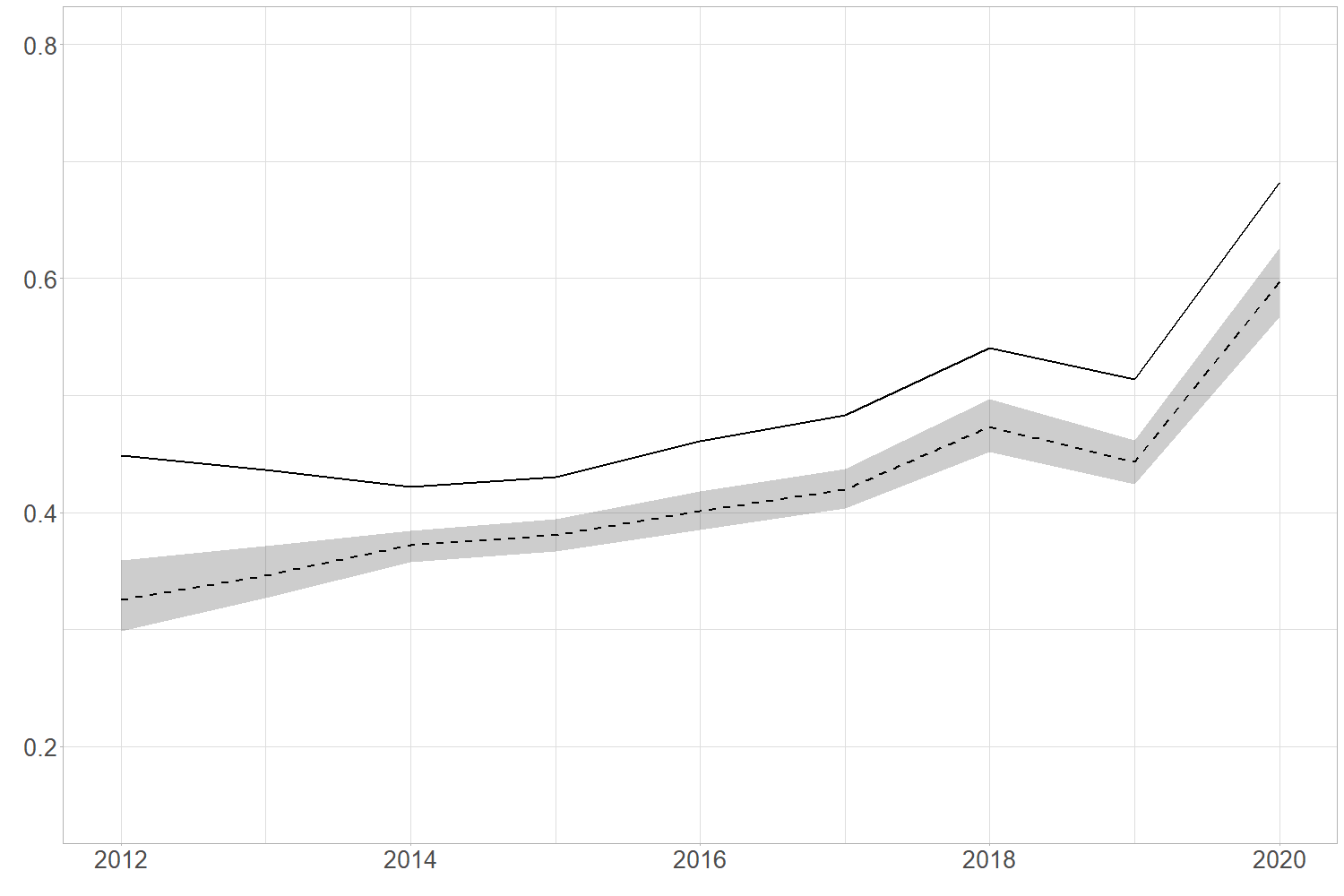}}
\qquad
\subfloat[][B\&A, Economics, Finance - females]{
    \includegraphics[width=.5\linewidth]{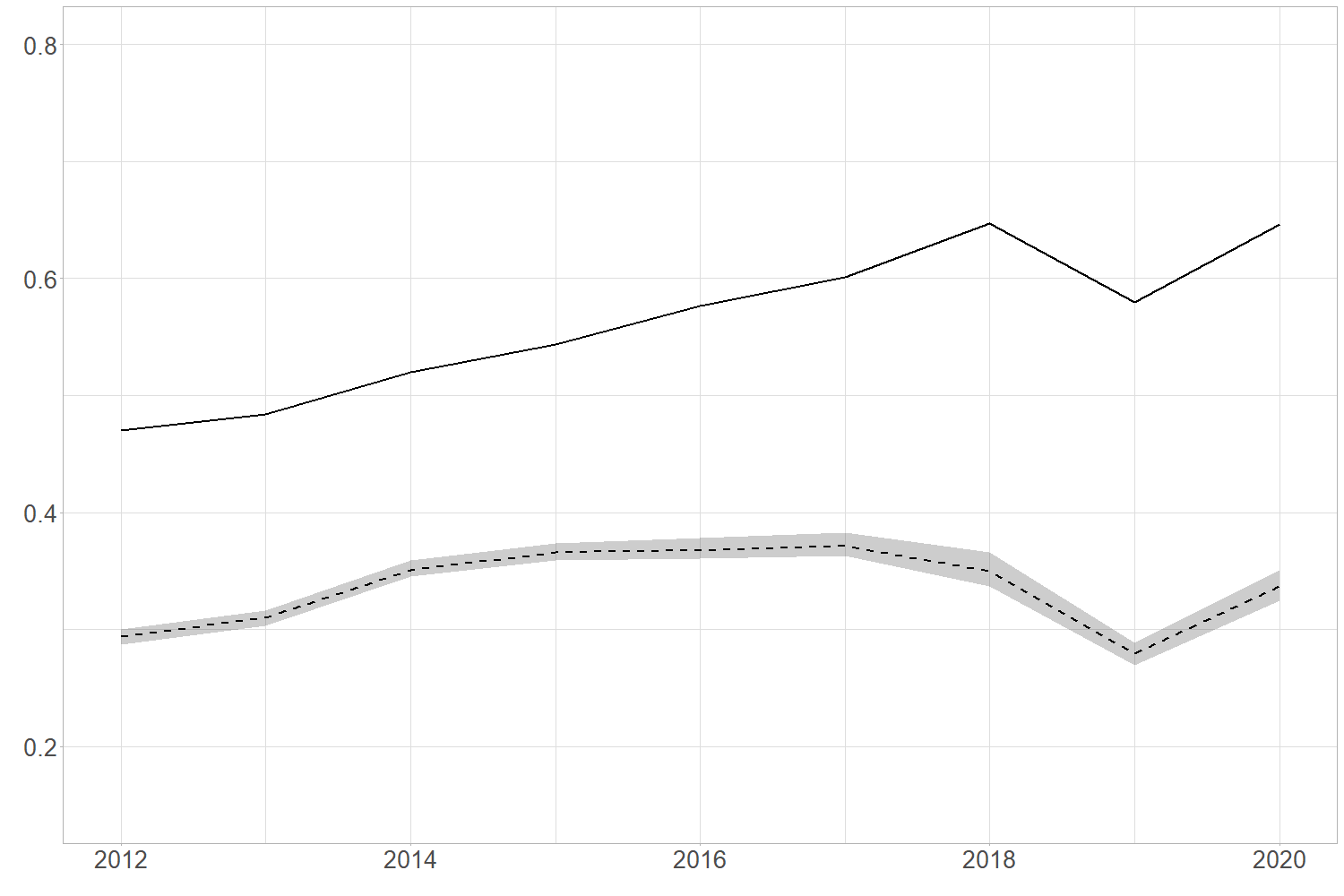}}
\subfloat[][B\&A, Economics, Finance - males]{
    \includegraphics[width=.5\linewidth]{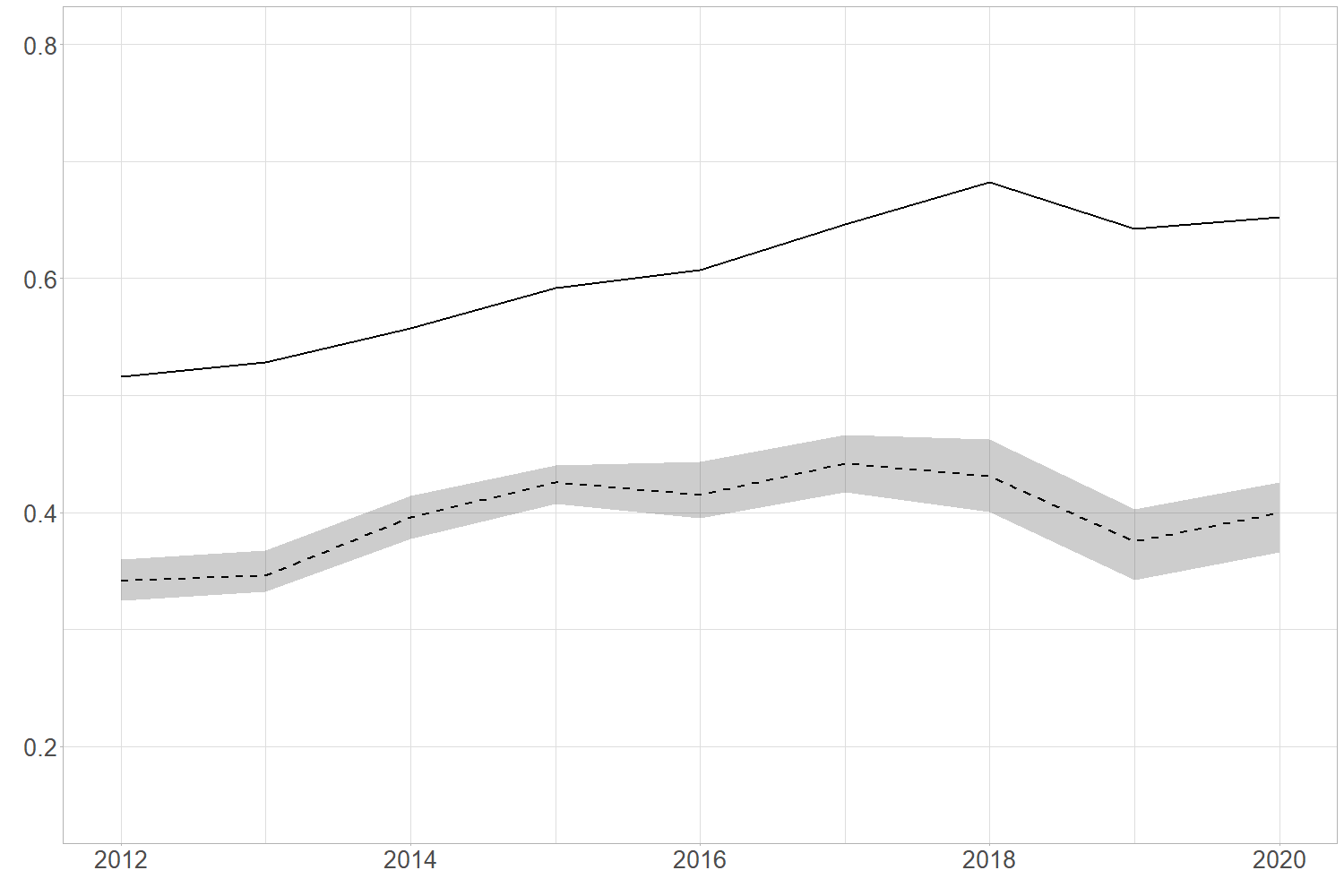}}
\qquad
\subfloat[][Communication and Publishing - females]{
    \includegraphics[width=.5\linewidth]{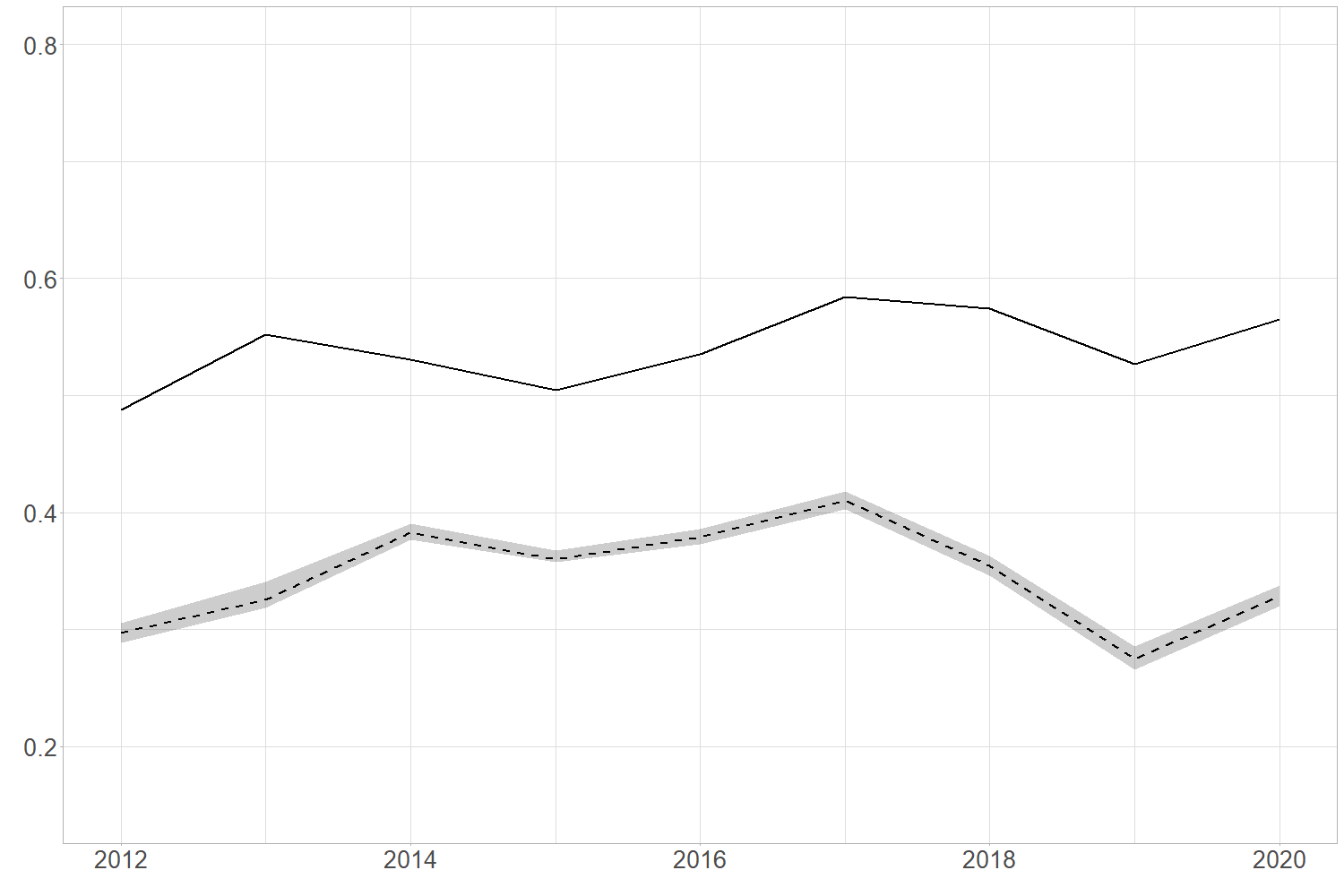}}
\subfloat[][Communication and Publishing - males]{
    \includegraphics[width=.5\linewidth]{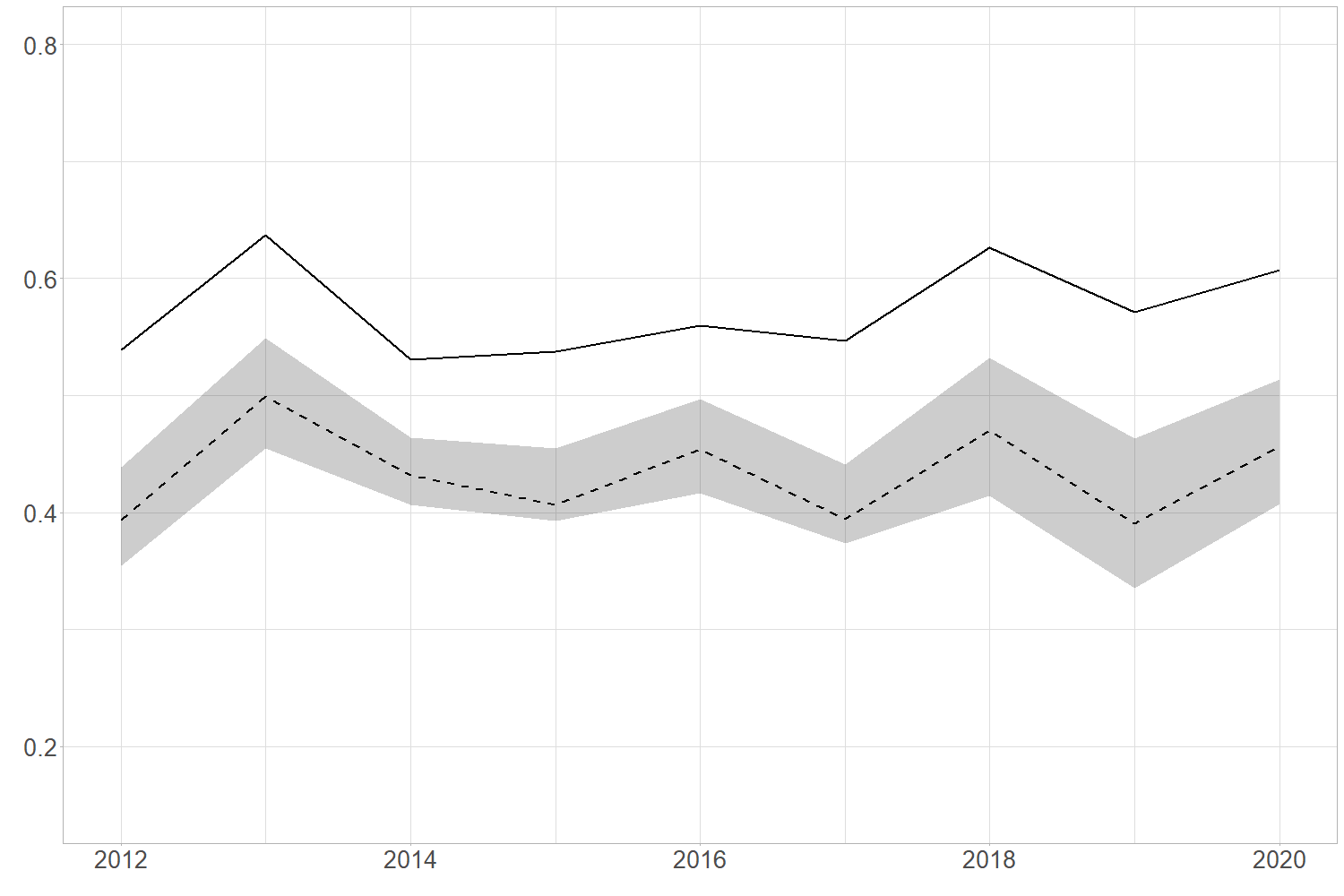}}
\qquad
\subfloat[][Industrial and Information Engineering - females]{
    \includegraphics[width=.5\linewidth]{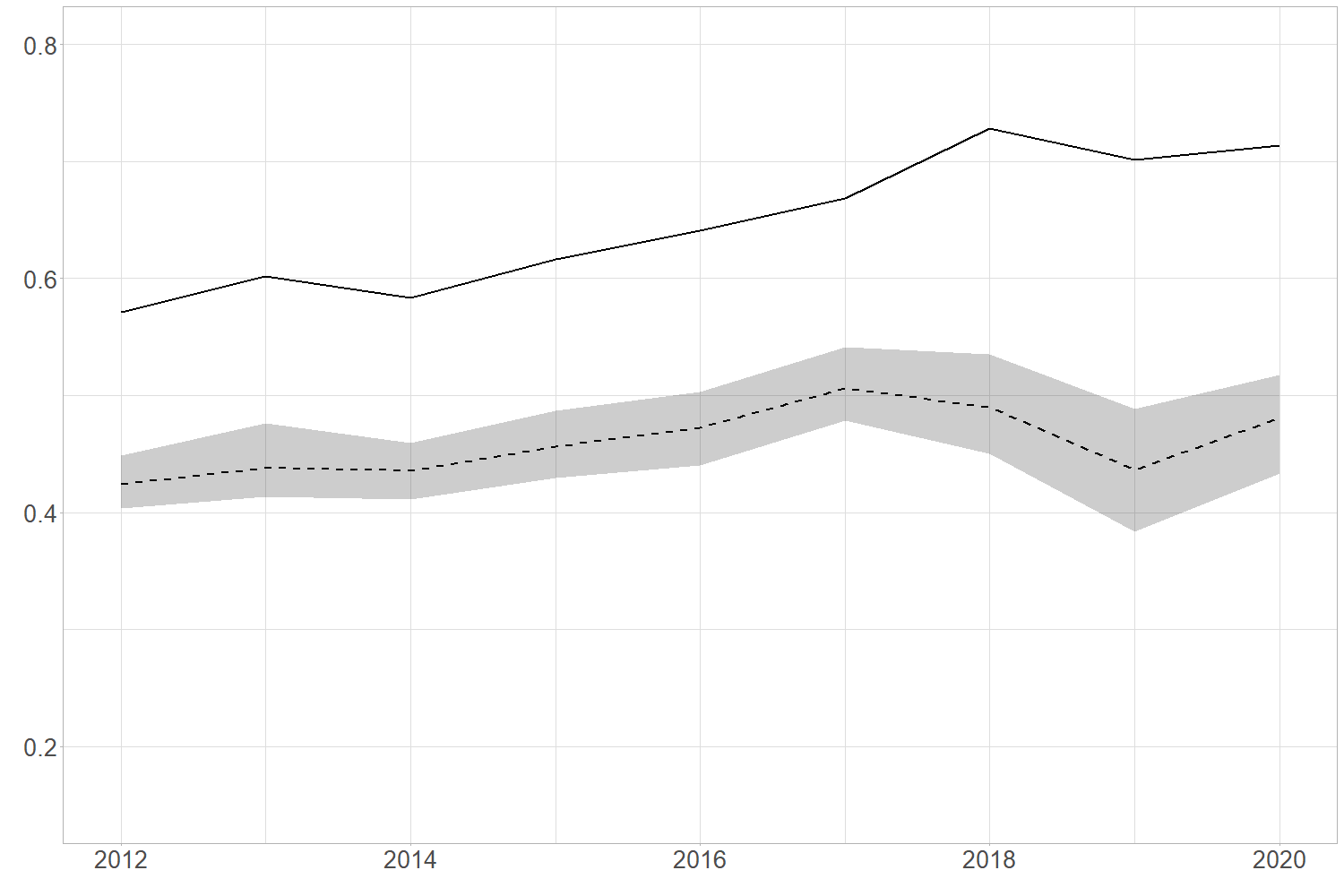}}
\subfloat[][Industrial and Information Engineering - males]{
    \includegraphics[width=.5\linewidth]{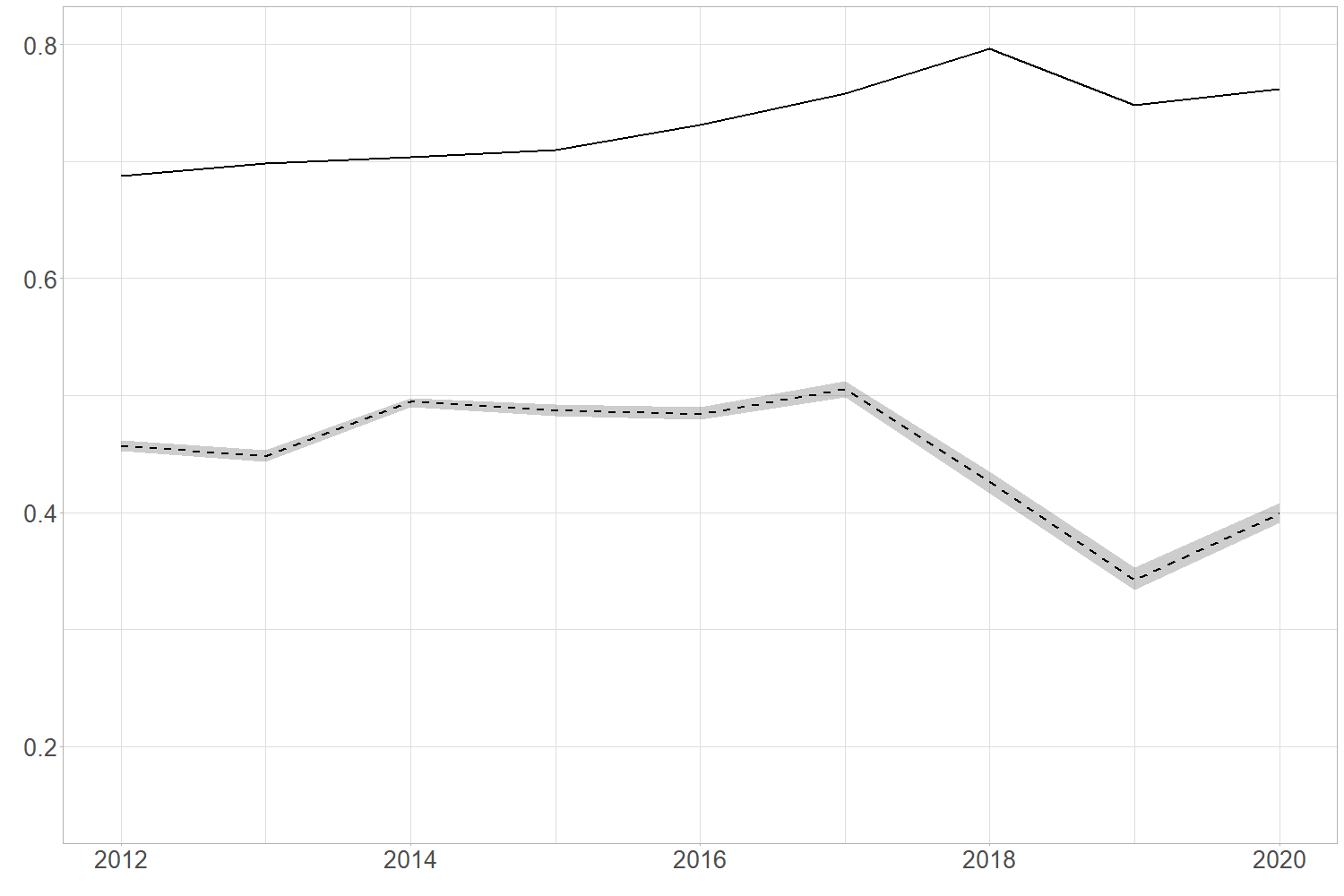}}
\end{figure}
\begin{figure}\ContinuedFloat
\centering
\subfloat[][Law and Legal sciences - males]{
    \includegraphics[width=.5\linewidth]{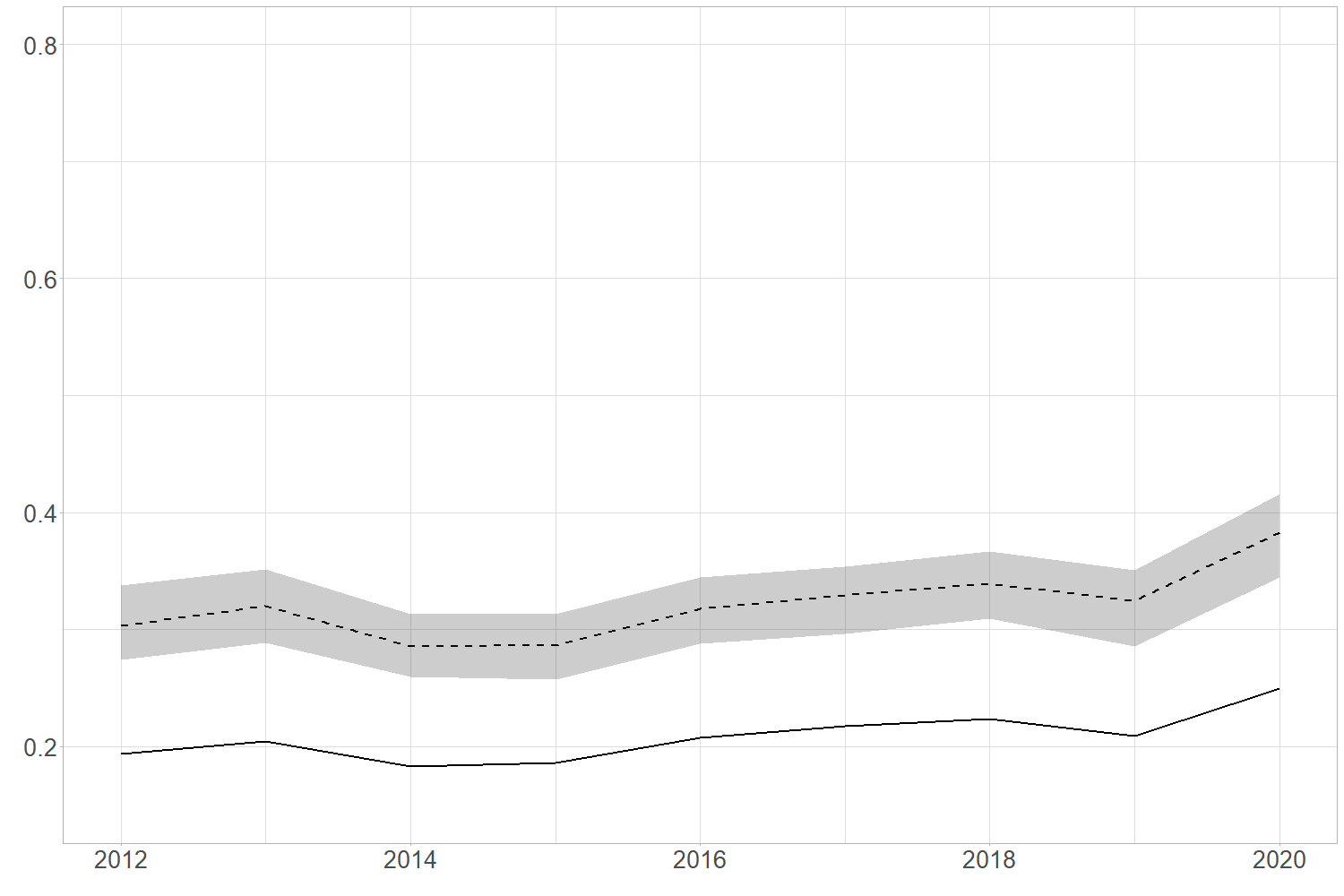}}
    \subfloat[][Medicine, Dentistry, Pharmacy - males]{
    \includegraphics[width=.5\linewidth]{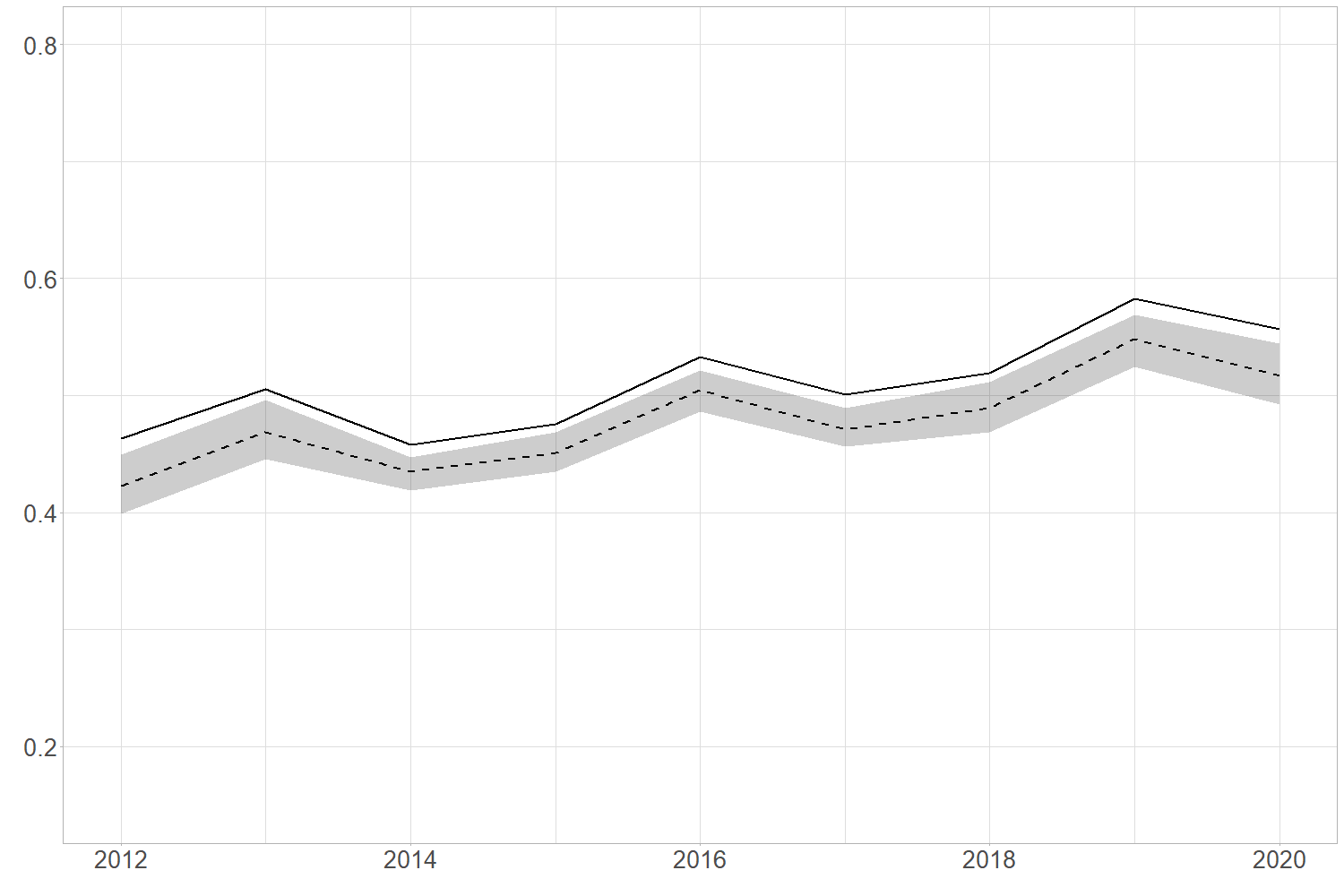}}
\qquad
\subfloat[][Literature and Humanities - females]{
    \includegraphics[width=.5\linewidth]{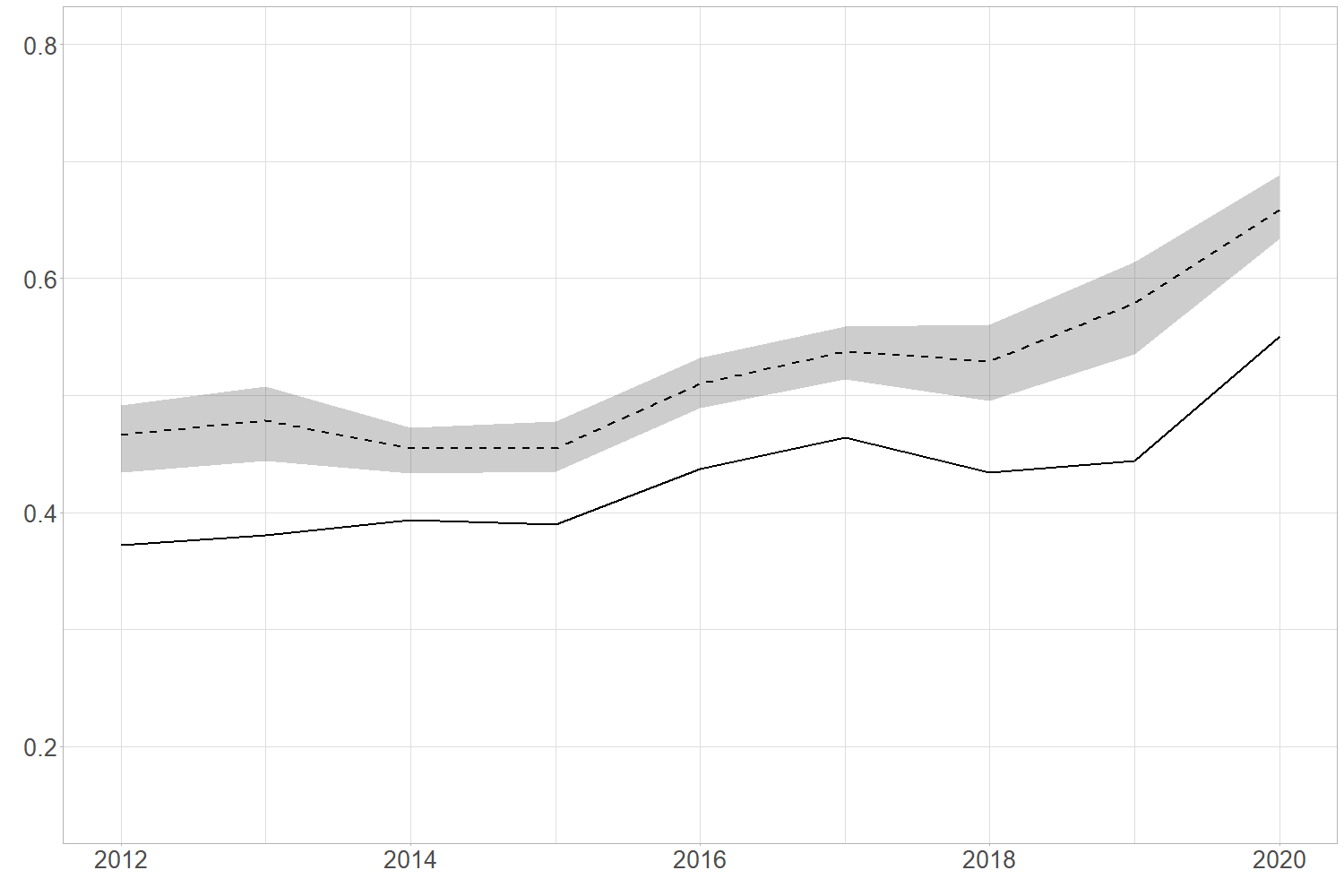}}
\subfloat[][Literature and Humanities - males]{
    \includegraphics[width=.5\linewidth]{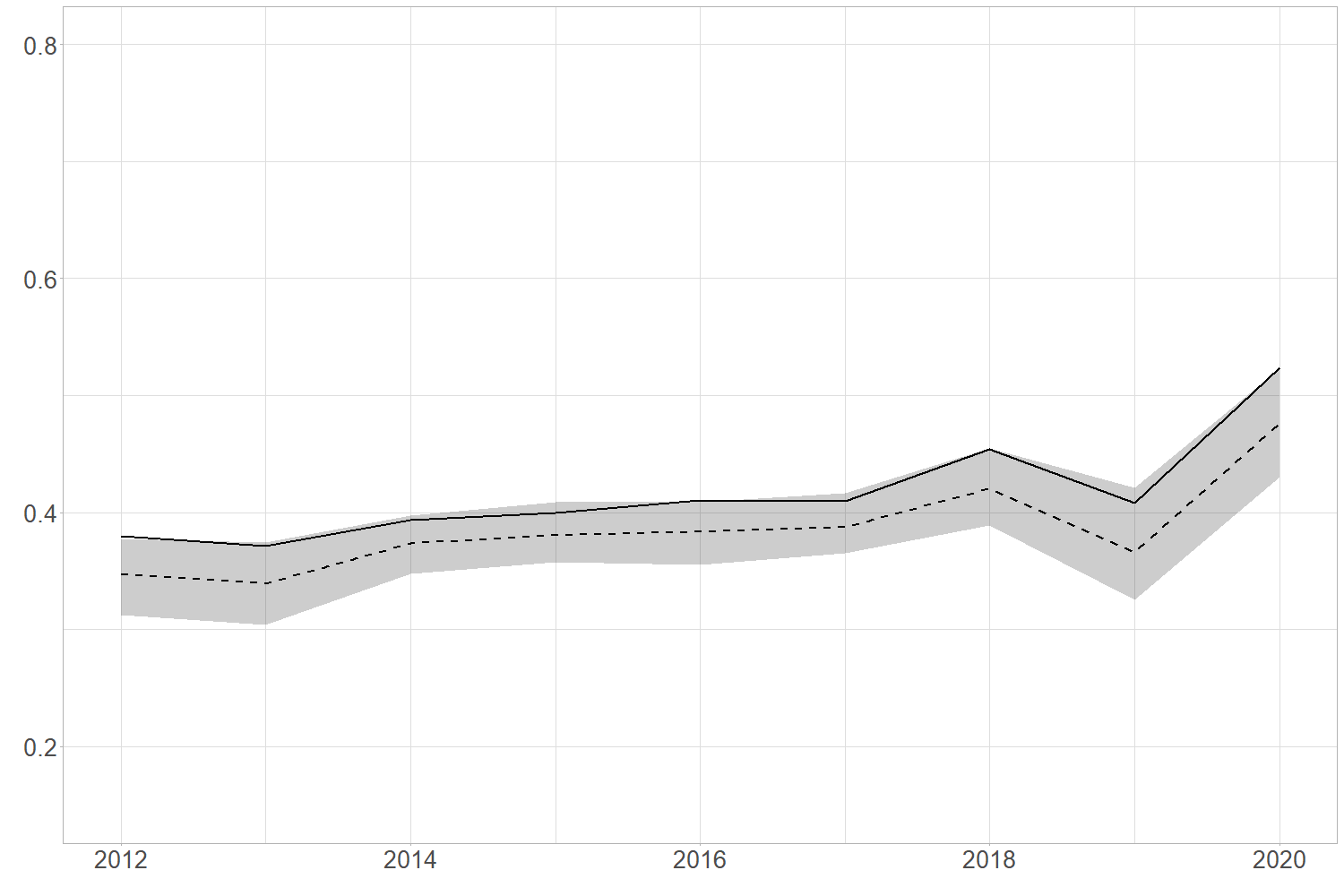}}

\qquad
\subfloat[][Political Science - females]{
    \includegraphics[width=.5\linewidth]{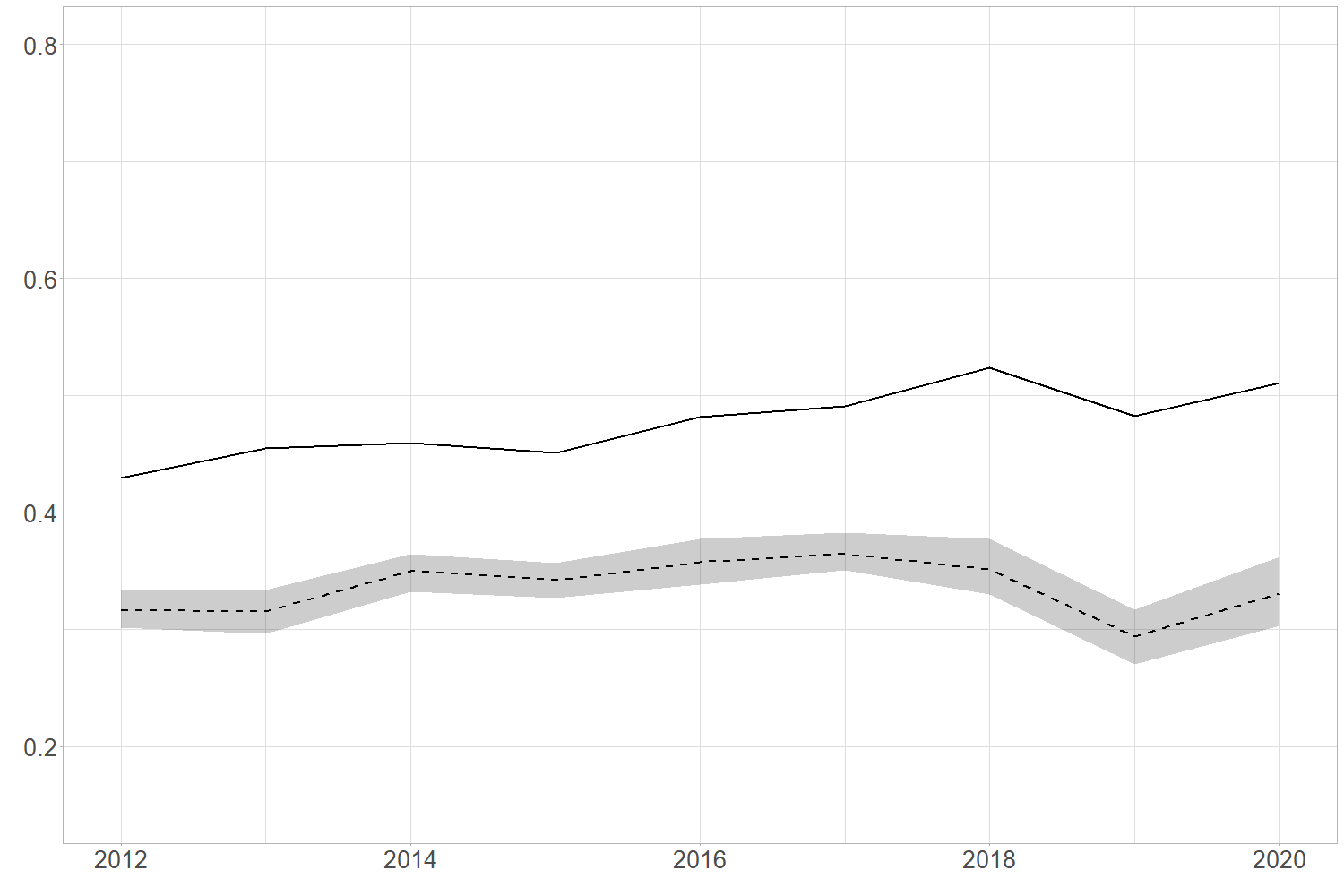}}
\subfloat[][Political Science - males]{
    \includegraphics[width=.5\linewidth]{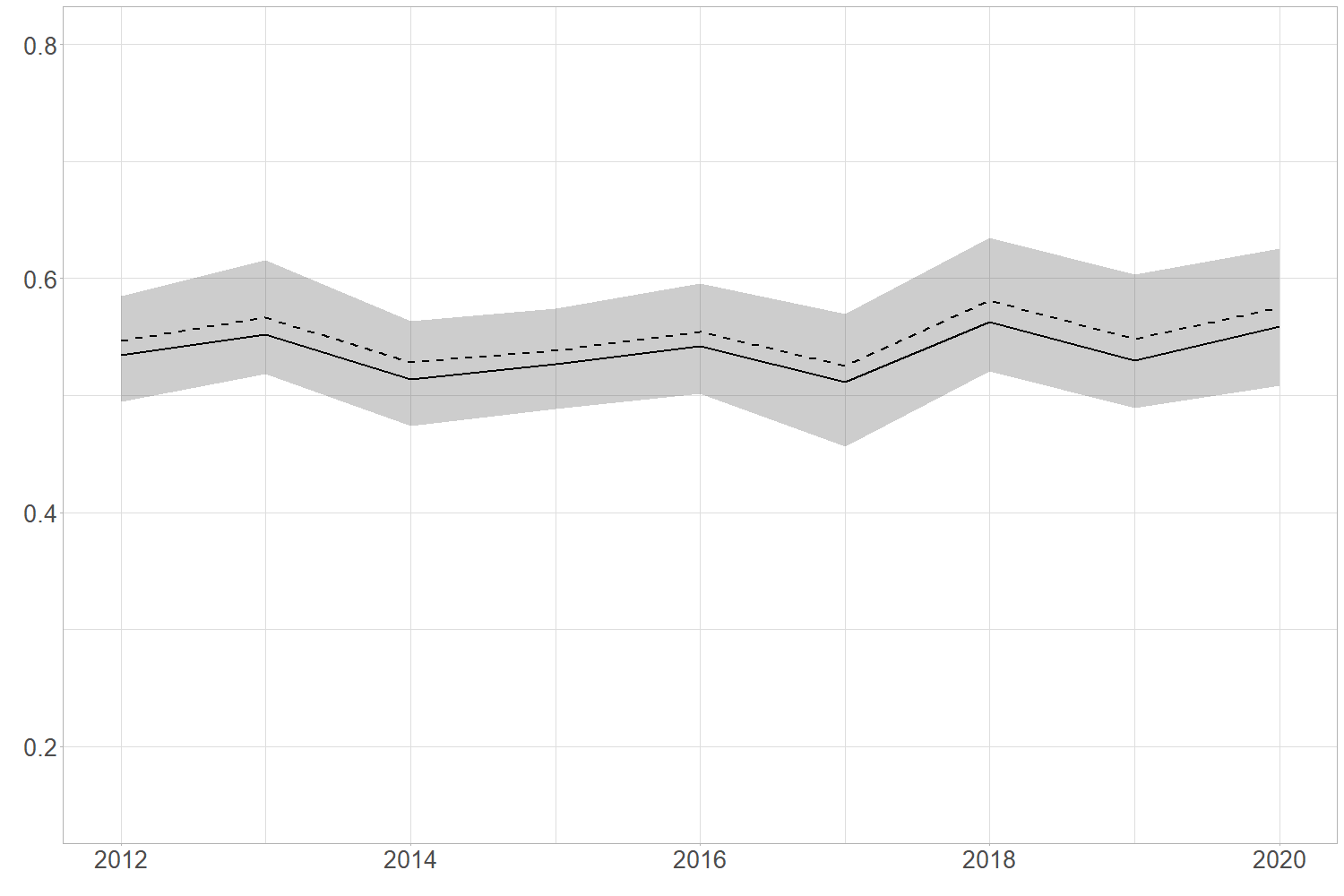}}
\qquad
\subfloat[][Science and IT - females]{
    \includegraphics[width=.5\linewidth]{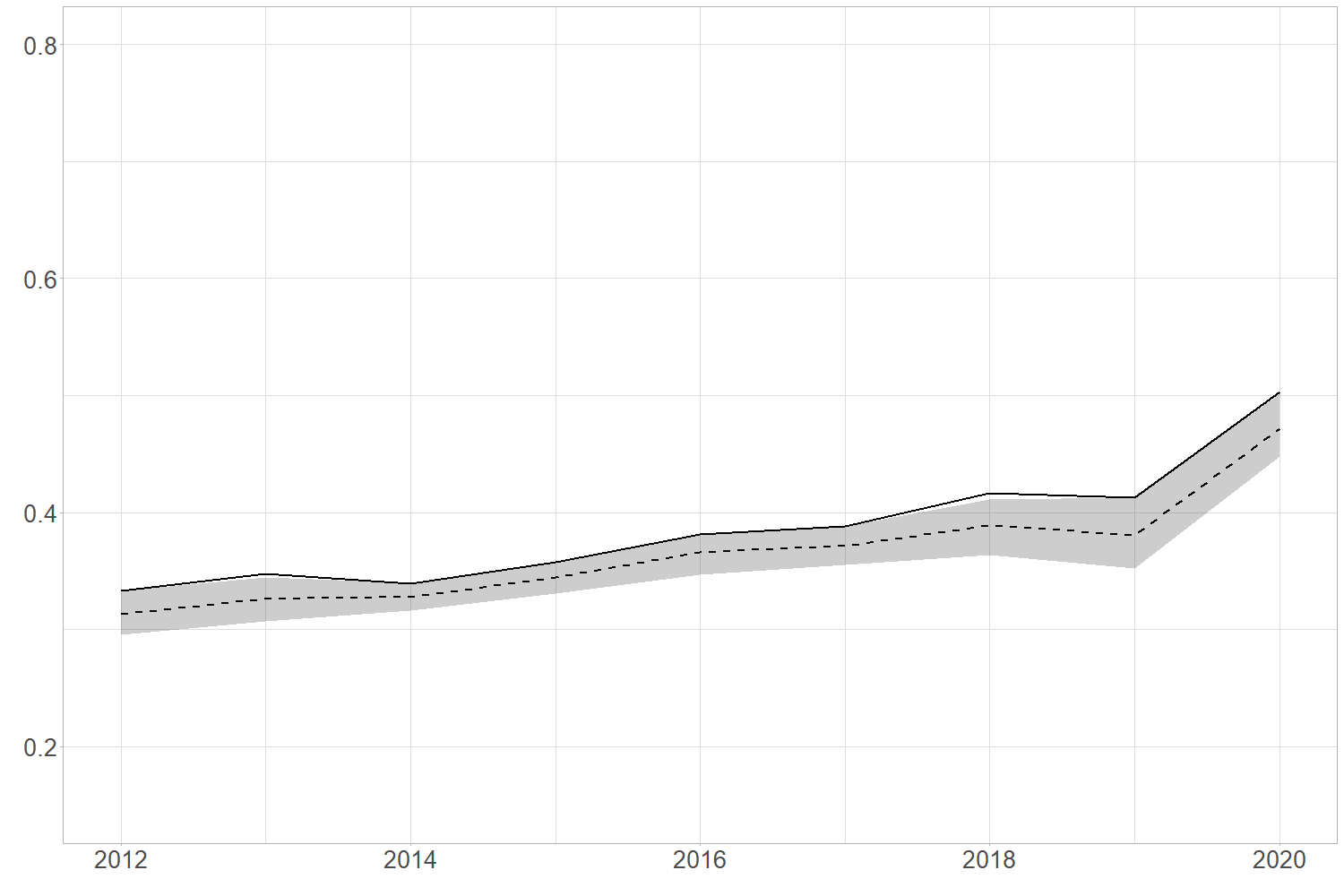}}
\subfloat[][Science and IT - males]{
    \includegraphics[width=.5\linewidth]{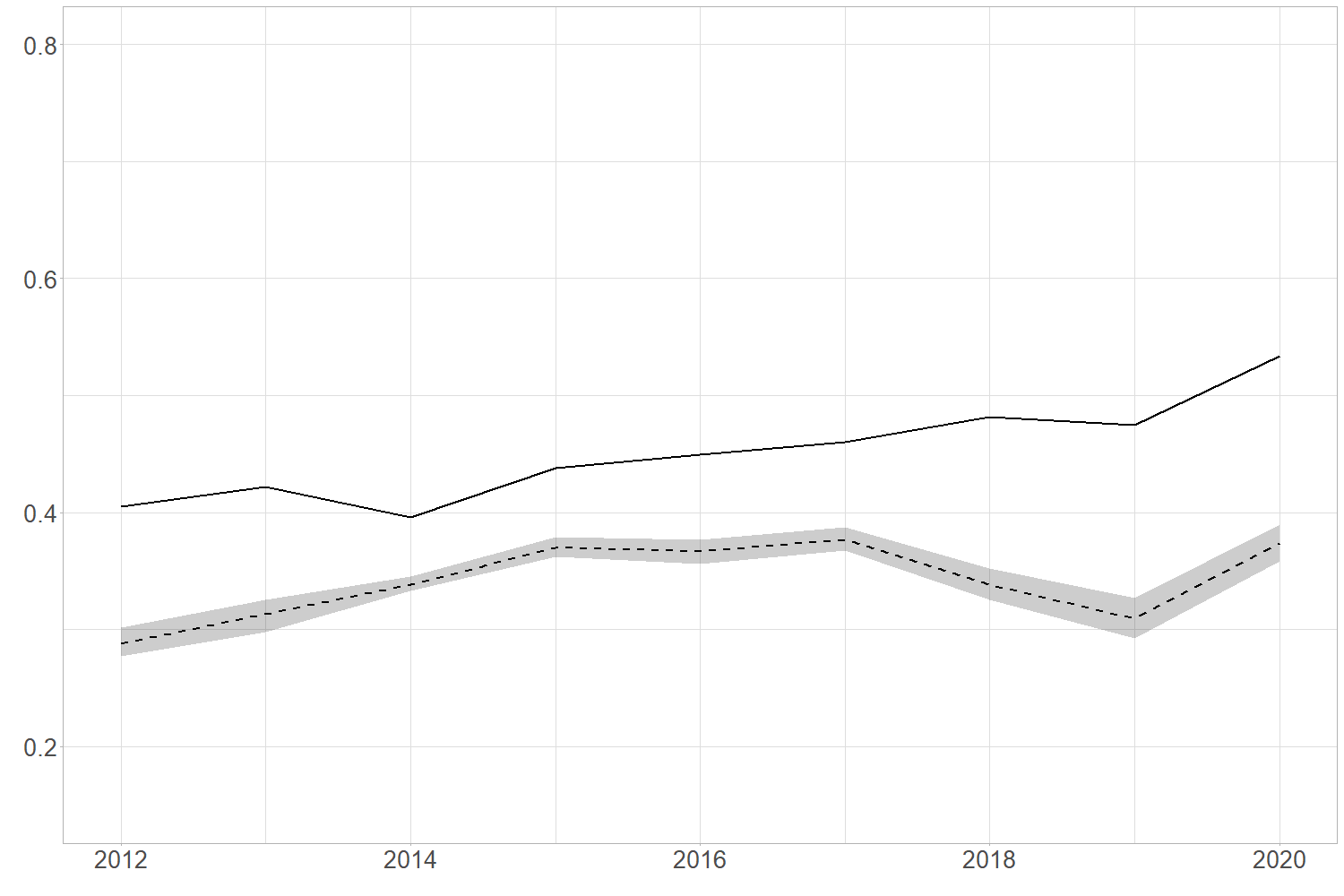}}
\caption{Posterior mean (dashed line) and 95\% highest posterior density interval of the females' employment rates for the years between 2012 and 2020, by degree program, estimated at step (iii), versus the employment rates computed using raw Almalaurea data (solid line).}
\label{fig:step3}
\end{figure}

\clearpage
\section{Sensitivity results}
\subsection{Sensitivity to prior specification at step (ii)}
We test the robustness of the odds ratio estimates at step (ii), namely for the year 2011, to different specifications of its prior standard deviation.
Figure \ref{fig:sens1} shows that the results are robust, with a few exceptions. 
As expected, when a substantial conflict between different data sources occurs, the impact of the prior is more evident. This is the case of females in ``Communication and Publishing'' and ``Political Science'', and of males in ``Industrial Engineering''.
In these cases, allowing for a wide prior affects the convergence of the MCMC. 
Generally speaking, one should not rely on results showing too large odds ratios, the interpretation of which cannot be precise.

\begin{figure}[!h]
    \centering
\subfloat[][Agriculture and Forestry, Veterinary - females]{
    \includegraphics[width=.5\linewidth]{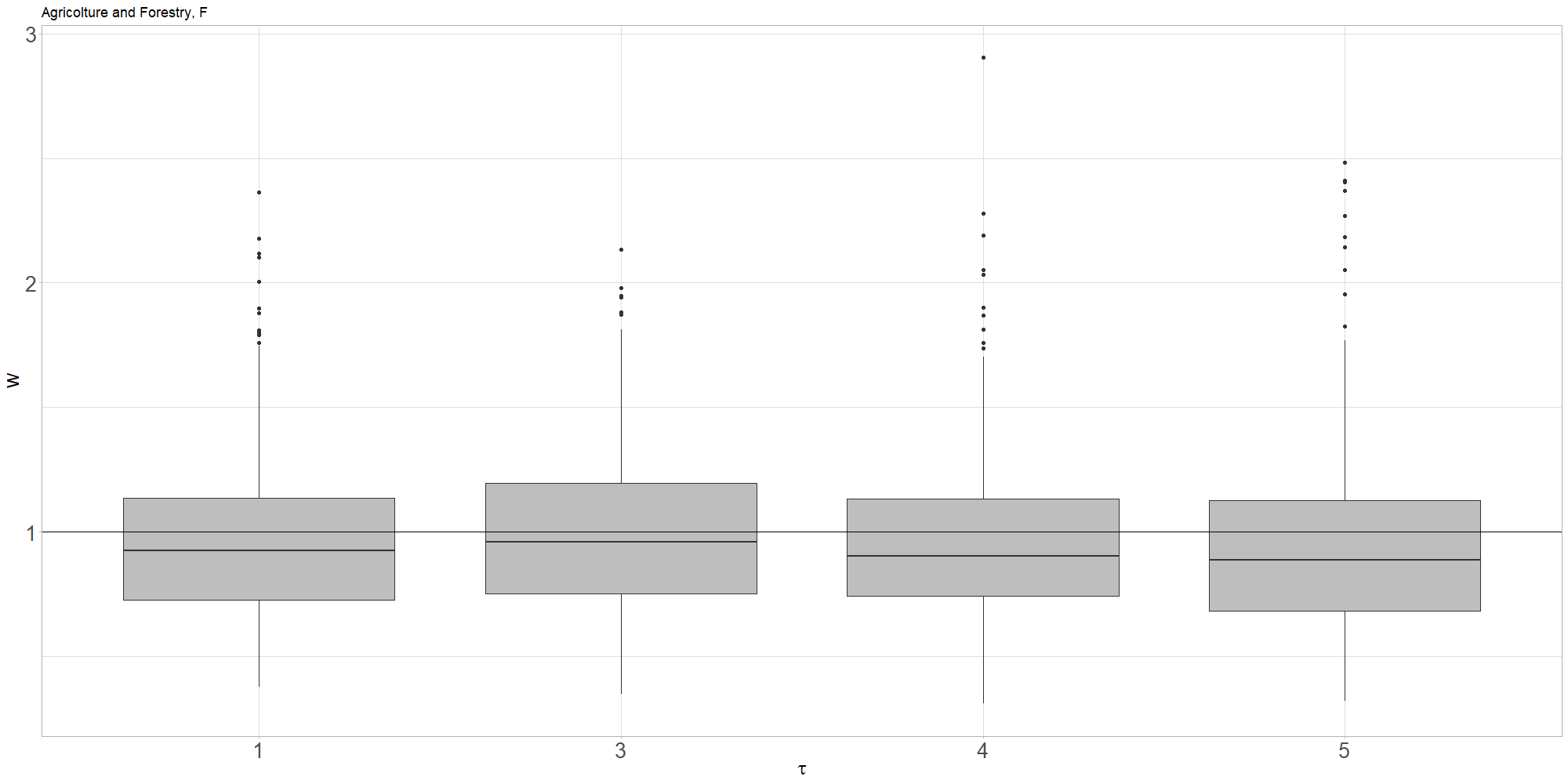}}
\subfloat[][Agriculture and Forestry, Veterinary - males]{
    \includegraphics[width=.5\linewidth]{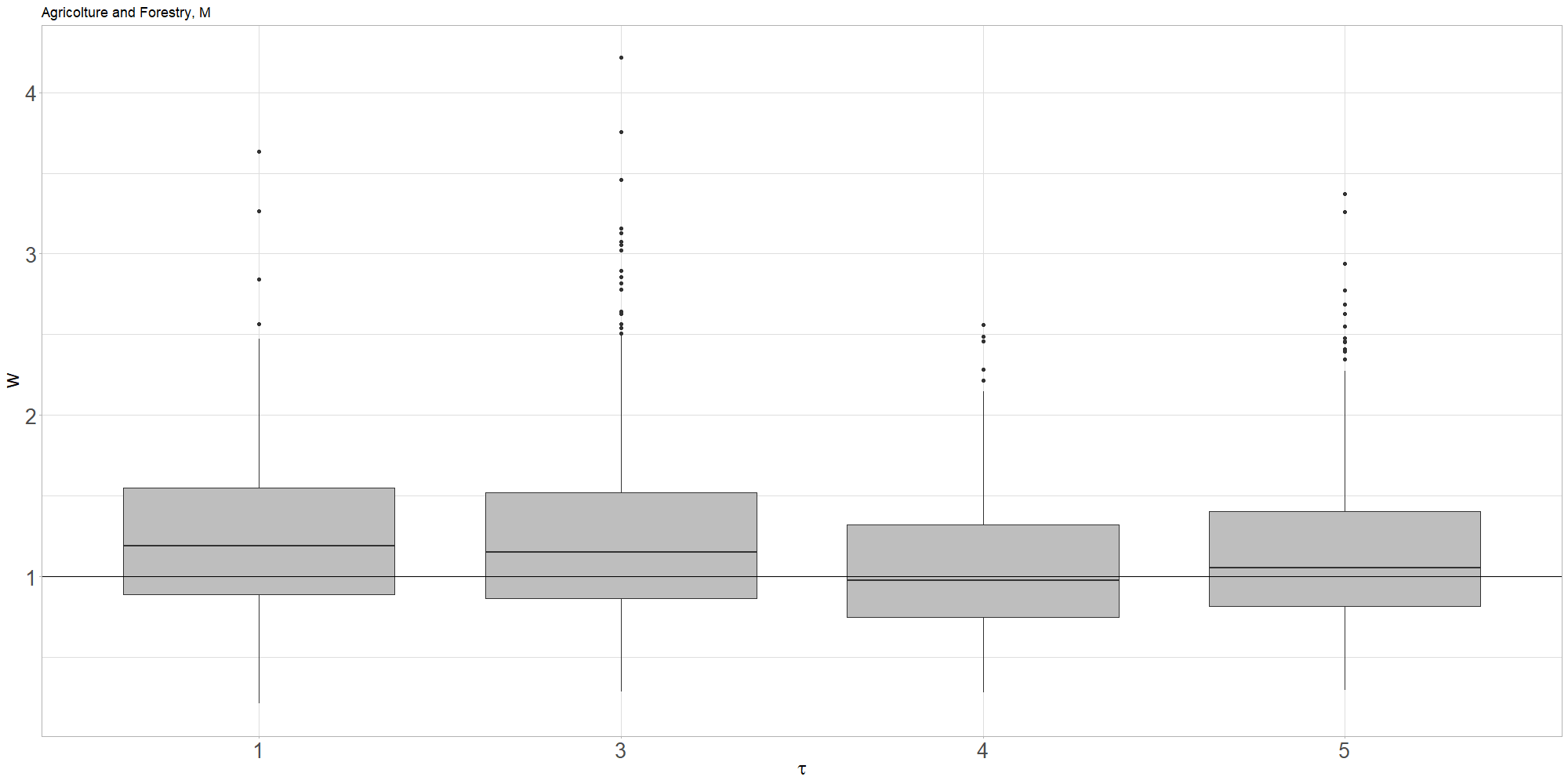}}
\qquad
\subfloat[][Architecture and Engineering - females]{
    \includegraphics[width=.5\linewidth]{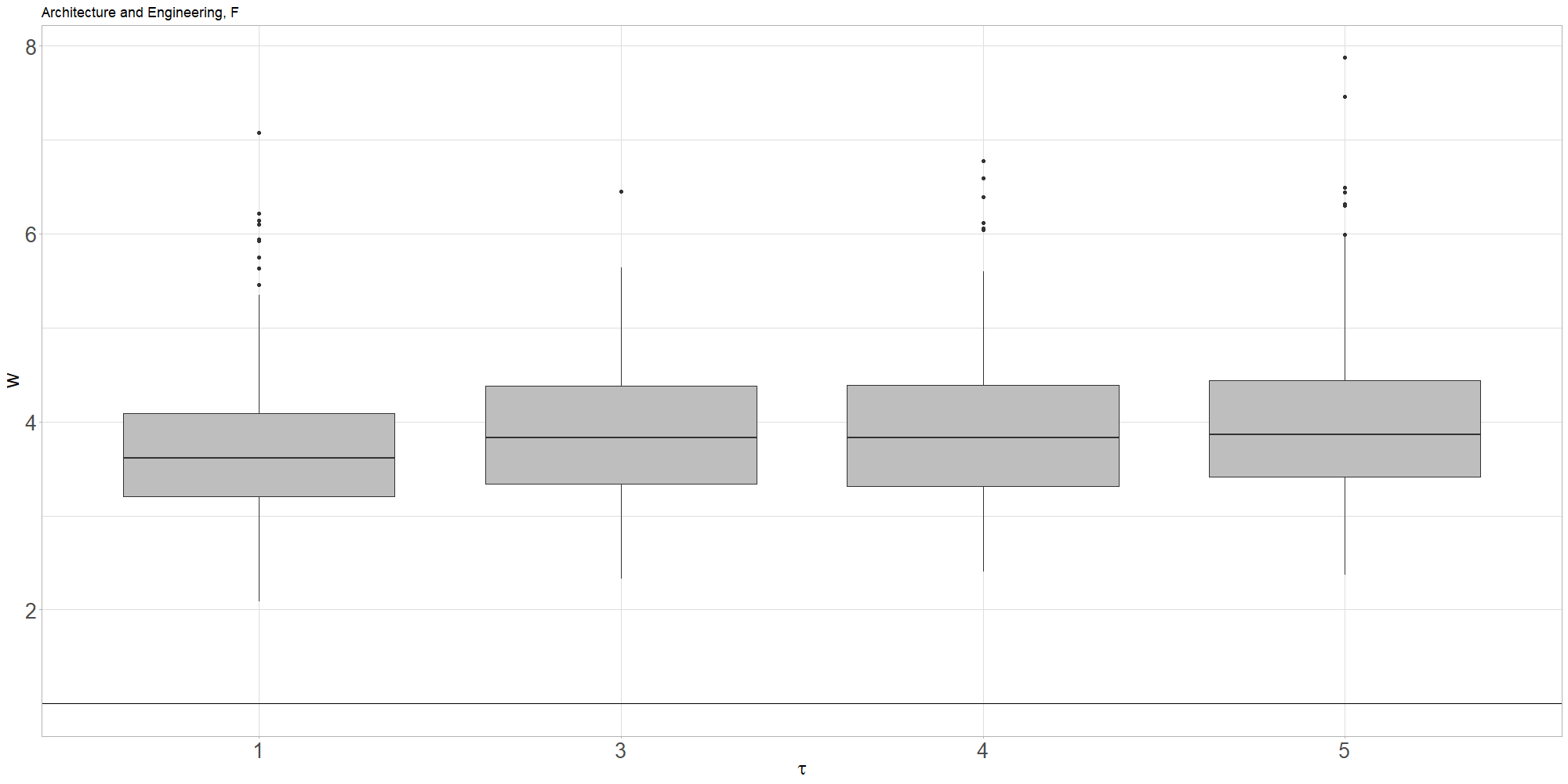}}
\subfloat[][Architecture and Engineering - males]{
    \includegraphics[width=.5\linewidth]{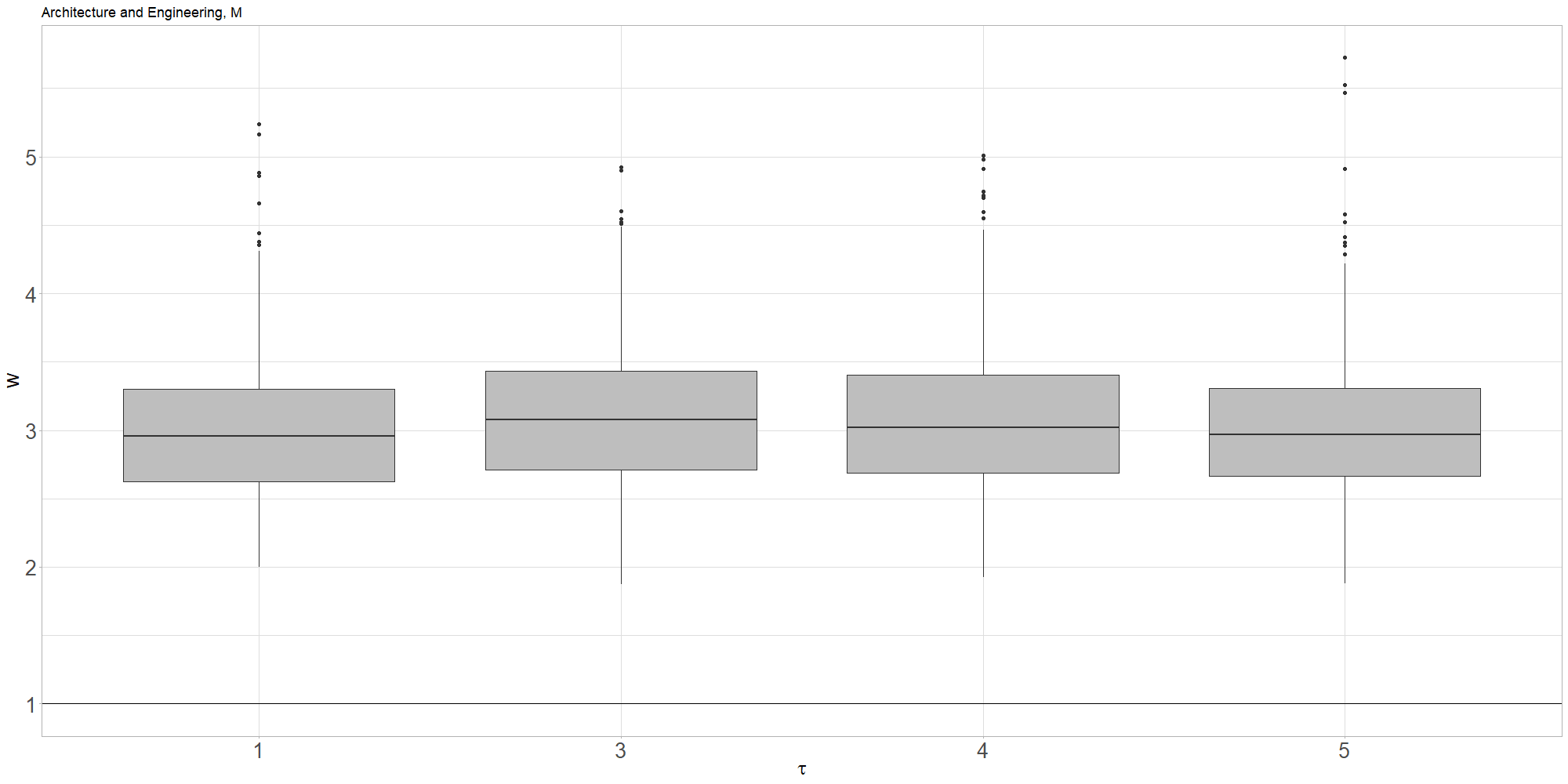}}
\qquad 
\subfloat[][B\&A, Economics, Finance - females]{
    \includegraphics[width=.5\linewidth]{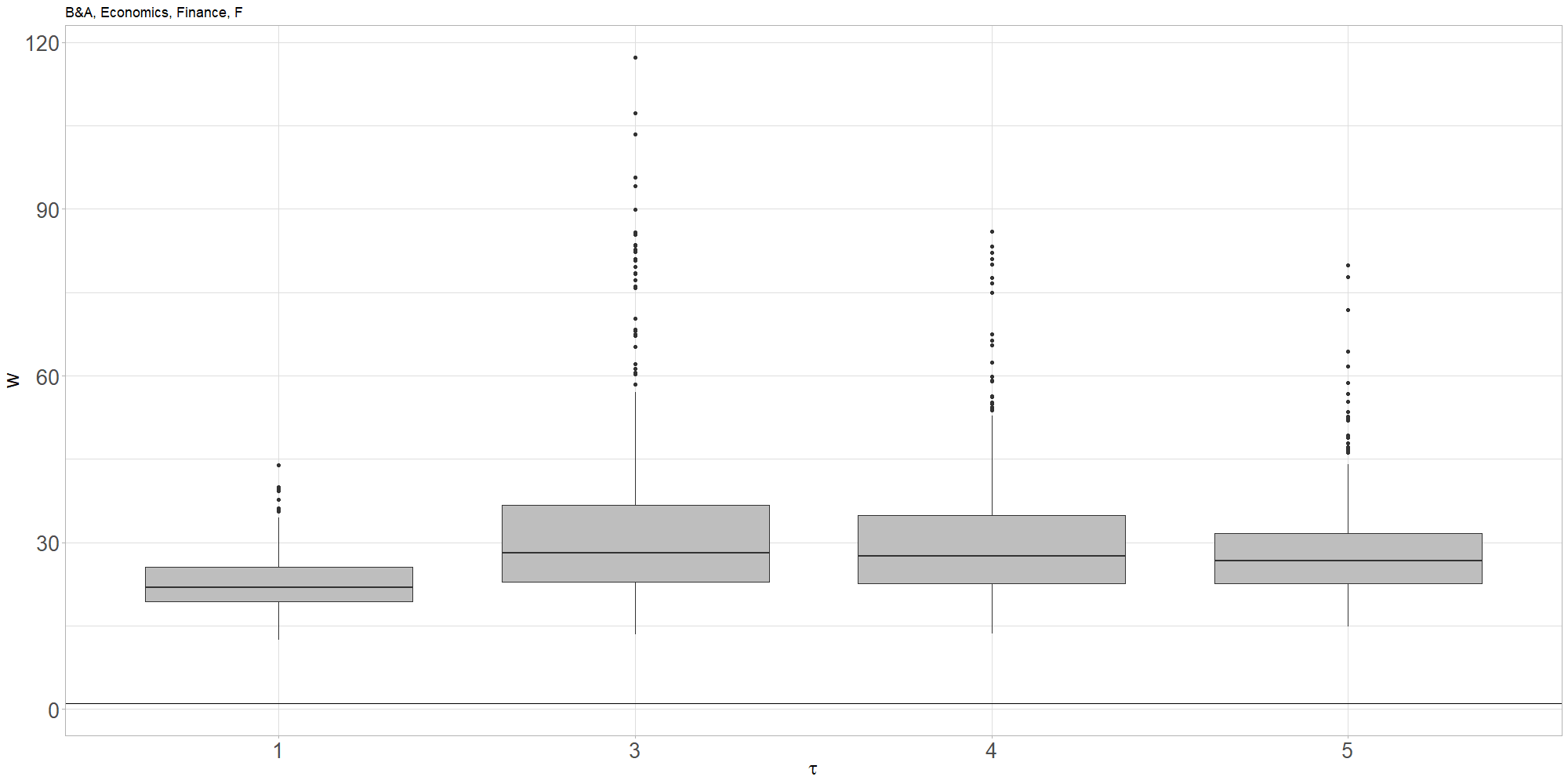}}
\subfloat[][B\&A, Economics, Finance - males]{
    \includegraphics[width=.5\linewidth]{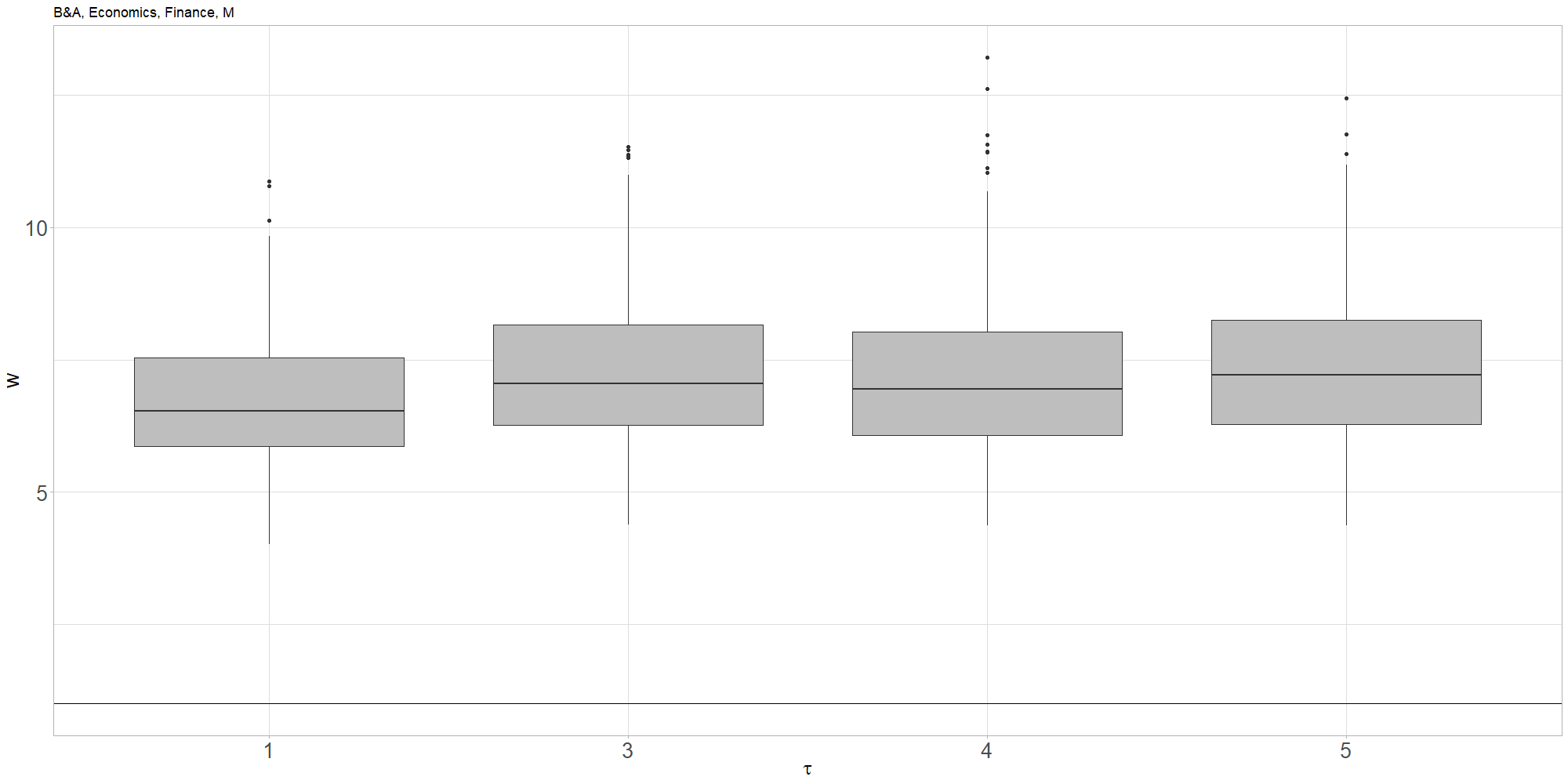}}
\end{figure}
\begin{figure}\ContinuedFloat
\centering
\subfloat[][Communication and Publishing - females]{
    \includegraphics[width=.5\linewidth]{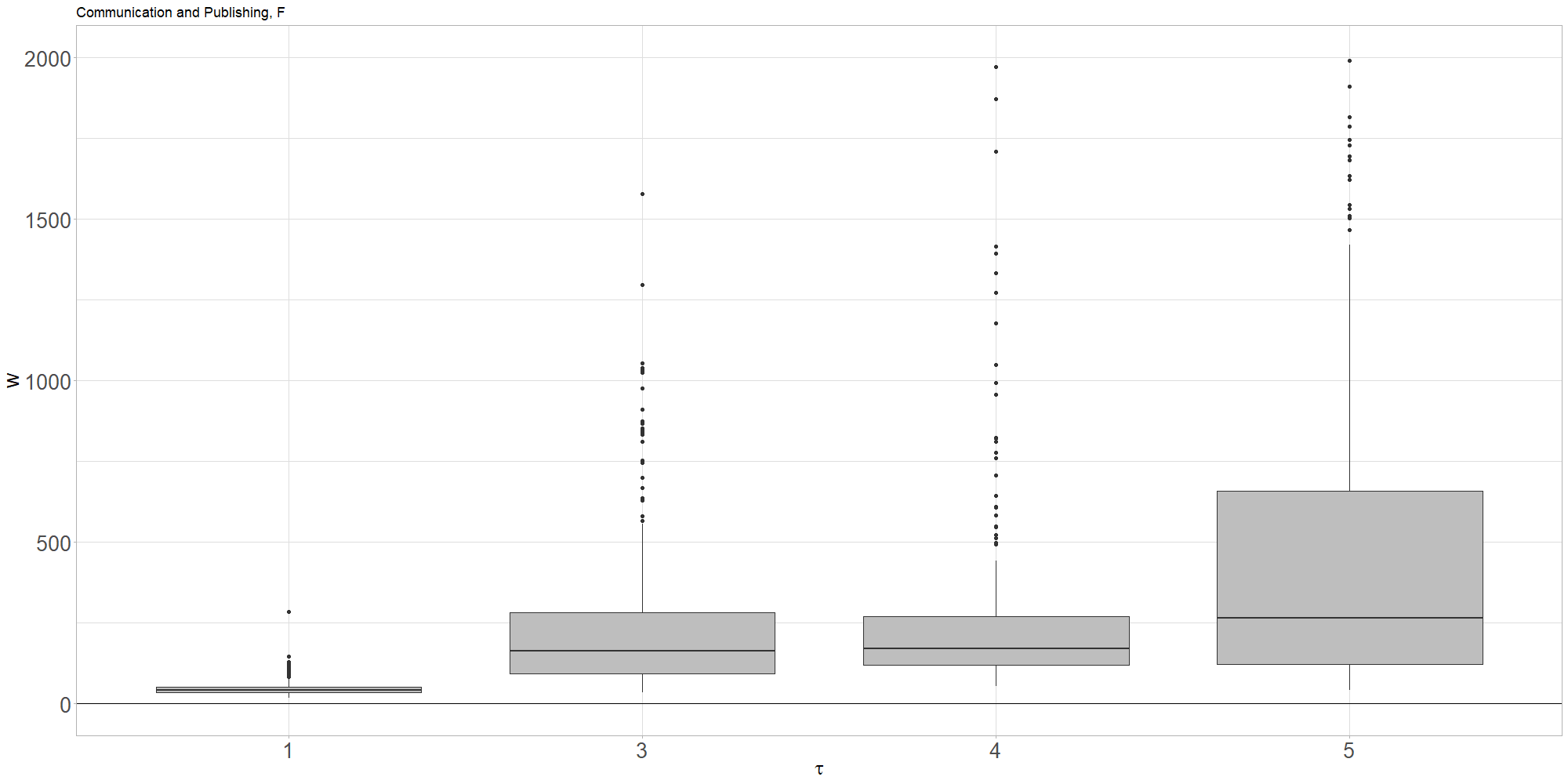}}
\subfloat[][Communication and Publishing - males]{
    \includegraphics[width=.5\linewidth]{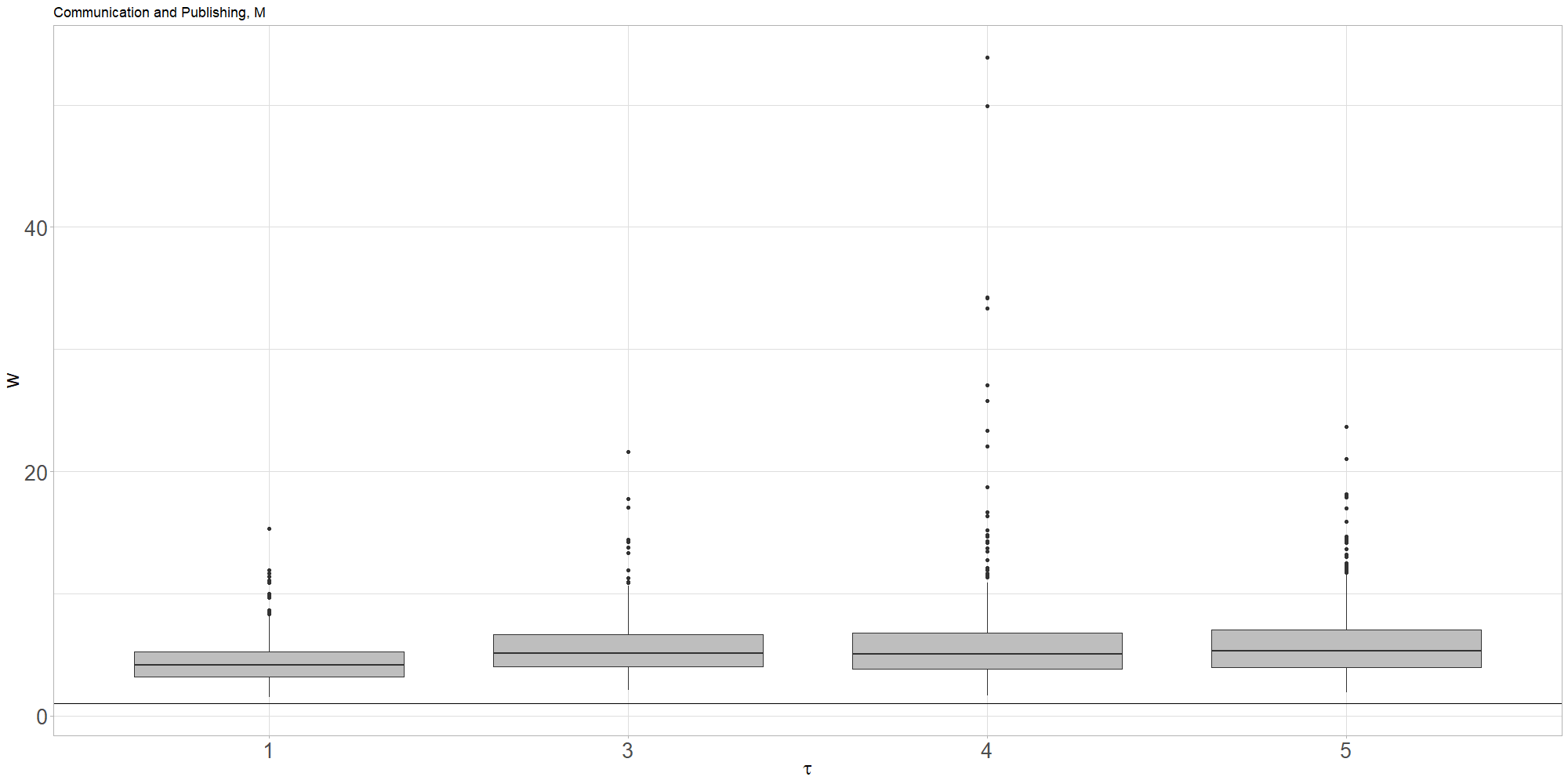}}
\end{figure}
\begin{figure}\ContinuedFloat
\centering
\subfloat[][Industrial and Information Engineering - females]{
    \includegraphics[width=.5\linewidth]{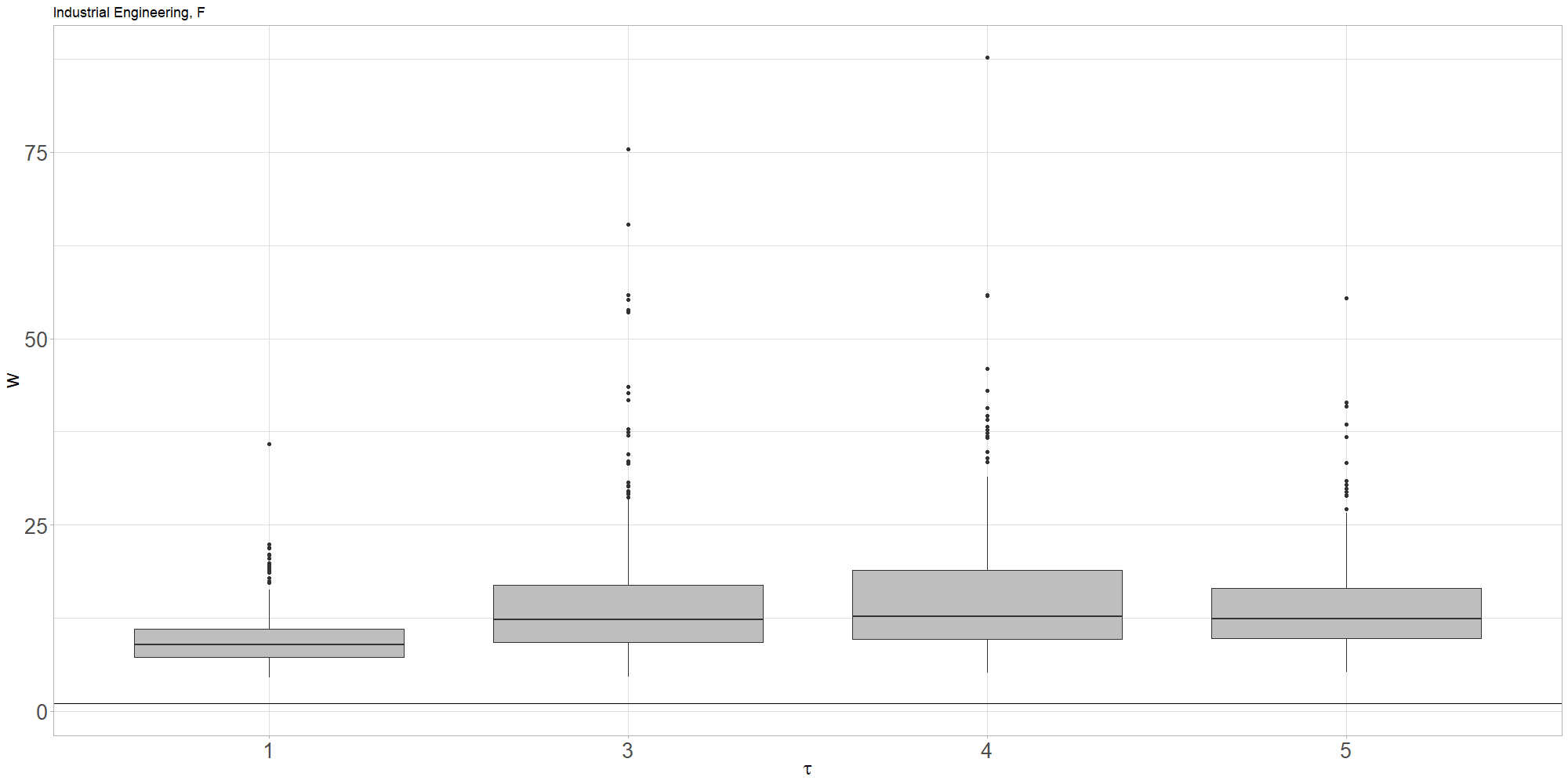}}
\subfloat[][Industrial and Information Engineering - males]{
    \includegraphics[width=.5\linewidth]{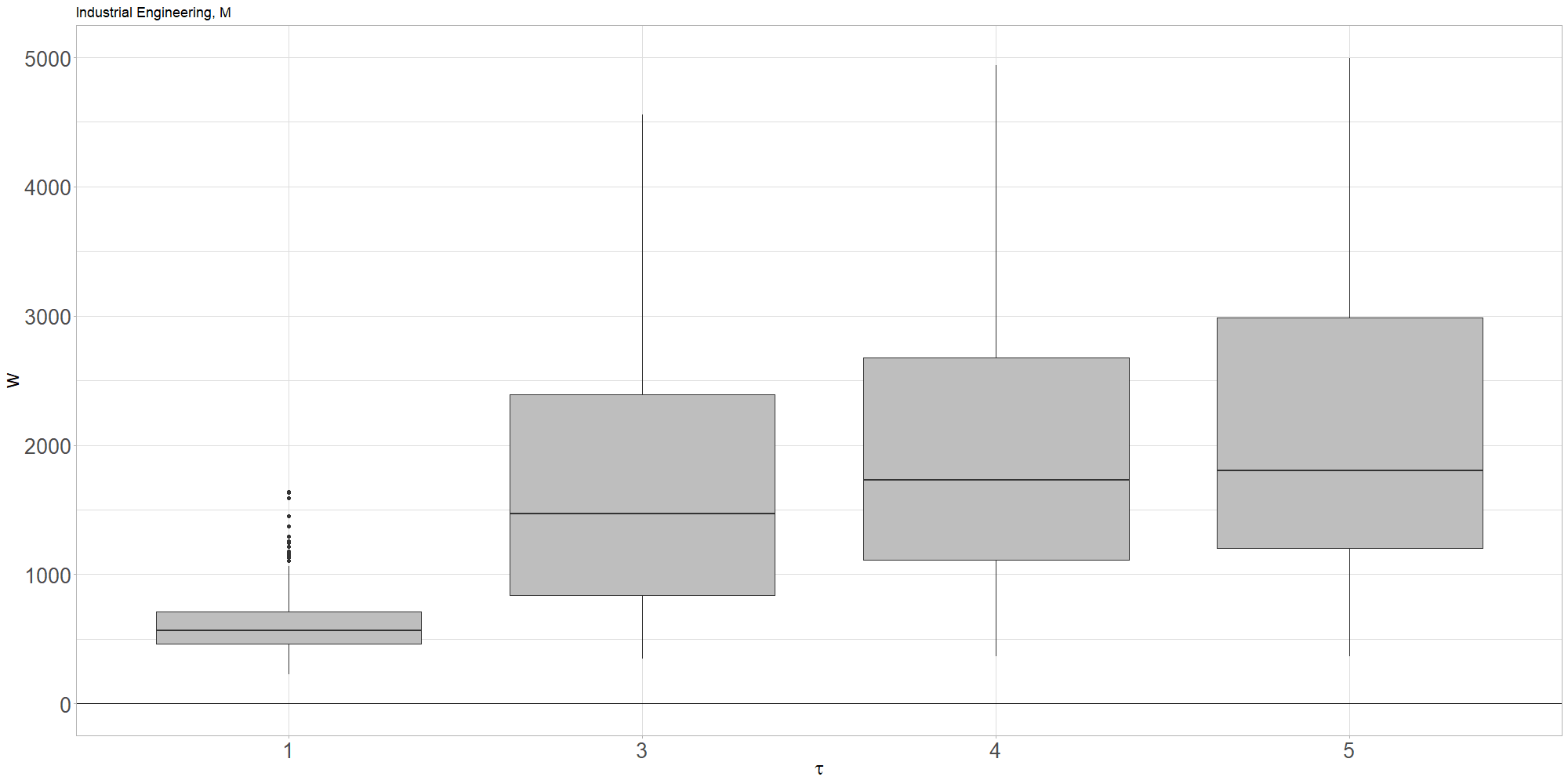}}
    \qquad
\subfloat[][Law and Legal sciences - females]{
    \includegraphics[width=.5\linewidth]{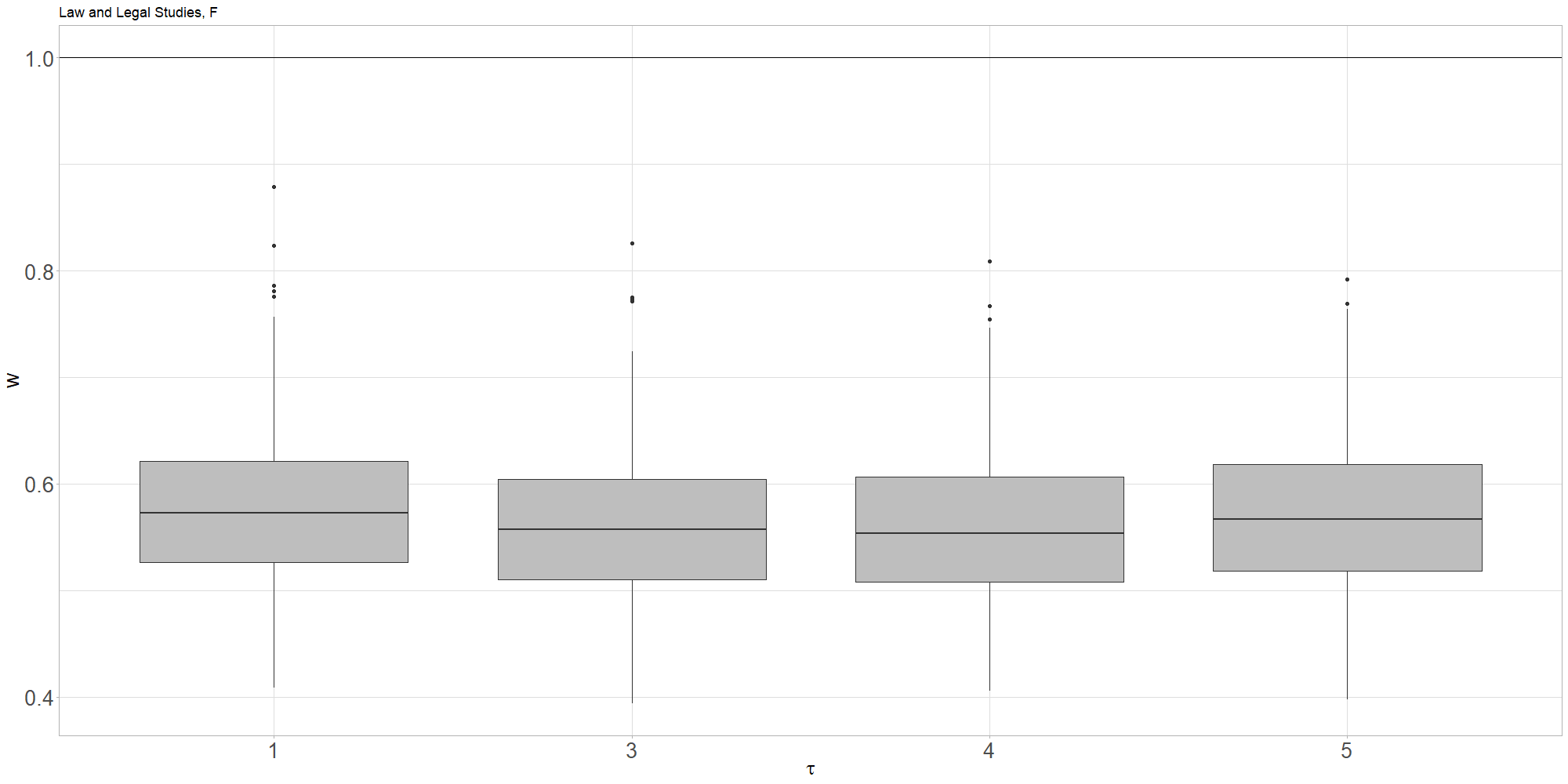}}
\subfloat[][Law and Legal sciences - males]{
    \includegraphics[width=.5\linewidth]{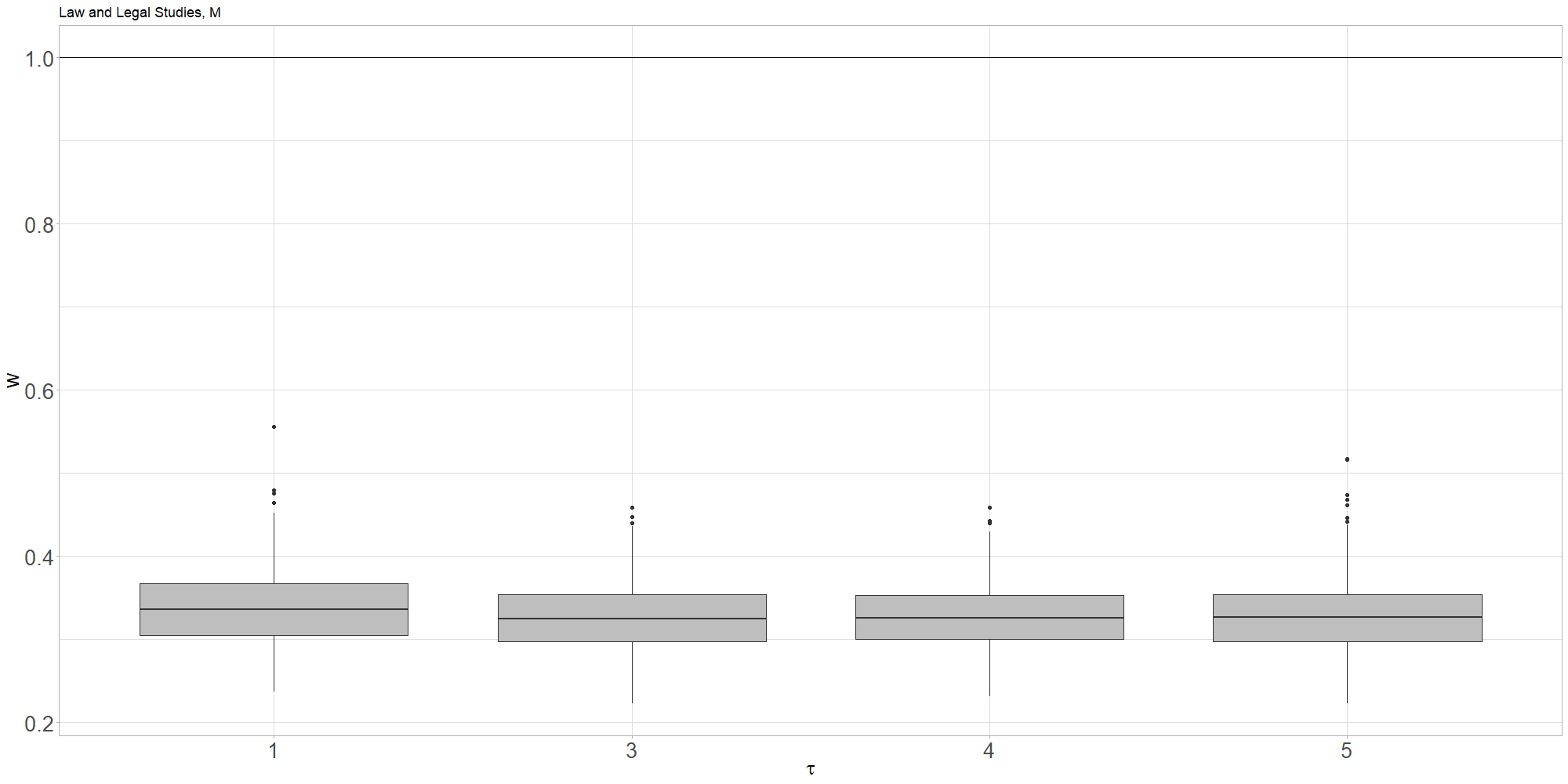}}
    \qquad
\subfloat[][Literature and Humanities - females]{
    \includegraphics[width=.5\linewidth]{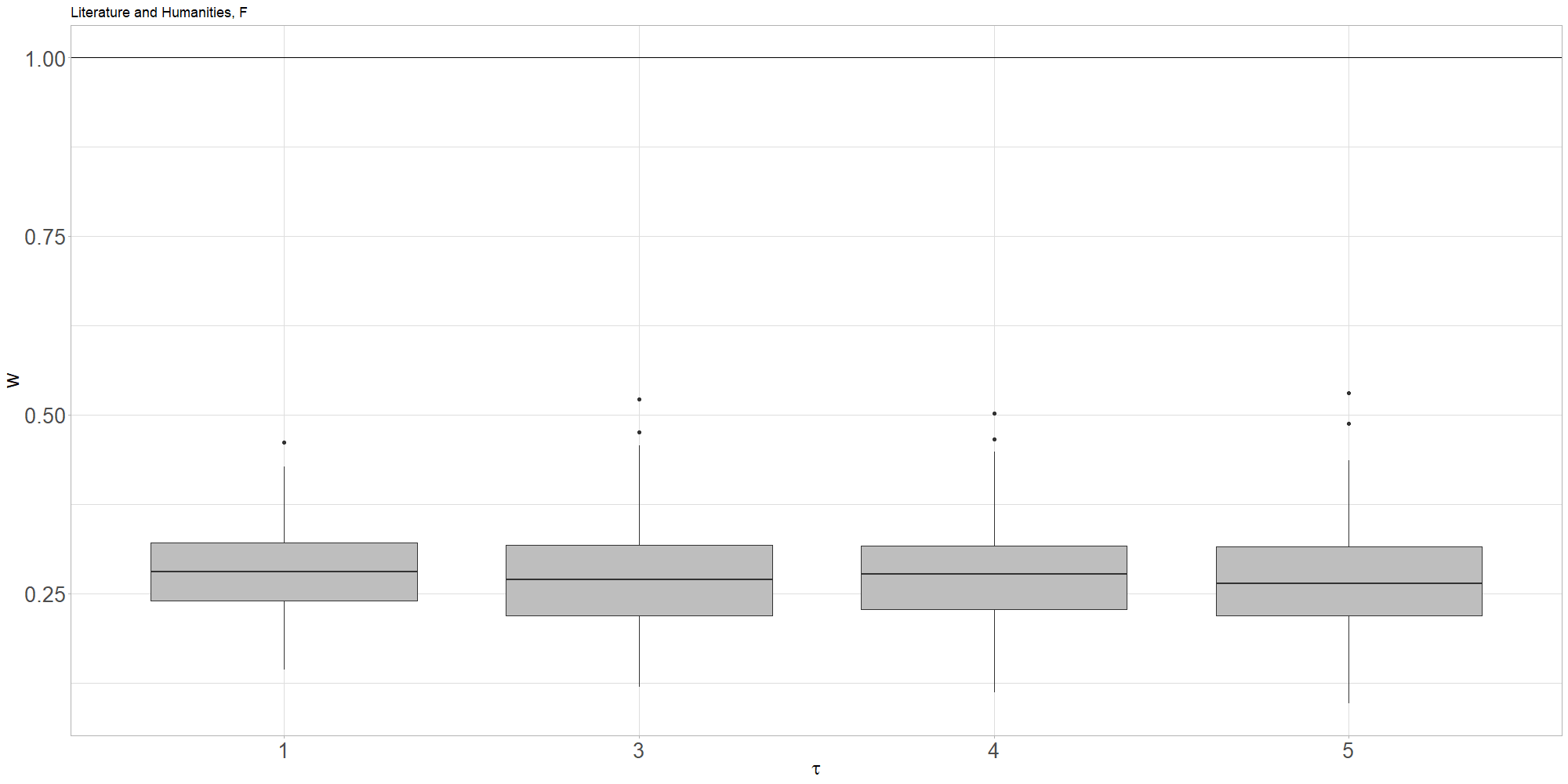}}
\subfloat[][Literature and Humanities - males]{
    \includegraphics[width=.5\linewidth]{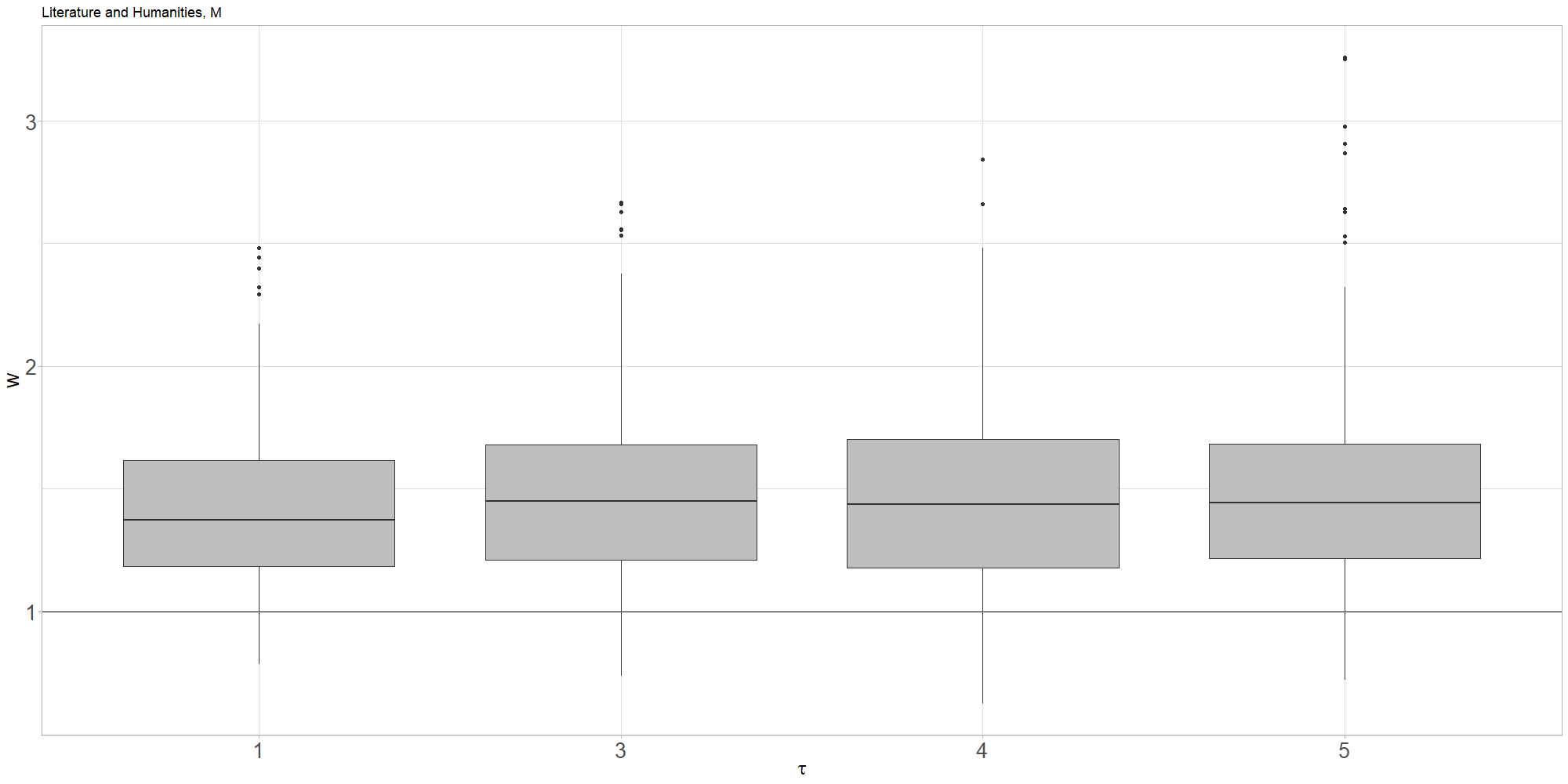}}
    \qquad
            \subfloat[][Medicine, Dentistry, Pharmacy - females]{
    \includegraphics[width=.5\linewidth]{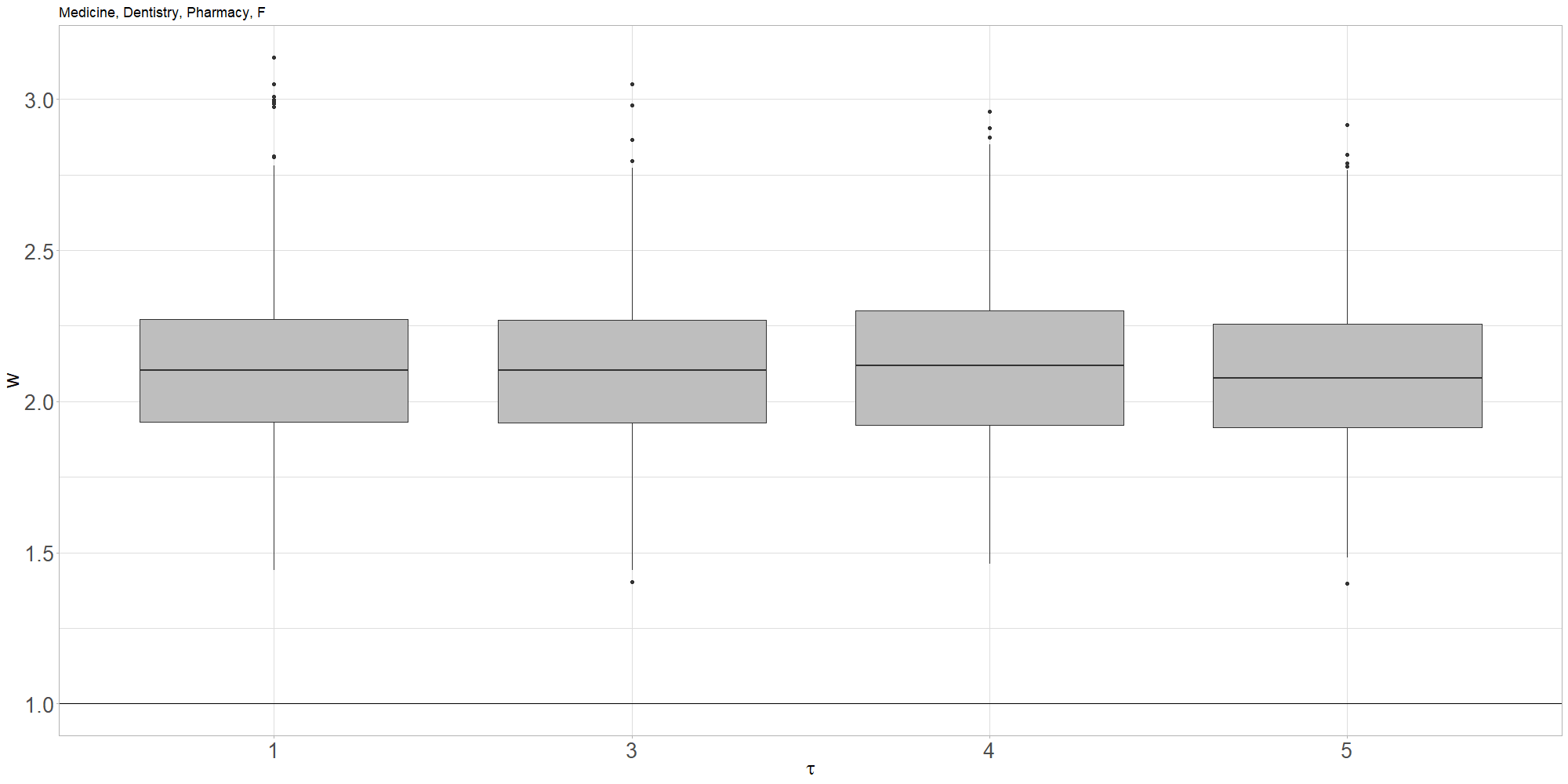}}
    \subfloat[][Medicine, Dentistry, Pharmacy - males]{
    \includegraphics[width=.5\linewidth]{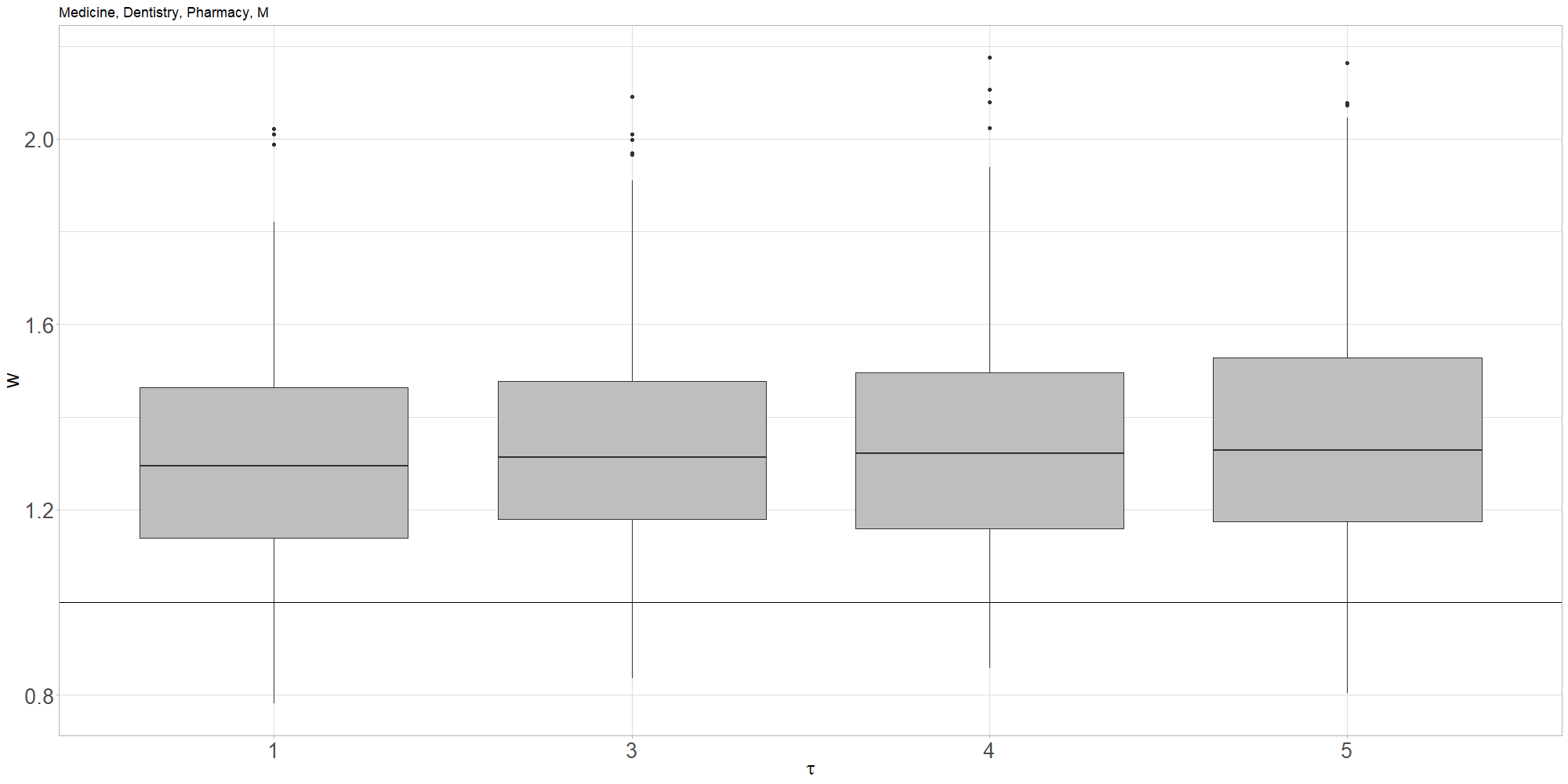}}
    \qquad

\end{figure}
\begin{figure}\ContinuedFloat
\centering
\qquad
\subfloat[][Political Science - females]{
    \includegraphics[width=.5\linewidth]{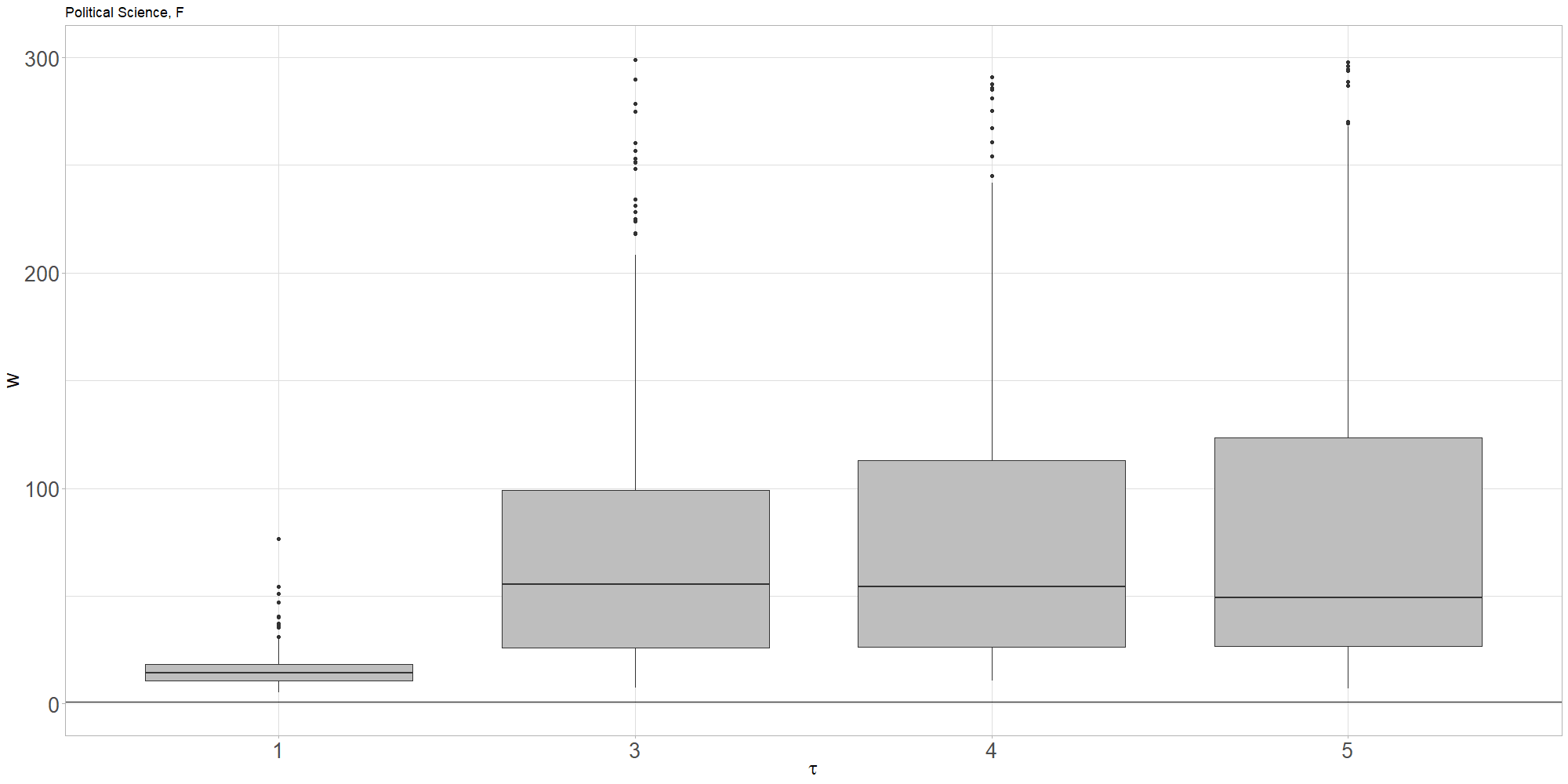}}
\subfloat[][Political Science - males]{
    \includegraphics[width=.5\linewidth]{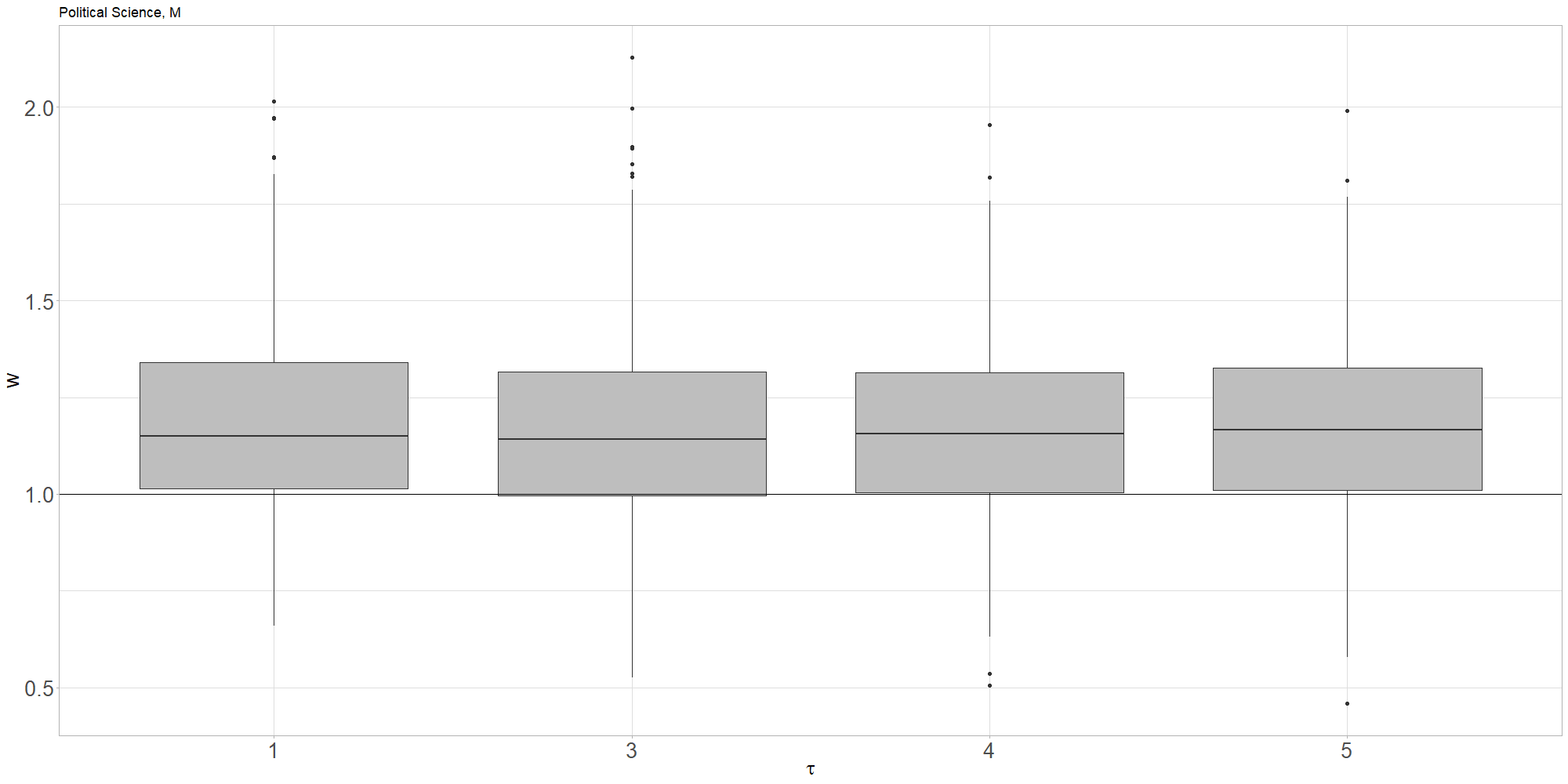}}
\qquad
\subfloat[][Science and IT - females]{
    \includegraphics[width=.5\linewidth]{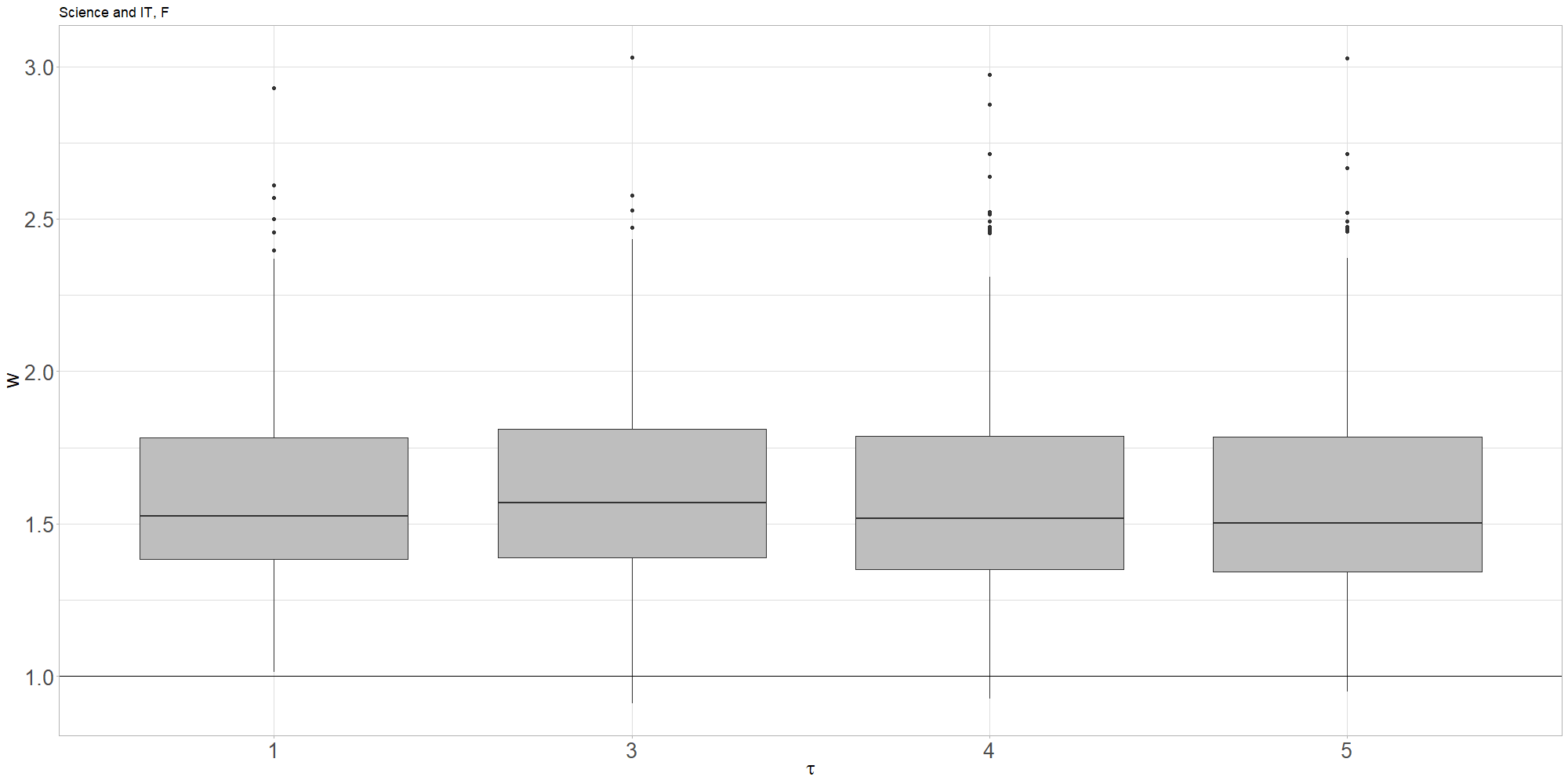}}
\subfloat[][Science and IT - males]{
    \includegraphics[width=.5\linewidth]{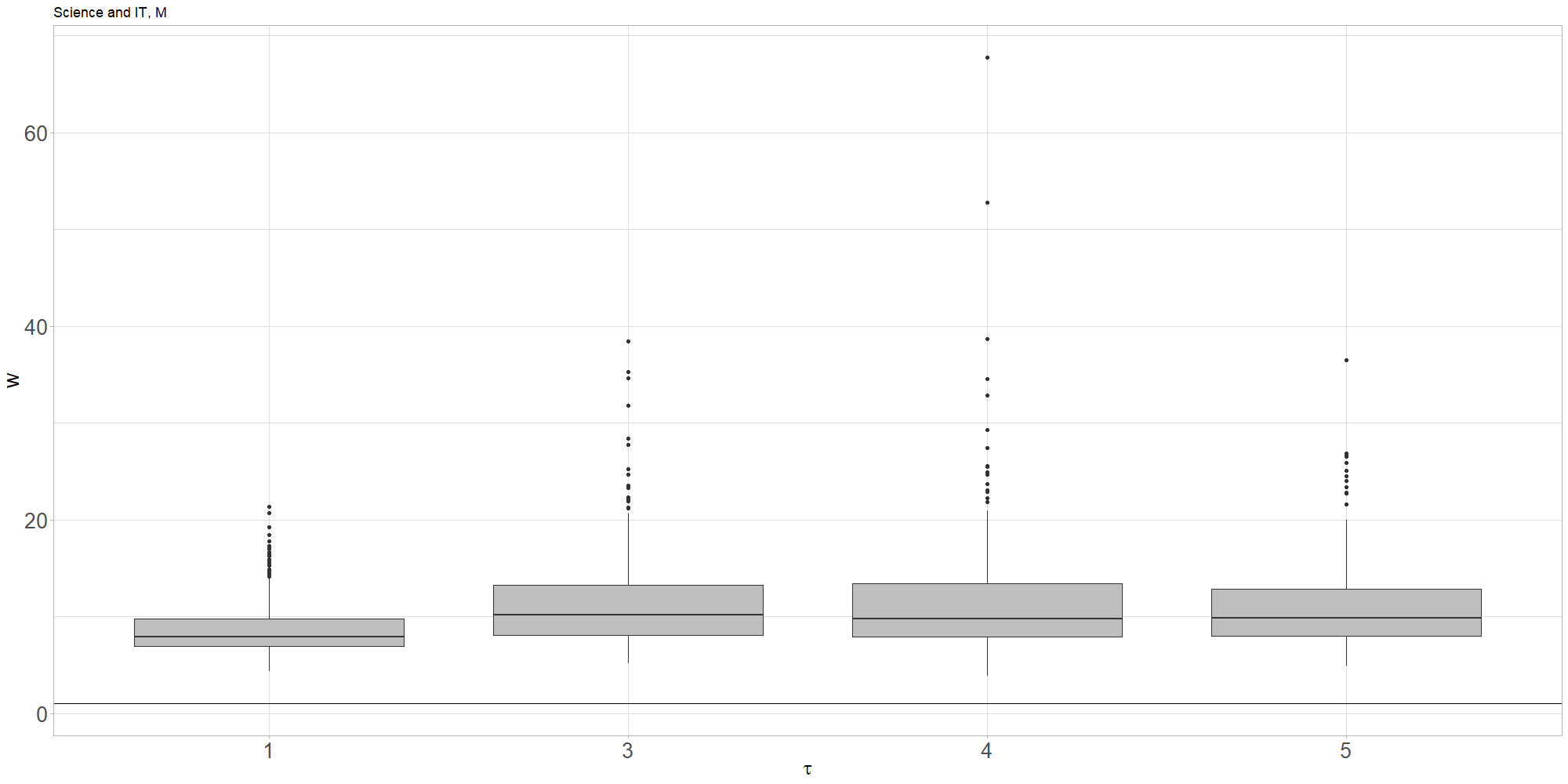}}
\caption{Posterior distribution of $w^{11}$ estimated at step (ii) for different values of $\tau^2$, by degree program and gender.}
\label{fig:sens1}
\end{figure}

\clearpage
\subsection{Robustness in time}
We test the sensitivity of the employment rates estimates for the 2012-2020 cohorts to different prior specifications.
In particular, we assume 
\begin{equation}
    \log(w_{hi}^{A\text{year}}) \sim \text{N}({q_{\alpha, hi}^{A2012}}^*, {\tau_{hi}^2}^*),
\end{equation} 
where ${q_{\alpha, hi}^{A2012}}^*$ is the posterior $\alpha$-th quantile of $w_{hi}^{A2012}$, with $\alpha = 0.25, 0.75$.
For the two specifications, figure \ref{fig:sens} shows the $95\%$ highest posterior density intervals of the employment rates.
The results are robust to the different specifications.
For those groups whose Almalaurea rates were included in the intervals only for some years, namely ``Literature and Humanities" (males) and ``Science and IT" (females), now the interval estimated using the prior centered on the first quartile always covers them. 
Using the same prior, also the Almalaurea rates for males in ``Medicine, Dentistry and Pharmacy" are covered by the intervals.

\begin{figure}[!htbp]
    \centering
\subfloat[][Agriculture and Forestry, Veterinary - females]{
    \includegraphics[width=.5\linewidth]{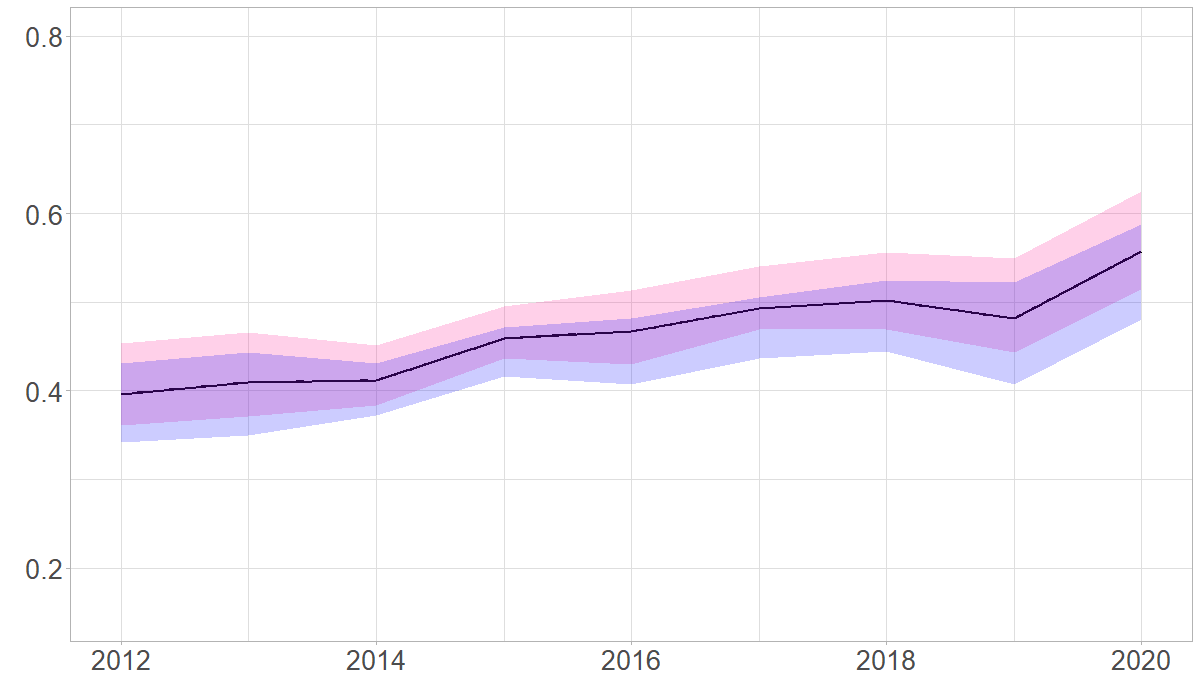}}
\subfloat[][Agriculture and Forestry, Veterinary - males]{
    \includegraphics[width=.5\linewidth]{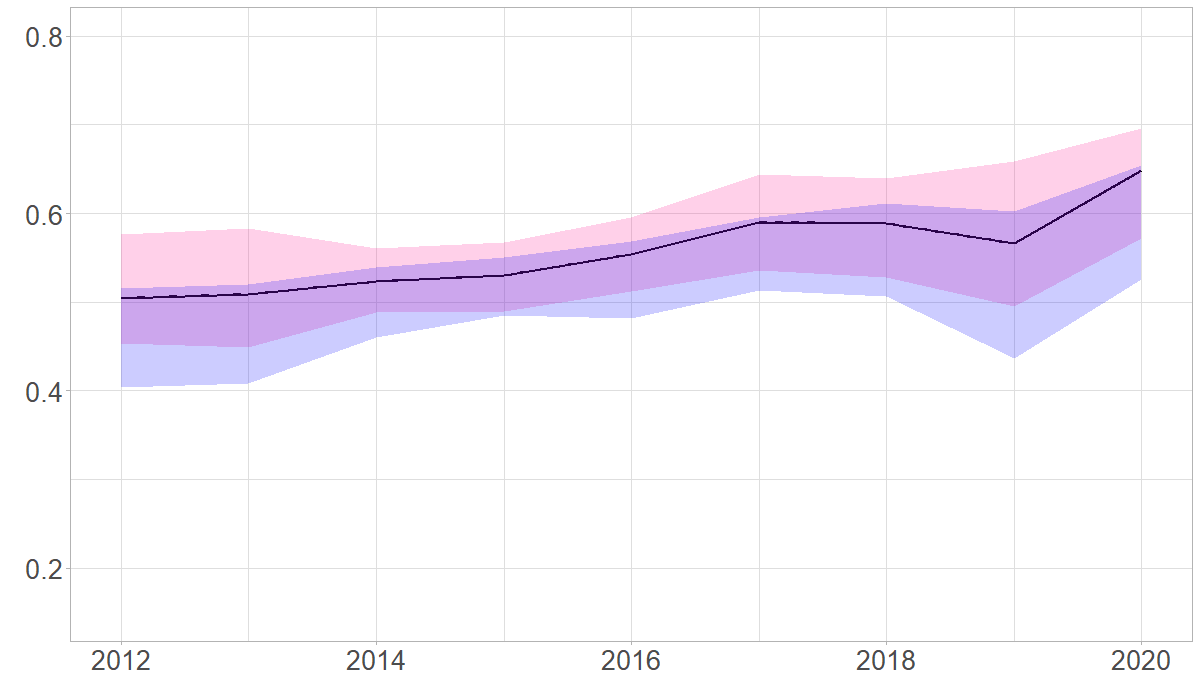}}
\qquad
\subfloat[][Architecture and Engineering - females]{
    \includegraphics[width=.5\linewidth]{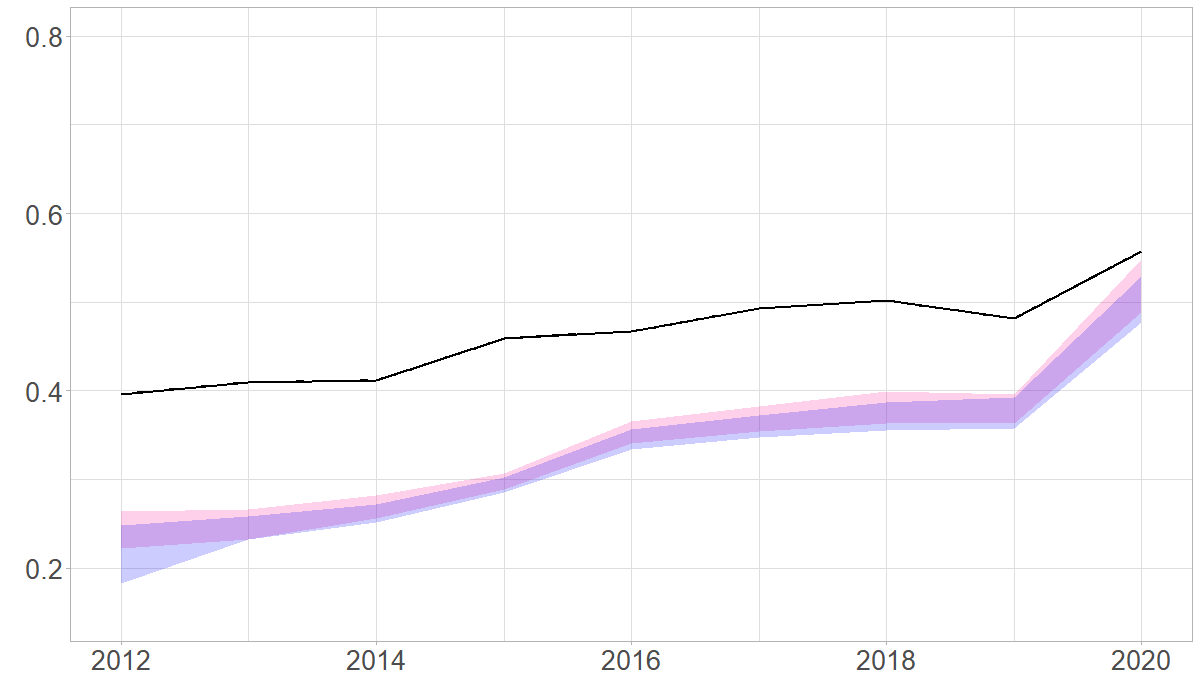}}
\subfloat[][Architecture and Engineering - males]{
    \includegraphics[width=.5\linewidth]{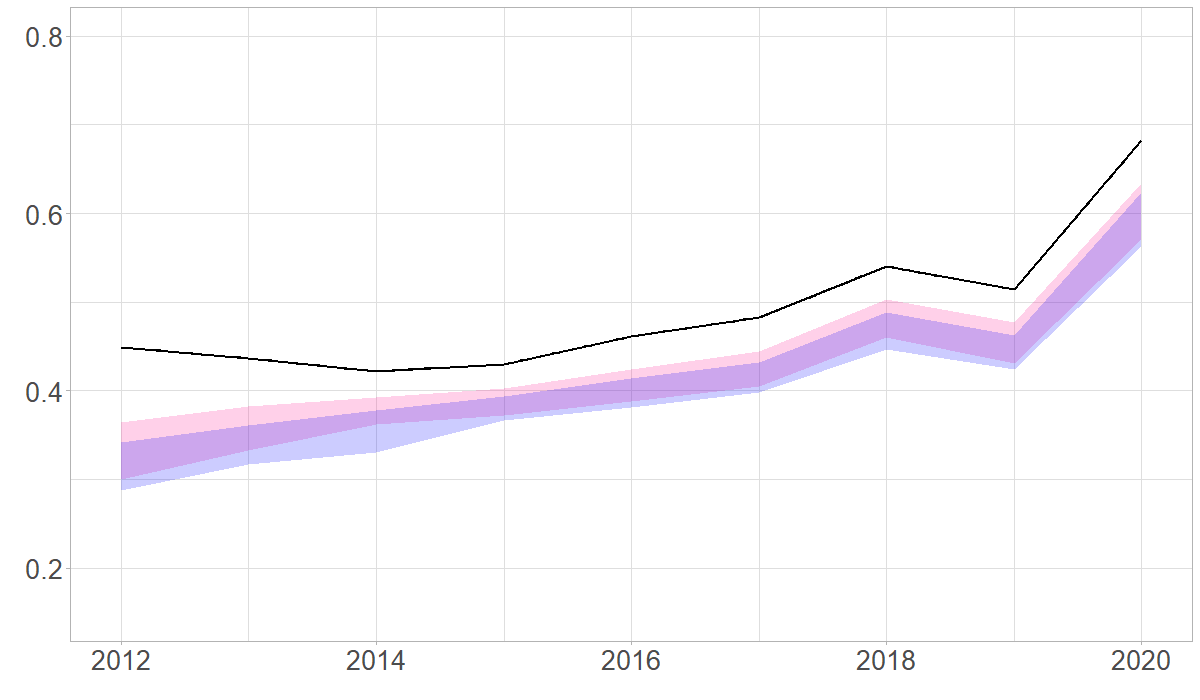}}
\qquad
\subfloat[][B\&A, Economics, Finance - females]{
    \includegraphics[width=.5\linewidth]{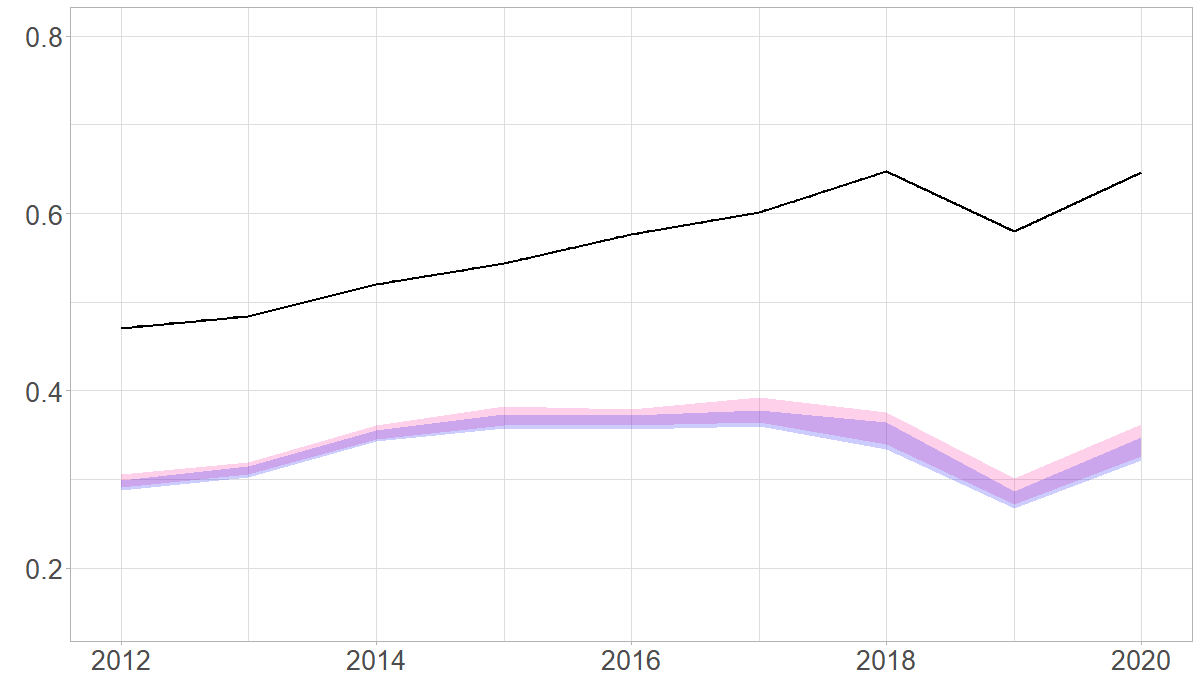}}
\subfloat[][B\&A, Economics, Finance - males]{
    \includegraphics[width=.5\linewidth]{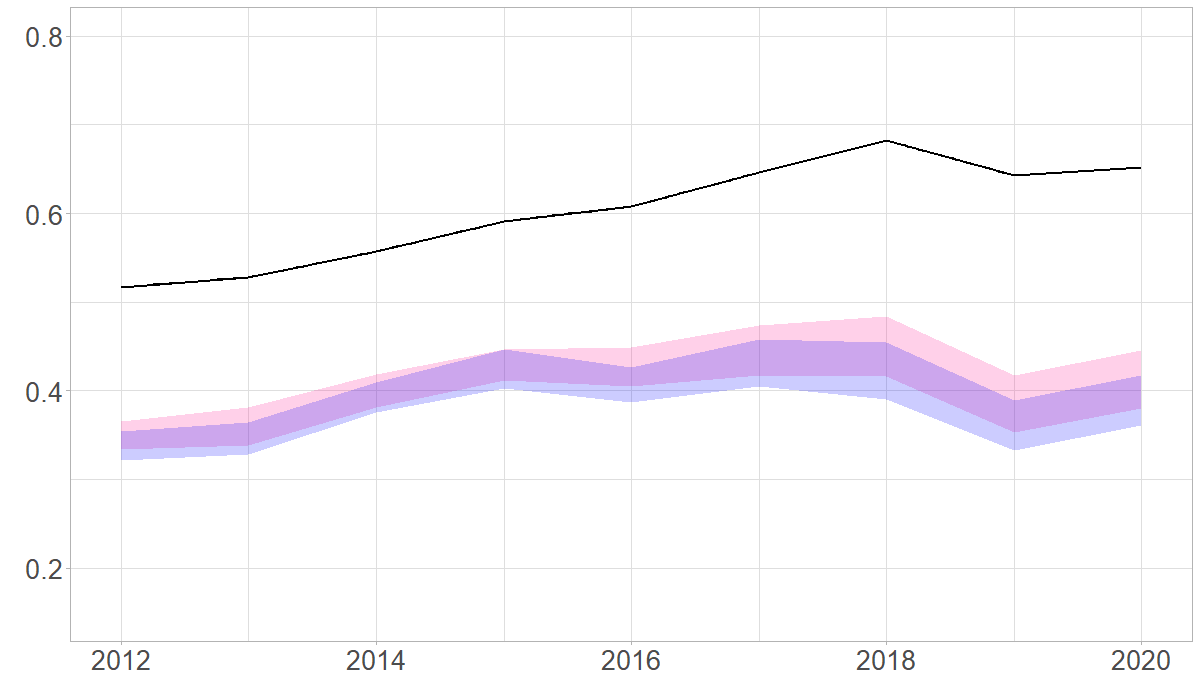}}
\end{figure}
\begin{figure}\ContinuedFloat
\centering
\subfloat[][Communication and Publishing - females]{
    \includegraphics[width=.5\linewidth]{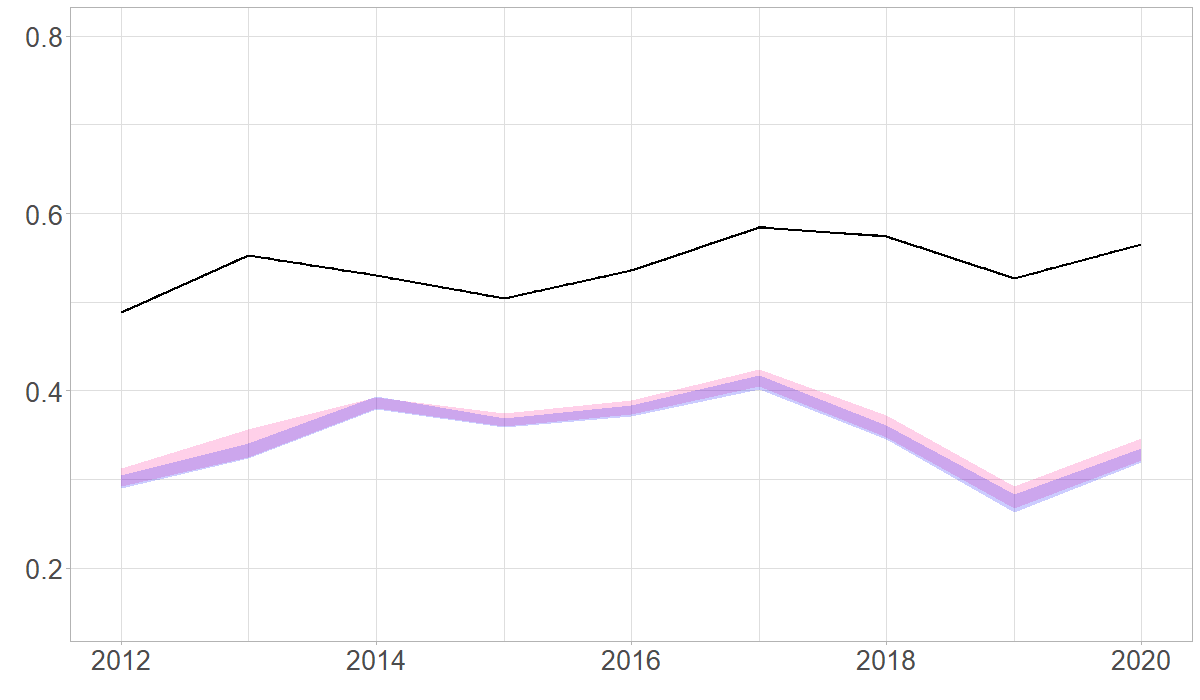}}
\subfloat[][Communication and Publishing - males]{
    \includegraphics[width=.5\linewidth]{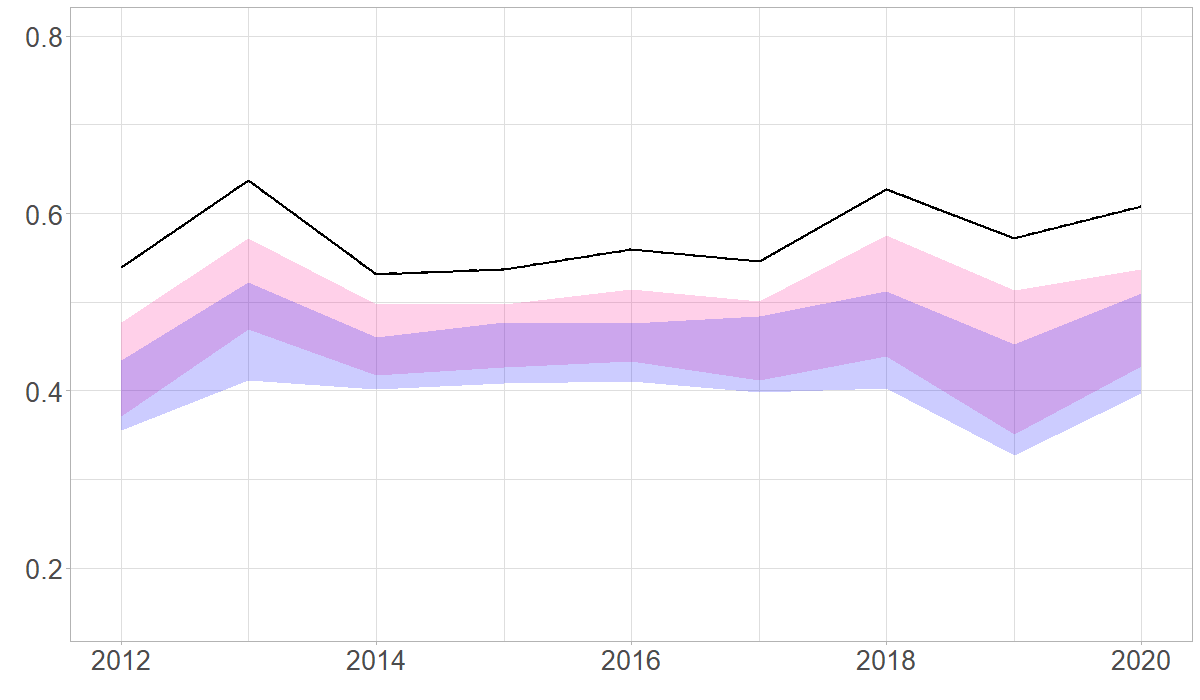}}
\qquad
\subfloat[][Industrial and Information Engineering - females]{
    \includegraphics[width=.5\linewidth]{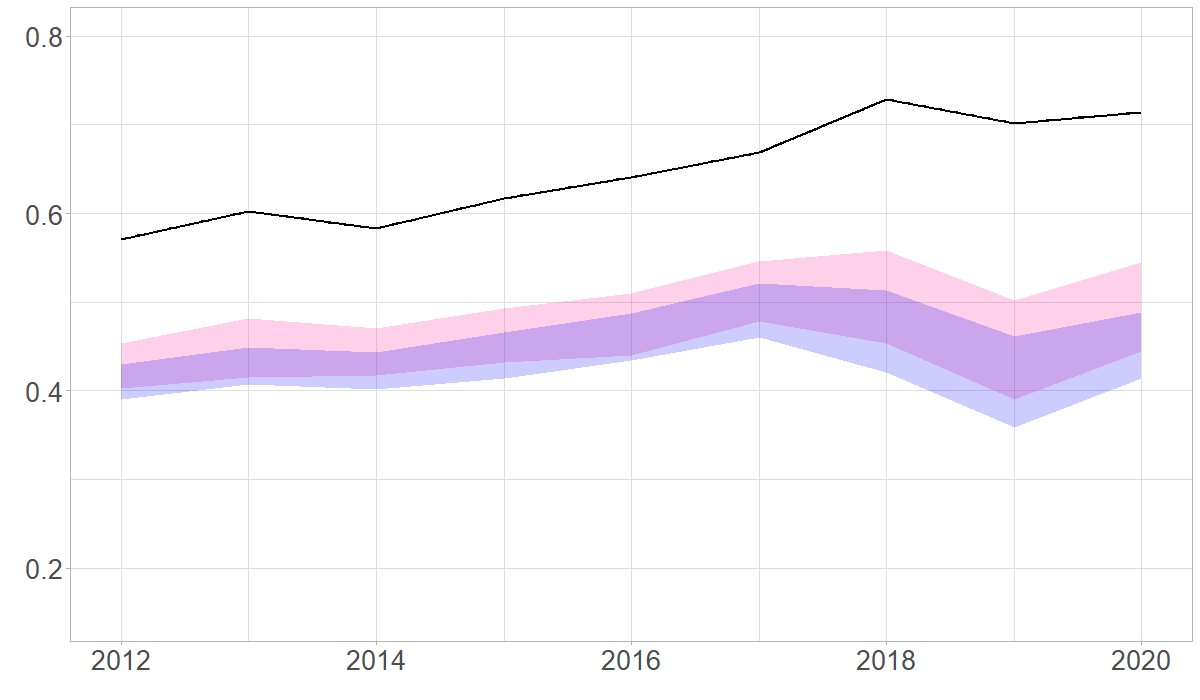}}
\subfloat[][Industrial and Information Engineering - males]{
    \includegraphics[width=.5\linewidth]{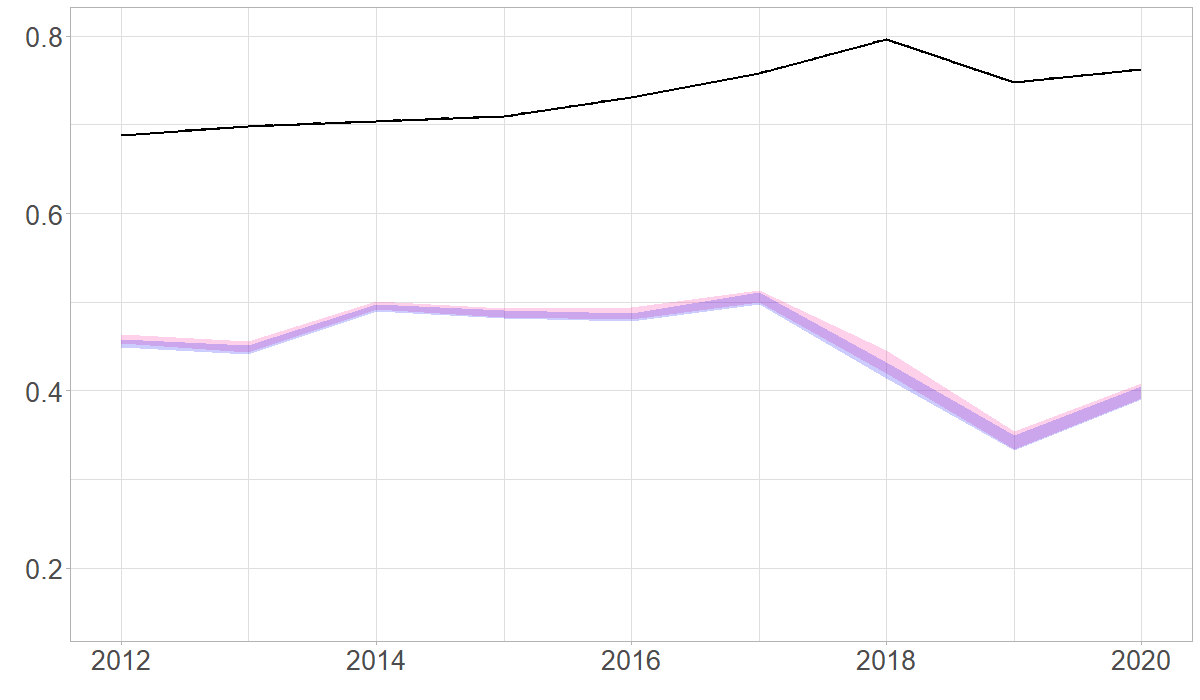}}
\qquad
\subfloat[][Law and Legal sciences - females]{
    \includegraphics[width=.5\linewidth]{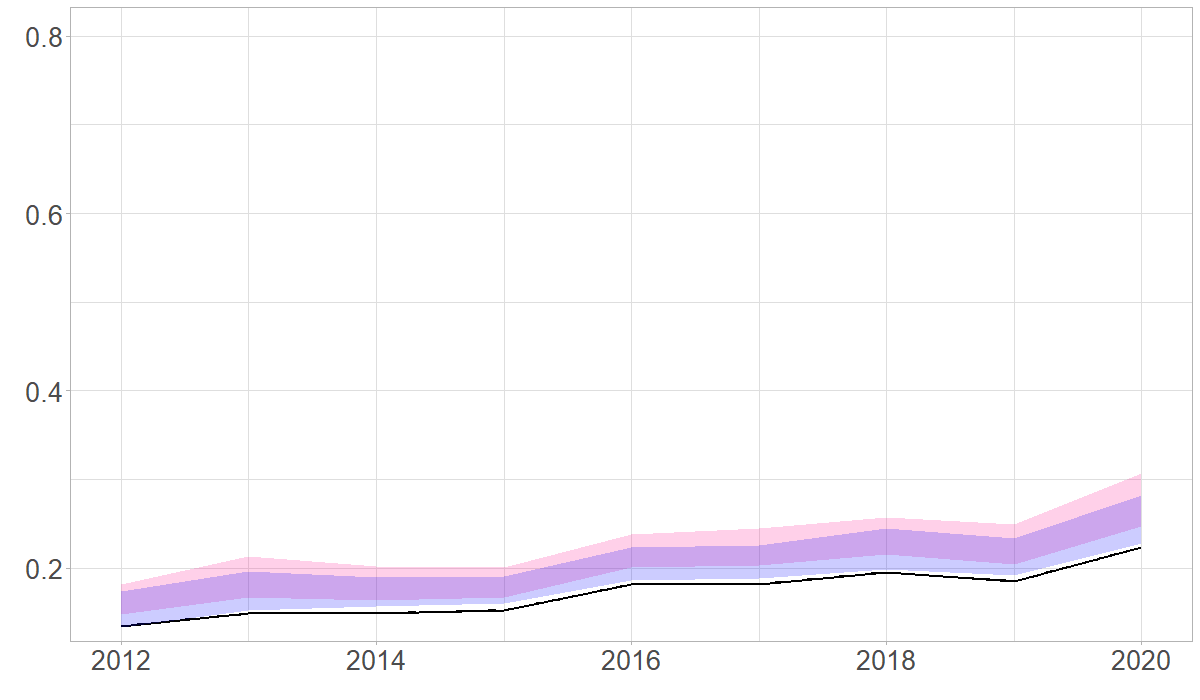}}
\subfloat[][Law and Legal sciences - males]{
    \includegraphics[width=.5\linewidth]{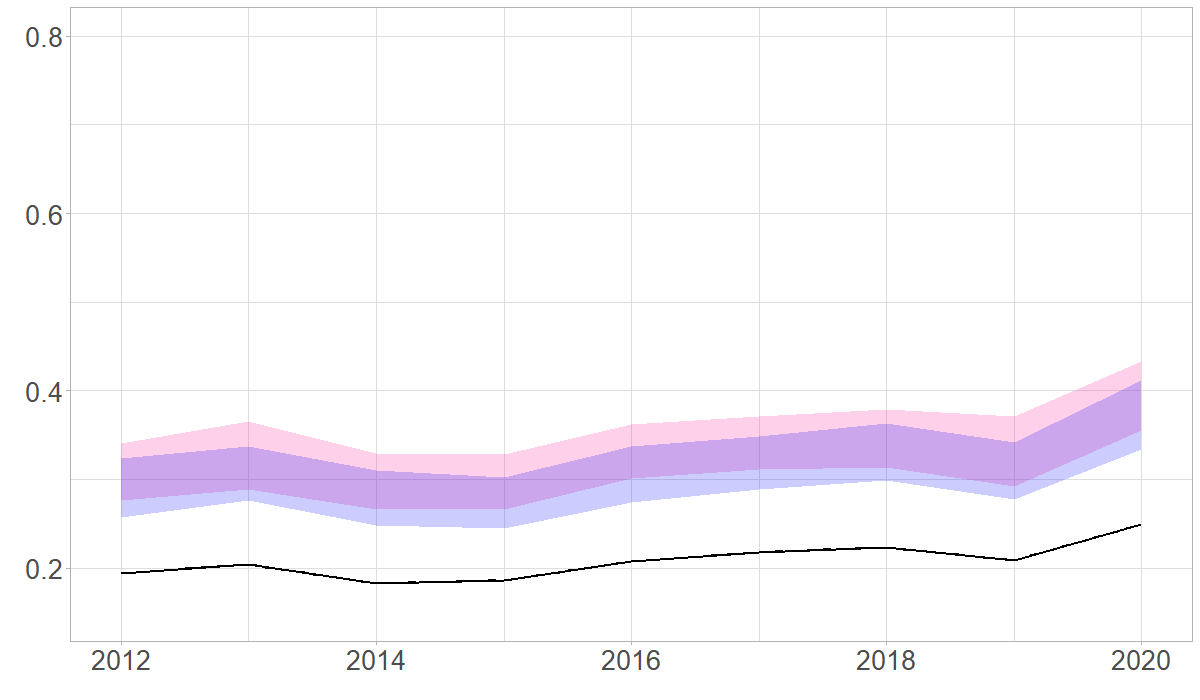}}
\qquad
\subfloat[][Literature and Humanities - females]{
    \includegraphics[width=.5\linewidth]{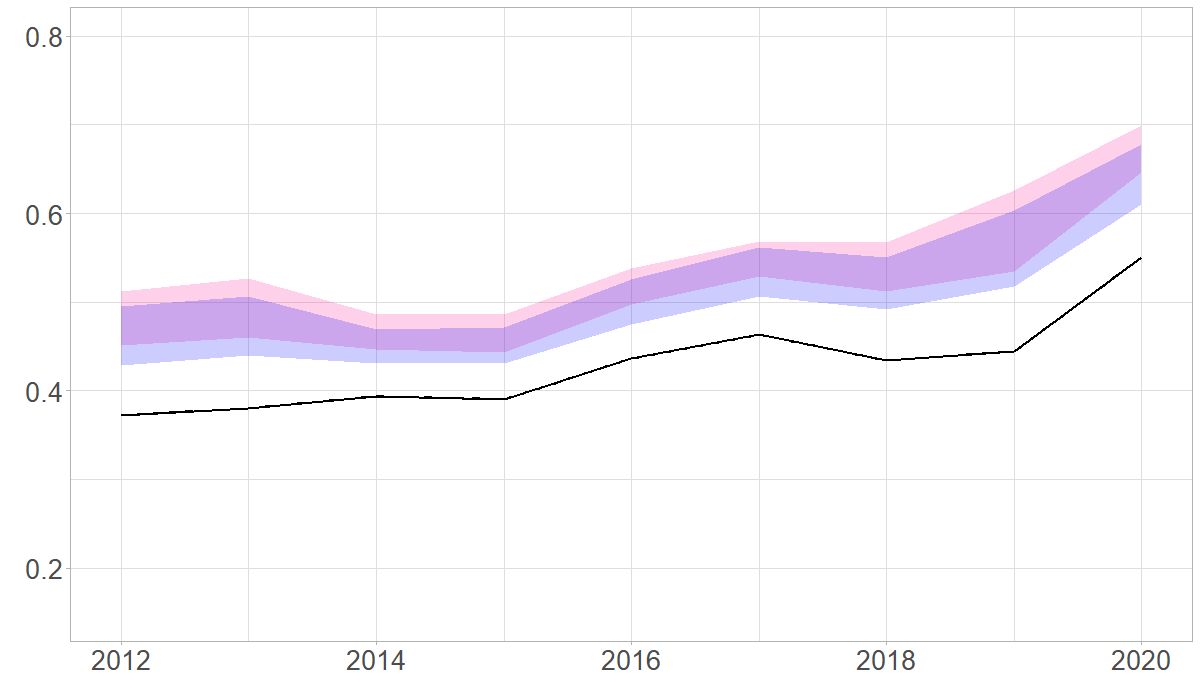}}
\subfloat[][Literature and Humanities - males]{
    \includegraphics[width=.5\linewidth]{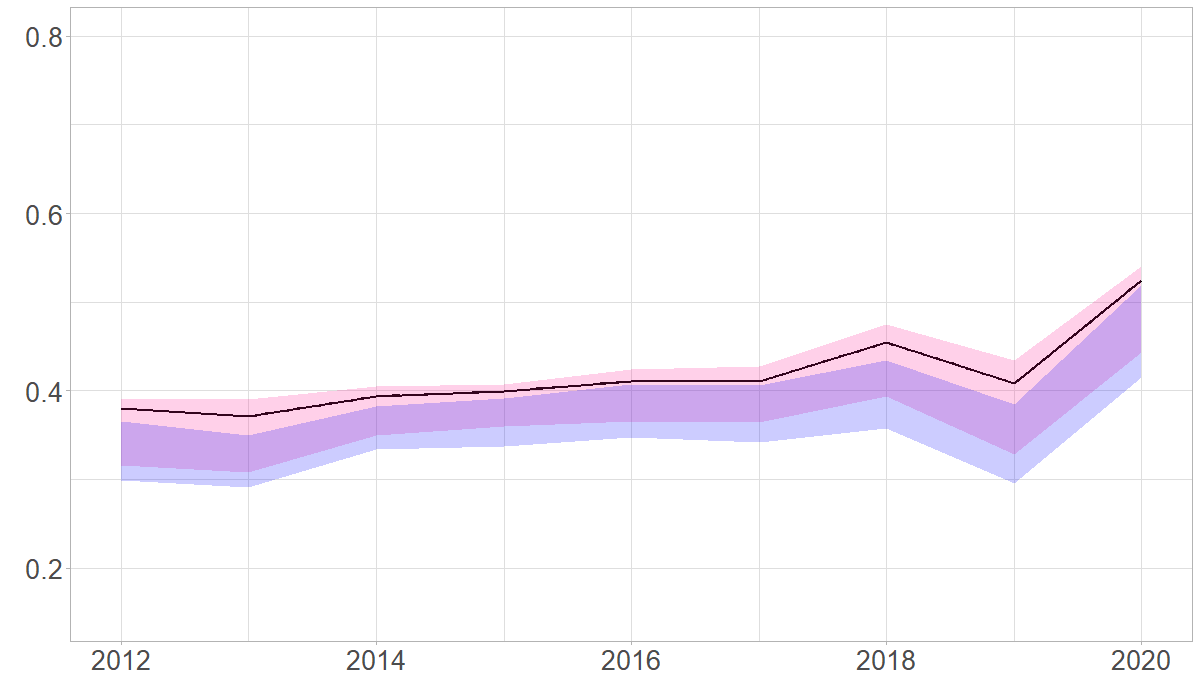}}
\end{figure}
\begin{figure}\ContinuedFloat
\centering
\subfloat[][Medicine, Dentistry, Pharmacy - females]{
    \includegraphics[width=.5\linewidth]{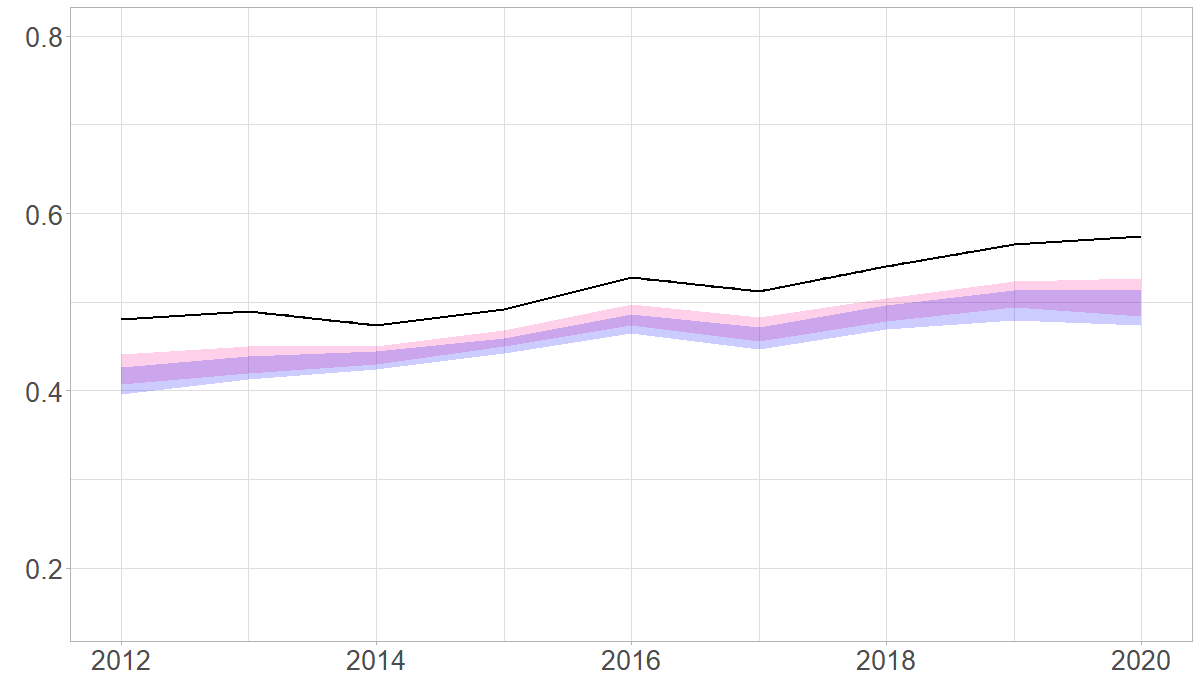}}
\subfloat[][Medicine, Dentistry, Pharmacy - males]{
    \includegraphics[width=.5\linewidth]{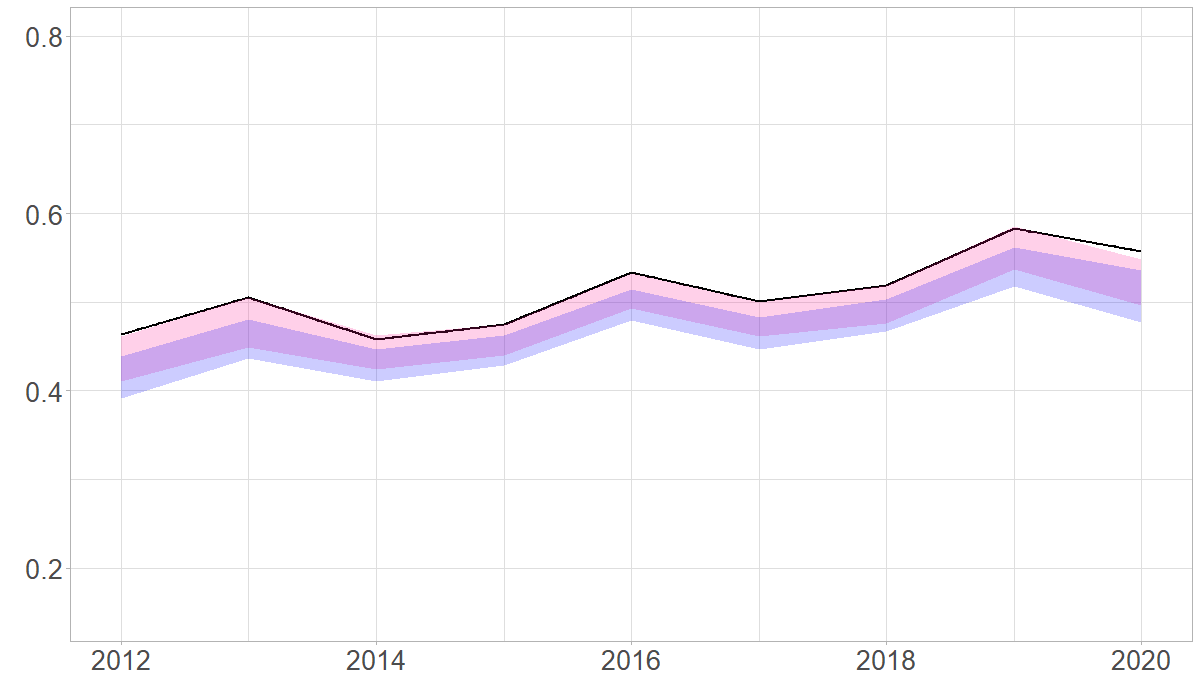}}
\qquad
\subfloat[][Political Science - females]{
    \includegraphics[width=.5\linewidth]{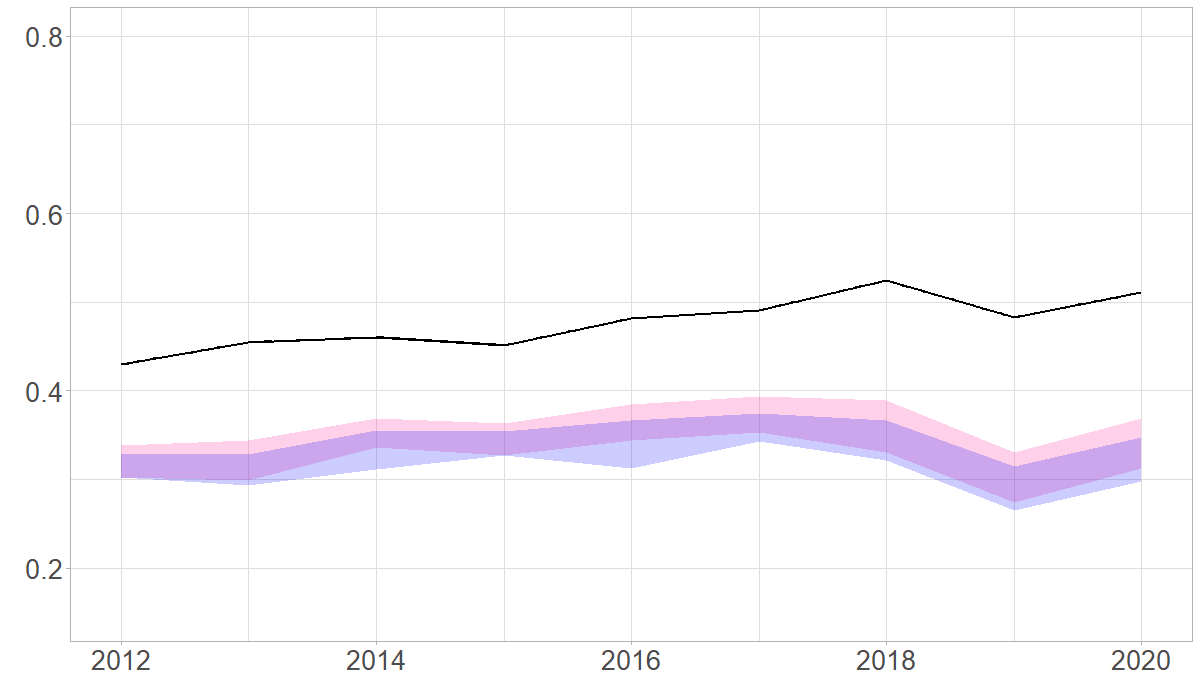}}
\subfloat[][Political Science - males]{
    \includegraphics[width=.5\linewidth]{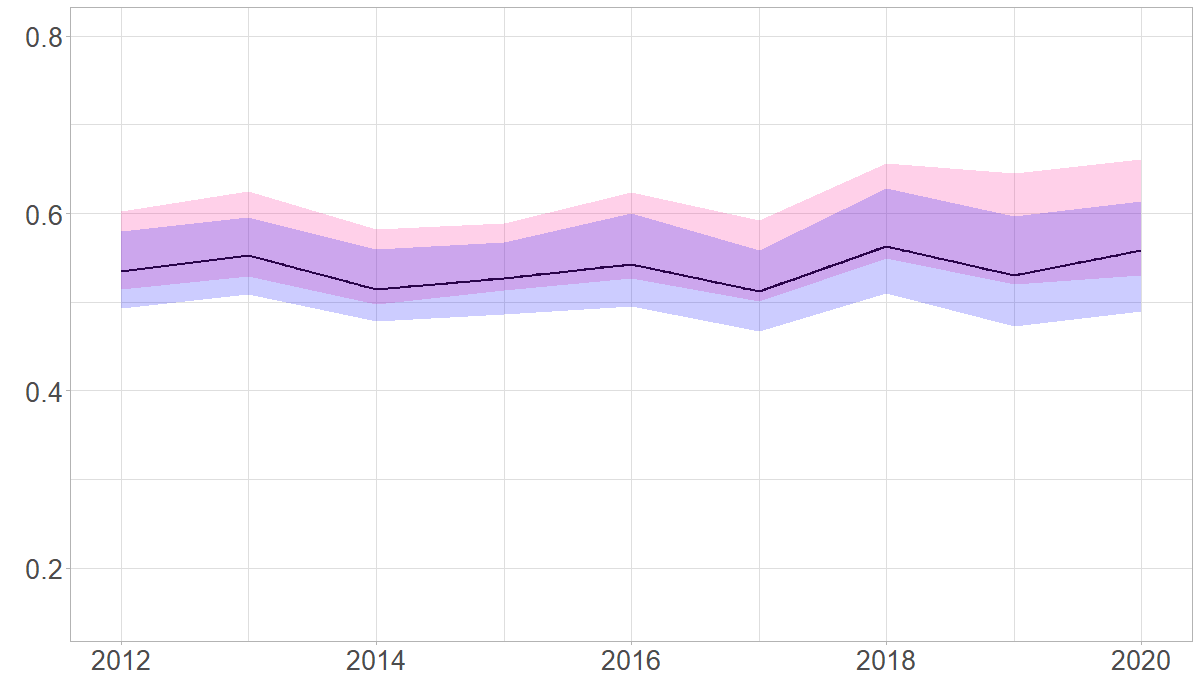}}
\qquad
\subfloat[][Science and IT - females]{
    \includegraphics[width=.5\linewidth]{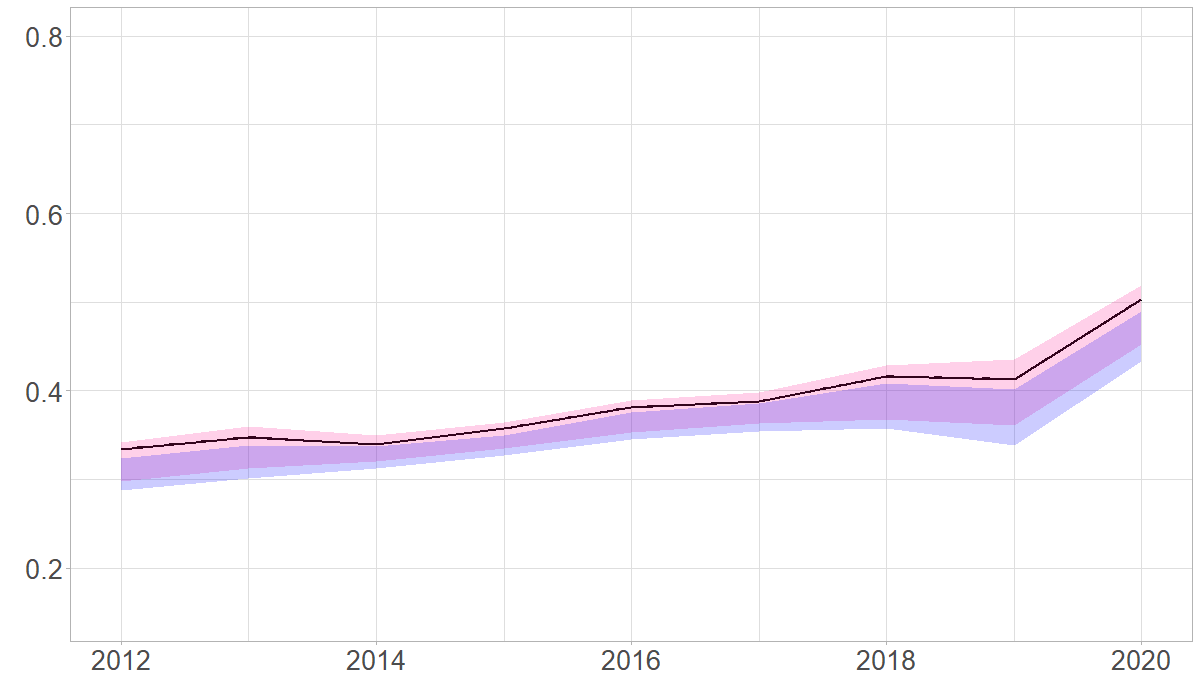}}
\subfloat[][Science and IT - males]{
    \includegraphics[width=.5\linewidth]{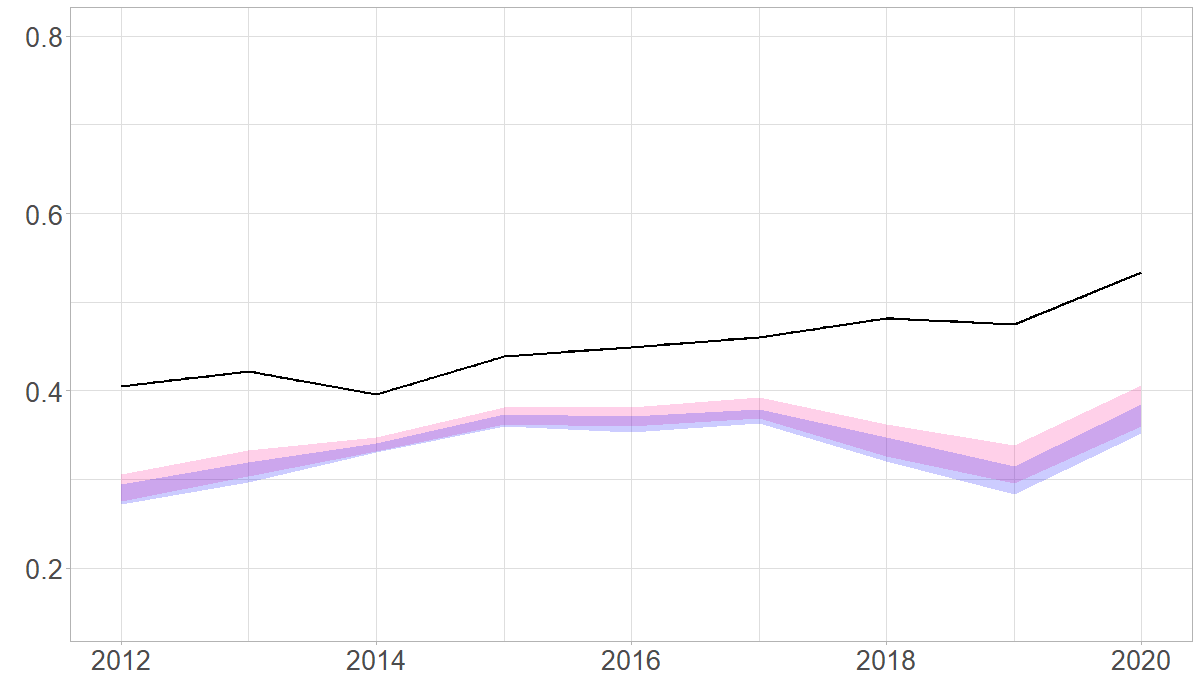}}
\caption{95\% highest posterior density intervals of the employment rates of the 2012-2020 cohorts, by degree program and gender, versus the employment rates computed using raw Almalaurea data (solid line). 
Results obtained with the log odds prior centered on the first (pink) and third (blue) quartiles of the $w^{A2012}_{hi}$ posterior.}
\label{fig:sens}
\end{figure}

\section*{References}
\begin{itemize}
    \item[[1]] G. Clart\'e, C. P. Robert, R. J. Ryder, and J. Stoehr. Componentwise approximate Bayesian computation via Gibbs-like steps. \underline{Biometrika}, 108(3):591–607, 2021.
    \item[[2]] A. Fog. Sampling methods for Wallenius' and Fisher's noncentral hypergeometric distributions. \underline{Communications in Statistics—Simulation and Computation}, 37(2):241–257, 2008.
    \item[[3]] C. Grazian, F. Leisen, and B. Liseo. Modelling preference data with the Wallenius distribution. \underline{Journal of the Royal Statistical Society: Series A (Statistics in Society)}, 182(2):541–558, 2019.
    \item[[4]] G. Karabatsos and F. Leisen. An approximate likelihood perspective on ABC methods. \underline{Statistics Surveys}, 12:66–104, 2018
    \item[[5]] J. K. Pritchard, M. T. Seielstad, A. Perez-Lezaun, and M. W. Feldman. Population growth of human Y chromosomes: a study of Y chromosome microsatellites. \underline{Molecular biology} \underline{and evolution}, 16(12):1791–1798, 1999.
    \item[[6]] S. A. Sisson, Y. Fan, and M. Beaumont. \underline{Handbook of approximate Bayesian computation}. CRC Press, 2018.
    \item[[7]] S. Tavar\'e, D. J. Balding, R. C. Griffiths, and P. Donnelly. Inferring coalescence times from DNA sequence data. Genetics, 145(2):505–518, 1997.
\end{itemize}
\end{document}